# A kind of model of quantum electrodynamics with nonlocal interaction

T. Mei

(Department of Journal, Central China Normal University, Wuhan, Hubei PRO, People's Republic of China

E-Mail:   meitao@mail.ccnu.edu.cn     meitaowh@public.wh.hb.cn )

**Abstract:** We present a kind of model of quantum electrodynamics with nonlocal interaction, all the action and the equations of motion of charged particle and electromagnetic field are given. The main characteristics of the theory are: the model obeys the action principle; free charged particle and free electromagnetic field obey the Dirac equation and the Maxwell equation of free fields, respectively; for the case of interaction, both the equations of motion of charged particle and electromagnetic field lead to the normal current conservation naturally; the theory is Lorentz invariant and gauge invariant, and returns to the conventional local QED under appropriate limit conditions. Taking advantage of the Yang-Feldman equations and the Lehmann-Symanzik-Zimmermann formalism, we establish the corresponding quantum theory.

There are a large number of literatures on nonlocal field theory. In this paper, we present a kind of model of quantum electrodynamics with nonlocal interaction. The model shows that the gauge principle holds not only for local interaction but also for nonlocal interaction for the system of charged particle and electromagnetic field.

All symbols and conventions of this paper follow Ref.[1], for example, $g^{\alpha\beta} = \mathrm{diag}\,(+1, -1, -1, -1)$.

## 1  A scalar function $f(x)$ and two vector functions $U^\mu(x)$ and $V^\mu(x)$

We take into account a scalar function $f(x)$ and two vector functions $U^\mu(x)$ and $V^\mu(x)$. For $f(x)$ we ask that

① The corresponding inverse function $\widetilde{f}(x)$ exists and satisfies

$$\int \mathrm{d}^4 z\, \widetilde{f}(x-z) f(y-z) = \delta^4(x-y). \tag{1-1}$$

If the Fourier transform of $f(x)$ exists:

$$f(x) = \int \frac{d^4k}{(2\pi)^4} \hat{f}(k) e^{ik \cdot x}, \tag{1-2}$$

then it is easy to prove that $\widetilde{f}(x) = \int \frac{d^4k}{(2\pi)^4} \frac{1}{\hat{f}(k)} e^{-ik \cdot x}$ satisfies (1-1).

② Under some appropriate limit conditions, $f(x)$ satisfies

$$f(x) \to \delta^4(x). \tag{1-3}$$

For $U^\mu(x)$ and $V^\mu(x)$ we introduce

$$U^\mu_\nu(x) \equiv \frac{\partial U^\mu(x)}{\partial x^\nu}, \quad V^\mu_\nu(x) \equiv \frac{\partial V^\mu(x)}{\partial x^\nu}, \tag{1-4}$$

It is obvious that the following formulas hold:

$$\frac{\partial U^\alpha_\beta(x)}{\partial x^\gamma} = \frac{\partial^2 U^\alpha(x)}{\partial x^\beta \partial x^\gamma} = \frac{\partial U^\alpha_\gamma(x)}{\partial x^\beta}, \quad \frac{\partial V^\alpha_\beta(x)}{\partial x^\gamma} = \frac{\partial^2 V^\alpha(x)}{\partial x^\beta \partial x^\gamma} = \frac{\partial V^\alpha_\gamma(x)}{\partial x^\beta}. \tag{1-5}$$

For $U^\mu(x)$ and $V^\mu(x)$ we ask that

① Both the two corresponding determinants satisfy

$$U(x) \equiv \left\| U^\mu_\nu(x) \right\| \neq 0, \quad V(x) \equiv \left\| V^\mu_\nu(x) \right\| \neq 0; \tag{1-6}$$

Hence, the corresponding inverse functions $\widetilde{U}^\mu_\nu(x) \equiv \frac{\partial x^\mu}{\partial U^\nu(x)}$ and $\widetilde{V}^\mu_\nu(x) \equiv \frac{\partial x^\mu}{\partial V^\nu(x)}$ exist and satisfy

$$U^\mu_\lambda(x) \widetilde{U}^\lambda_\nu(x) = \delta^\mu_\nu, \quad U^\lambda_\nu(x) \widetilde{U}^\mu_\lambda(x) = \delta^\mu_\nu; \quad V^\mu_\lambda(x) \widetilde{V}^\lambda_\nu(x) = \delta^\mu_\nu, \quad V^\lambda_\nu(x) \widetilde{V}^\mu_\lambda(x) = \delta^\mu_\nu. \tag{1-7}$$

② Under some appropriate limit conditions, $U^\mu(x)$ and $V^\mu(x)$ satisfies

$$U^\mu(x) \to x^\mu, \quad V^\mu(x) \to x^\mu. \tag{1-8}$$

In sum, what we make demands on a scalar function $f(x)$ and two vector functions $U^\mu(x)$ and $V^\mu(x)$ are (1-1), (1-3), (1-6) and (1-8). These demands are not harsh and there are many functions satisfying these conditions. For example, the function

$$\int \frac{d^4k}{(2\pi)^4} e^{ik \cdot x} \left( a_1 + \frac{a_2}{1 + a_3 (k^2)^2} + a_4 \exp\left(-a_5 (k^2)^2\right) + \frac{a_6}{1 + a_7 (k_\lambda x^\lambda)^2} + a_8 \exp\left(-a_9 (k_\lambda x^\lambda)^2\right) \right)$$

in which all $a_i$ ($i = 1, 2, \cdots, 9$) are constants satisfies (1-1) and (1-3) under the limit $a_1 \to 1$, $a_2 \to 0$, $a_4 \to 0$, $a_6 \to 0$ and $a_8 \to 0$, or under the limit $a_3 \to 0$, $a_5 \to 0$, $a_7 \to 0$, $a_9 \to 0$ and $a_1 + a_2 + a_4 + a_6 + a_8 \to 1$; the function

$$x^\mu \left( b_1 + \frac{b_2}{1+b_3(x^2)^2} + b_4 \exp\left(-b_5(x^2)^2\right) + b_6 \sin(b_7 x^2) \right)$$

in which all $b_i$ ($i=1,2,\cdots,7$) are constants satisfies (1-6) and (1-8) under the limit $b_1 \to 1$, $b_2 \to 0$, $b_4 \to 0$ and $b_6 \to 0$, or under the limit $b_3 \to 0$, $b_5 \to 0$, $b_7 \to 0$ and $b_1 + b_2 + b_4 + b_6 \to 1$; etc.

In this paper we consider two cases. In Case1, we assume that all the functions $f(x)$, $U^\mu(x)$ and $V^\mu(x)$ are independent of wave function $\psi(x)$ of charged particle and electromagnetic four-potential $A^\mu(x)$; In Case2, we still assume that both the functions $f(x)$ and $U^\mu(x)$ are independent of $\psi(x)$ and $A^\mu(x)$, but $V^\mu(x)$ is a function of $A^\mu(x)$.

Some characteristics of the functions $f(x)$, $U^\mu(x)$ and $V^\mu(x)$ will be given and proved in place.

## 2 The case that all the functions $f(x)$, $U^\mu(x)$ and $V^\mu(x)$ are independent of $\psi(x)$ and $A^\mu(x)$

### 2.1 The action and its gauge invariance

As well known, the action of the conventional local QED is

$$S = S_{EM} + S_e + S_{I(CL)}, \qquad (2\text{-}1)$$

$$S_{EM} = -\frac{1}{4}\int d^4x \left(A_{\mu,\nu}(x) - A_{\nu,\mu}(x)\right)\left(A^{\mu,\nu}(x) - A^{\nu,\mu}(x)\right), \quad S_e = \int d^4x \overline{\psi}(x)\left(i\gamma^\mu \frac{\partial}{\partial x^\mu} - m\right)\psi(x), \qquad (2\text{-}2)$$

$$S_{I(CL)} = -e\int d^4x\, j^\mu(x) A_\mu(x). \qquad (2\text{-}3)$$

In (2-3), $j^\alpha(x) = \overline{\psi}(x)\gamma^\alpha \psi(x)$ is current density of charged particle.

Replacing $S_{I(CL)}$ given by (2-3), we employ

$$S_I = -e\int d^4x\, d^4y\, j^\alpha(x) U_\alpha^\beta(x) f\left(U^\lambda(x) - V^\lambda(y)\right) V(y) \widetilde{V}_\beta^\gamma(y) A_\gamma(y) \qquad (2\text{-}4)$$

as the interaction action of the system of charged particle and electromagnetic field. In (2-4), all the functions $f(x)$, $U^\mu(x)$, $U_\alpha^\beta(x)$, $V^\mu(y)$, $V(y)$ and $\widetilde{V}_\beta^\gamma(y)$ are introduced in Sect.1 and independent of $\psi(x)$ and $A^\mu(x)$.

The interaction expressed by (2-4) is nonlocal since the point $x^\mu$ in $j^\alpha(x)$ is different from the point $y^\mu$ in $A_\gamma(y)$. On the other hand, $S_I$ returns to $S_{I(CL)}$ under the limit conditions (1-3) and (1-8).

We now prove that the action (2-1), (2-2) and (2-4) is invariant under the following gauge transformation:

$$A'_\mu(x) = A_\mu(x) + \frac{\partial \theta(x)}{\partial x^\mu}, \tag{2-5}$$

$$\psi'(x) = e^{-ie\chi(x)}\psi(x), \tag{2-6}$$

$$\chi(x) = \int d^4y f\big(U^\lambda(x) - V^\lambda(y)\big) V(y)\theta(y), \tag{2-7}$$

where $\theta(x)$ is an arbitrary scalar function.

At first, under the transformation (2-5) and (2-6), $S'_{EM}$ and $S'_e$ in which the field variables are $A'_\mu(y)$ and $\psi'(x)$ become

$$\begin{aligned}
S'_{EM} + S'_e &= -\frac{1}{4}\int d^4x \big(A'_{\mu,\nu}(x) - A'_{\nu,\mu}(x)\big)\big(A'^{\mu,\nu}(x) - A'^{\nu,\mu}(x)\big) + \int d^4x \overline{\psi'}(x)\left(i\gamma^\mu \frac{\partial}{\partial x^\mu} - m\right)\psi'(x) \\
&= -\frac{1}{4}\int d^4x \big(A_{\mu,\nu}(x) - A_{\nu,\mu}(x)\big)\big(A^{\mu,\nu}(x) - A^{\nu,\mu}(x)\big) + \int d^4x \overline{\psi}(x) e^{ie\chi(x)}\left(i\gamma^\mu \frac{\partial}{\partial x^\mu} - m\right)\big(e^{-ie\chi(x)}\psi(x)\big) \\
&= -\frac{1}{4}\int d^4x \big(A_{\mu,\nu}(x) - A_{\nu,\mu}(x)\big)\big(A^{\mu,\nu}(x) - A^{\nu,\mu}(x)\big) + \int d^4x \overline{\psi}(x)\left(i\gamma^\mu \frac{\partial}{\partial x^\mu} - m\right)\psi(x) \\
&\quad + e\int d^4x \overline{\psi}(x)\gamma^\mu \frac{\partial \chi(x)}{\partial x^\mu}\psi(x) \\
&= S_{EM} + S_e + e\int d^4x\, j^\mu(x)\frac{\partial \chi(x)}{\partial x^\mu}, \tag{2-8}
\end{aligned}$$

On the other hand, notice that under the transformation (2-5) and (2-6), all the functions $f(x)$, $U^\mu(x)$, $U^\beta_\alpha(x)$, $V^\mu(y)$, $V(y)$ and $\widetilde{V}^\gamma_\beta(y)$ are invariant since they are independent of $\psi(x)$ and $A^\mu(x)$, and

$$j'^\alpha(x) = \overline{\psi'}(x)\gamma^\alpha \psi'(x) = \overline{\psi}(x)\gamma^\alpha \psi(x) = j^\alpha(x), \tag{2-9}$$

$S'_I$ in which the field variables are $A'_\mu(y)$ and $\psi'(x)$ thus becomes

$$\begin{aligned}
S'_I &= -e\int d^4x d^4y\, j'^\alpha(x) U^\beta_\alpha(x) f\big(U^\lambda(x) - V^\lambda(y)\big) V(y)\widetilde{V}^\gamma_\beta(y) A'_\gamma(y) \\
&= -e\int d^4x d^4y\, j^\alpha(x) U^\beta_\alpha(x) f\big(U^\lambda(x) - V^\lambda(y)\big) V(y)\widetilde{V}^\gamma_\beta(y)\left(A_\gamma(y) + \frac{\partial \theta(y)}{\partial y^\gamma}\right) \\
&= -e\int d^4x d^4y\, j^\alpha(x) U^\beta_\alpha(x) f\big(U^\lambda(x) - V^\lambda(y)\big) V(y)\widetilde{V}^\gamma_\beta(y) A_\gamma(y) \\
&\quad - e\int d^4x d^4y\, j^\alpha(x) U^\beta_\alpha(x) f\big(U^\lambda(x) - V^\lambda(y)\big) V(y)\widetilde{V}^\gamma_\beta(y)\frac{\partial \theta(y)}{\partial y^\gamma} \\
&= S_I - e\int d^4x d^4y\, j^\alpha(x) U^\beta_\alpha(x) f\big(U^\lambda(x) - V^\lambda(y)\big)\left(\frac{\partial\big(V(y)\widetilde{V}^\gamma_\beta(y)\theta(y)\big)}{\partial y^\gamma} - \frac{\partial\big(V(y)\widetilde{V}^\gamma_\beta(y)\big)}{\partial y^\gamma}\theta(y)\right);
\end{aligned}$$

In the Appendix A of this paper we shall prove

$$\frac{\partial\big(V(y)\widetilde{V}^\rho_\sigma(y)\big)}{\partial y^\rho} = 0, \tag{2-10}$$

substituting (2-10) to the above expression of $S'_I$ and using integration by parts, $S'_I$ becomes

$$S'_I = S_I + e\int d^4x d^4y\, j^\alpha(x) U^\beta_\alpha(x)\frac{\partial f\big(U^\lambda(x) - V^\lambda(y)\big)}{\partial y^\gamma} V(y)\widetilde{V}^\gamma_\beta(y)\theta(y);$$

In the Appendix A of this paper we shall prove

$$\frac{\partial f(U^\lambda(x)-V^\lambda(y))}{\partial y^\alpha} = -\widetilde{U}^\gamma_\beta(x)\frac{\partial f(U^\lambda(x)-V^\lambda(y))}{\partial x^\gamma}V^\beta_\alpha(y), \qquad (2\text{-}11)$$

substituting (2-11) to the above expression of $S'_I$ and using (1-7), $S'_I$ becomes

$$S'_I = S_I - e\int d^4x d^4y j^\alpha(x) U^\beta_\alpha(x)\widetilde{U}^\sigma_\rho(x)\frac{\partial f(U^\lambda(x)-V^\lambda(y))}{\partial x^\sigma}V^\rho_\gamma(y)V(y)\widetilde{V}^\gamma_\beta(y)\theta(y)$$

$$= S_I - e\int d^4x d^4y j^\alpha(x) \frac{\partial f(U^\lambda(x)-V^\lambda(y))}{\partial x^\alpha}V(y)\theta(y);$$

Hence, combining (2-8) with the above expression of $S'_I$, we have

$$S'_{EM} + S'_e + S'_I = S_{EM} + S_e + S_I$$
$$+ e\int d^4x j^\mu(x)\frac{\partial \chi(x)}{\partial x^\mu} - e\int d^4x j^\alpha(x)\frac{\partial}{\partial x^\alpha}\int d^4y f(U^\lambda(x)-V^\lambda(y))V(y)\theta(y),$$

thus, if the function $\chi(x)$ is introduced by (2-7), then we have $S'_{EM} + S'_e + S'_I = S_{EM} + S_e + S_I$, this means that the action (2-1), (2-2) and (2-4) is invariant under the gauge transformation (2-5) ~ (2-7).

**2.2 The equations of motion of charged particle and electromagnetic field and whose gauge invariance**

By the variational equation $\dfrac{\delta S}{\delta \overline{\psi}(x)} = 0$ we obtain the equation of motion of charged particle

$$\left(i\gamma^\mu\frac{\partial}{\partial x^\mu} - m\right)\psi(x) = e\gamma^\mu \Phi_\mu(x)\psi(x), \qquad (2\text{-}12)$$

$$\Phi_\mu(x) = U^\alpha_\mu(x)\int d^4y f(U^\lambda(x)-V^\lambda(y))V(y)\widetilde{V}^\beta_\alpha(y)A_\beta(y). \qquad (2\text{-}13)$$

The classical limit of the equation (2-4) is

$$m\frac{d^2x^\rho}{d\tau^2} = e\left(\Phi^{\rho,\sigma}(x)-\Phi^{\sigma,\rho}(x)\right)\frac{dx_\sigma}{d\tau} = eg^{\rho\mu}\left(\Phi_{\mu,\nu}(x)-\Phi_{\nu,\mu}(x)\right)\frac{dx^\nu}{d\tau}, \qquad (2\text{-}14)$$

where the term $eg^{\rho\mu}\left(\Phi_{\mu,\nu}(x)-\Phi_{\nu,\mu}(x)\right)\dfrac{dx^\nu}{d\tau}$ is a generalization of the Lorentz force. According to the formulas (1-5), (2-10), (A-3) and (A-4) we can prove

$$\Phi_{\mu,\nu}(x) - \Phi_{\nu,\mu}(x) = \frac{\partial \Phi_\mu(x)}{\partial x^\nu} - \frac{\partial \Phi_\nu(x)}{\partial x^\mu}$$
$$= U^\alpha_\mu(x)U^\beta_\nu(x)\int d^4y f(U^\lambda(x)-V^\lambda(y))V(y)\widetilde{V}^\rho_\alpha(y)\widetilde{V}^\sigma_\beta(y)\left(\frac{\partial A_\rho(y)}{\partial y^\sigma}-\frac{\partial A_\sigma(y)}{\partial y^\rho}\right). \qquad (2\text{-}15)$$

By the same approach proving that the action (2-1), (2-2) and (2-4) is gauge invariant showed in Sect. 2.1 we can prove that the equation of motion (2-12) ~ (2-13) is invariant under the gauge transformation (2-5) ~ (2-7). On the other hand, it is obvious that (2-14) and (2-15) is gauge invariant.

By the variational equation $\frac{\delta S}{\delta A_\mu(y)}=0$ we obtain the equation of motion of electromagnetic field

$$\frac{\partial}{\partial y^\nu}\left(A^{\mu,\nu}(y)-A^{\nu,\mu}(y)\right)=eW^\mu(y), \tag{2-16}$$

$$W^\mu(y)=V(y)\widetilde{V}_\nu^\mu(y)\int d^4x\, j^\alpha(x)U_\alpha^\nu(x)f\left(U^\lambda(x)-V^\lambda(y)\right). \tag{2-17}$$

It is obvious that (2-16) and (2-17) is invariant under the gauge transformation (2-5) ~ (2-7).

Under the limit conditions (1-3) and (1-8), (2-12), (2-13) and (2-16), (2-17) return to two equations $\left(i\gamma^\mu\frac{\partial}{\partial x^\mu}-m\right)\psi(x)=e\gamma^\mu A_\mu(x)\psi(x)$ and $\frac{\partial}{\partial y^\nu}\left(A^{\mu,\nu}(y)-A^{\nu,\mu}(y)\right)=ej^\mu(y)$ in the conventional local QED, respectively.

The equation of motion of electromagnetic field (2-16) can be written to different forms. We first calculate $U(z)\widetilde{U}_\rho^\sigma(z)\int d^4y\, \widetilde{f}\left(U^\lambda(z)-V^\lambda(y)\right)V_\mu^\rho(y)$ for (2-16), where the function $\widetilde{f}(x)$ is the inverse function of $f(x)$ and satisfies (1-1), and have

$$\begin{aligned}&U(z)\widetilde{U}_\rho^\sigma(z)\int d^4y\, \widetilde{f}\left(U^\lambda(z)-V^\lambda(y)\right)V_\mu^\rho(y)\frac{\partial}{\partial y^\nu}\left(A^{\mu,\nu}(y)-A^{\nu,\mu}(y)\right)\\ &=U(z)\widetilde{U}_\rho^\sigma(z)\int d^4y\, \widetilde{f}\left(U^\lambda(z)-V^\lambda(y)\right)V_\mu^\rho(y)eW^\mu(y);\end{aligned} \tag{2-18}$$

For the "left" expression of (2-18), notice that $\frac{\partial V_\mu^\rho(y)}{\partial y^\nu}\left(A^{\mu,\nu}(y)-A^{\nu,\mu}(y)\right)=\frac{\partial^2 V^\rho(y)}{\partial y^\mu \partial y^\nu}\left(A^{\mu,\nu}(y)-A^{\nu,\mu}(y)\right)=0$, and using integration by parts and (2-11), and we have

$$\begin{aligned}&U(z)\widetilde{U}_\rho^\sigma(z)\int d^4y\, \widetilde{f}\left(U^\lambda(z)-V^\lambda(y)\right)V_\mu^\rho(y)\frac{\partial}{\partial y^\nu}\left(A^{\mu,\nu}(y)-A^{\nu,\mu}(y)\right)\\ &=U(z)\widetilde{U}_\rho^\sigma(z)\int d^4y\, \widetilde{f}\left(U^\lambda(z)-V^\lambda(y)\right)\frac{\partial}{\partial y^\nu}\left[V_\mu^\rho(y)\left(A^{\mu,\nu}(y)-A^{\nu,\mu}(y)\right)\right]\\ &=-U(z)\widetilde{U}_\rho^\sigma(z)\int d^4y\, \frac{\partial \widetilde{f}\left(U^\lambda(z)-V^\lambda(y)\right)}{\partial y^\nu}V_\mu^\rho(y)\left(A^{\mu,\nu}(y)-A^{\nu,\mu}(y)\right)\\ &=U(z)\widetilde{U}_\rho^\sigma(z)\widetilde{U}_\alpha^\beta(z)\int d^4y\, \frac{\partial \widetilde{f}\left(U^\lambda(z)-V^\lambda(y)\right)}{\partial z^\beta}V_\nu^\alpha(y)V_\mu^\rho(y)\left(A^{\mu,\nu}(y)-A^{\nu,\mu}(y)\right)\\ &=\widetilde{U}_\rho^\sigma(z)\frac{\partial}{\partial z^\beta}\left[U(z)\widetilde{U}_\alpha^\beta(z)\int d^4y\, \widetilde{f}\left(U^\lambda(z)-V^\lambda(y)\right)V_\nu^\alpha(y)V_\mu^\rho(y)\left(A^{\mu,\nu}(y)-A^{\nu,\mu}(y)\right)\right];\end{aligned}$$

In the Appendix A of this paper we shall prove

$$\int d^4z\, U(x)V(z)\widetilde{f}\left(U^\rho(x)-V^\rho(z)\right)f\left(U^\sigma(y)-V^\sigma(z)\right)=\delta^4(x-y), \tag{2-19}$$

using (2-19) we calculate the "right" expression of (2-18) and obtain

$$\begin{aligned}&U(z)\widetilde{U}_\rho^\sigma(z)\int d^4y\, \widetilde{f}\left(U^\lambda(z)-V^\lambda(y)\right)V_\mu^\rho(y)eW^\mu(y)\\ &=U(z)\widetilde{U}_\rho^\sigma(z)\int d^4y\, \widetilde{f}\left(U^\lambda(z)-V^\lambda(y)\right)eV(y)V_\mu^\rho(y)\widetilde{V}_\nu^\mu(y)\int d^4x\, j^\alpha(x)U_\alpha^\nu(x)f\left(U^\lambda(x)-V^\lambda(y)\right)\\ &=e\widetilde{U}_\rho^\sigma(z)j^\alpha(z)U_\alpha^\rho(z)=ej^\sigma(z).\end{aligned}$$

Substituting the above two results to (2-18), we obtain the different forms of the equation of motion of electromagnetic field:

$$U(x)\tilde{U}_\alpha^\mu(x)\tilde{U}_\beta^\nu(x)\frac{\partial \tilde{F}^{\alpha\beta}(x)}{\partial x^\nu} = -ej^\mu(x), \quad \tilde{U}_\alpha^\mu(x)\frac{\partial}{\partial x^\nu}\left(U(x)\tilde{U}_\beta^\nu(x)\tilde{F}^{\alpha\beta}(x)\right) = -ej^\mu(x), \quad (2\text{-}20)$$

where

$$\tilde{F}^{\alpha\beta}(x) = -\int d^4y\, \tilde{f}\left(U^\lambda(x) - V^\lambda(y)\right) V_\rho^\alpha(y) V_\sigma^\beta(y) \left(A^{\rho,\sigma}(y) - A^{\sigma,\rho}(y)\right); \quad (2\text{-}21)$$

$\tilde{F}^{\alpha\beta}(x)$ defined by (2-21) is antisymmetric in the pair of indices:

$$\tilde{F}^{\alpha\beta}(x) = -\tilde{F}^{\beta\alpha}(x). \quad (2\text{-}22)$$

**2.3 The current conservation**

We can prove immediately that the equation (2-12) leads to the current conservation equation

$$\frac{\partial j^\mu(x)}{\partial x^\mu} = \frac{\partial \left(\overline{\psi}(x)\gamma^\mu\psi(x)\right)}{\partial x^\mu} = 0. \quad (2\text{-}23)$$

On the other hand, for the equation of motion of electromagnetic field (2-16), using (2-10), (2-11) and integration by parts we have

$$0 = \frac{\partial}{\partial y^\mu}\frac{\partial}{\partial y^\nu}\left(A^{\mu,\nu}(y) - A^{\nu,\mu}(y)\right) = e\frac{\partial W^\mu(y)}{\partial y^\mu}$$

$$= e\frac{\partial}{\partial y^\mu}\left[V(y)\tilde{V}_\beta^\mu(y)\int d^4x\, j^\alpha(x) U_\alpha^\beta(x) f\left(U^\lambda(x) - V^\lambda(y)\right)\right]$$

$$= e\frac{\partial\left(V(y)\tilde{V}_\beta^\mu(y)\right)}{\partial y^\mu}\int d^4x\, j^\alpha(x) U_\alpha^\beta(x) f\left(U^\lambda(x) - V^\lambda(y)\right)$$

$$+ eV(y)\tilde{V}_\beta^\mu(y)\frac{\partial}{\partial y^\mu}\int d^4x\, j^\alpha(x) U_\alpha^\beta(x) f\left(U^\lambda(x) - V^\lambda(y)\right)$$

$$= eV(y)\tilde{V}_\beta^\mu(y)\int d^4x\, j^\alpha(x) U_\alpha^\beta(x) \frac{\partial f\left(U^\lambda(x) - V^\lambda(y)\right)}{\partial y^\mu}$$

$$= -eV(y)\tilde{V}_\beta^\mu(y)\int d^4x\, j^\alpha(x) U_\alpha^\beta(x) \tilde{U}_\rho^\sigma(x)\frac{\partial f\left(U^\lambda(x) - V^\lambda(y)\right)}{\partial x^\sigma} V_\mu^\rho(y)$$

$$= -eV(y)\int d^4x\, j^\alpha(x) \frac{\partial f\left(U^\lambda(x) - V^\lambda(y)\right)}{\partial x^\alpha}$$

$$= eV(y)\int d^4x\, \frac{\partial j^\alpha(x)}{\partial x^\alpha} f\left(U^\lambda(x) - V^\lambda(y)\right).$$

And, further, by calculating $U(z)\int d^4y\, \tilde{f}\left(U^\lambda(z) - V^\lambda(y)\right)$ for the above expression and using (2-19), we obtain $\dfrac{\partial j^\alpha(z)}{\partial z^\alpha} = 0$.

On the other hand, from the first equation in (2-20) and notice that $\tilde{F}^{\alpha\beta}(x)$ satisfies (2-22), we can obtain directly

$$\frac{\partial j^\mu(x)}{\partial x^\mu} = -\frac{1}{e}\frac{\partial}{\partial x^\mu}\left(U(x)\widetilde{U}^\mu_\alpha(x)\widetilde{U}^\nu_\beta(x)\frac{\partial \widetilde{F}^{\alpha\beta}(x)}{\partial x^\nu}\right) = -\frac{1}{e}U(x)\widetilde{U}^\mu_\alpha(x)\frac{\partial}{\partial x^\mu}\left(\widetilde{U}^\nu_\beta(x)\frac{\partial \widetilde{F}^{\alpha\beta}(x)}{\partial x^\nu}\right)$$

$$= -\frac{1}{e}U(x)\widetilde{U}^\mu_\alpha(x)\frac{\partial \widetilde{U}^\nu_\beta(x)}{\partial x^\mu}\frac{\partial \widetilde{F}^{\alpha\beta}(x)}{\partial x^\nu} - \frac{1}{e}U(x)\widetilde{U}^\mu_\alpha(x)\widetilde{U}^\nu_\beta(x)\frac{\partial^2 \widetilde{F}^{\alpha\beta}(x)}{\partial x^\mu \partial x^\nu}$$

$$= \frac{1}{e}U(x)\widetilde{U}^\mu_\alpha(x)\widetilde{U}^\nu_\rho(x)\widetilde{U}^\sigma_\beta(x)\frac{\partial U^\rho_\sigma(x)}{\partial x^\mu}\frac{\partial \widetilde{F}^{\alpha\beta}(x)}{\partial x^\nu}$$

$$= \frac{1}{2e}U(x)\widetilde{U}^\nu_\rho(x)\frac{\partial U^\rho_\sigma(x)}{\partial x^\mu}\left(\widetilde{U}^\mu_\alpha(x)\widetilde{U}^\sigma_\beta(x) + \widetilde{U}^\mu_\beta(x)\widetilde{U}^\sigma_\alpha(x)\right)\frac{\partial \widetilde{F}^{\alpha\beta}(x)}{\partial x^\nu} = 0.$$

In the above calculation, we have used the formula $\dfrac{\partial \widetilde{U}^\mu_\nu(y)}{\partial y^\rho} = -\widetilde{U}^\mu_\alpha(y)\widetilde{U}^\beta_\nu(y)\dfrac{\partial \widetilde{U}^\alpha_\beta(y)}{\partial y^\rho}$, which can be proved by the method proving (A-3); and taken account of $\dfrac{\partial U^\rho_\sigma(x)}{\partial x^\mu} = \dfrac{\partial U^\rho_\mu(x)}{\partial x^\sigma}$ due to (1-5), which leads to $\widetilde{U}^\mu_\alpha(x)\widetilde{U}^\sigma_\beta(x)\dfrac{\partial U^\rho_\sigma(x)}{\partial x^\mu} = \widetilde{U}^\mu_\alpha(x)\widetilde{U}^\sigma_\beta(x)\dfrac{\partial U^\rho_\mu(x)}{\partial x^\sigma} = \widetilde{U}^\mu_\beta(x)\widetilde{U}^\sigma_\alpha(x)\dfrac{\partial U^\rho_\sigma(x)}{\partial x^\mu}$.

Hence, similar to the conventional local QED, the equations of motion of electromagnetic field (2-16) and (2-17) leads to the current conservation equation (2-23) naturally.

## 3 Some remarks and the definition of transverse four-vector

### 3.1 Some remarks

What theory of quantum electrodynamics with nonlocal interaction we construct in Sect. 2 obeys the action principle, has gauge invariance, leads to the normal current conservation, and returns to the conventional local QED under the limit conditions (1-3) and (1-8). We can try to choose some special forms of the functions $f(x)$, $U^\mu(x)$ and $V^\mu(x)$ and obtain some concrete results of the theory.

A special case of $U^\mu(x) = x^\mu$, $V^\mu(x) = x^\mu$ and $f(x) = \int \dfrac{d^4 k}{(2\pi)^4}\dfrac{1}{1+r_0^4 k^4}e^{ik\cdot x}$, where $r_0$ is a constant, has been studied and, thus, a theory of classical electrodynamics without singularities has been obtained in Ref. [2]. This special choice, however, is incapable of removing the divergence of the vacuum polarization[3]. Hence, even if the functions $f(x)$, $U^\mu(x)$ and $V^\mu(x)$ being independent of $\psi(x)$ and $A^\mu(x)$ and making that there is not any divergent term in the classical and quantum theory existed, for which it seems as if we have no simple choice.

We now investigate the case that both two vector functions $U^\mu(x)$ and $V^\mu(x)$ are dependent of wave function $\psi(x)$ of charged particle and electromagnetic four-potential $A^\nu(x)$:

$$U^\mu(x) = U^\mu\big(x; \psi(x), A^\nu(x)\big), \quad V^\mu(x) = V^\mu\big(x; \psi(x), A^\nu(x)\big). \tag{3-1}$$

Although we assume that $f(x)$ is still independent of $\psi(x)$ and $A^\nu(x)$,

$f(U^\lambda(x) - V^\lambda(y))$ is now dependent of $\psi(x)$ and $A^\nu(x)$.

For this case, if we still choose (2-1), (2-2) and (2-4) as the action of the system, then it seems that it is quite difficult to find out two functions $U^\mu(x; \psi(x), A^\nu(x))$ and $V^\mu(x; \psi(x), A^\nu(x))$ such that the action (2-1), (2-2) and (2-4) is invariant under the gauge transformation (2-5) and (2-6).

For ensuring the gauge invariance, from (2-9) we see that $\overline{\psi}(x)\gamma^\mu\psi(x)$ is invariant under the transformation (2-6), Furthermore, all $\overline{\psi}(x)\Gamma_i\psi(x)$ ($i=1,2,\cdots,16$) as well, where $\Gamma_i$ ($i=1,2,\cdots,16$) denotes the sixteen matrices $I, \gamma^\mu, \sigma^{\alpha\beta}, \gamma^5\gamma^\mu, \gamma^5$; And then, we construct a general transverse four-vector $A^\mu_{\perp G}(x)$ corresponding to electromagnetic four-potential $A^\mu(x)$ (whose the exact definition will be discussed), which is invariant under the transformation (2-5). Hence, if $U^\mu(x)$ and $V^\mu(x)$ are functions of $\overline{\psi}(x)\Gamma_1\psi(x), \overline{\psi}(x)\Gamma_2\psi(x),\cdots$ and $A^\nu_{\perp G}(x)$:

$$U^\mu(x) = U^\mu\left(x; \overline{\psi}(x)\Gamma_1\psi(x), \overline{\psi}(x)\Gamma_2\psi(x), \cdots; A^\nu_{\perp G}(x)\right),$$
$$V^\mu(x) = V^\mu\left(x; \overline{\psi}(x)\Gamma_1\psi(x), \overline{\psi}(x)\Gamma_2\psi(x), \cdots; A^\nu_{\perp G}(x)\right),$$
(3-2)

then both $U^\mu(x)$ and $V^\mu(x)$ are also invariant under the gauge transformations (2-5) and (2-6); and, further, all $U^\mu_\nu(x)$, $U(x)$, $\tilde{U}^\mu_\nu(x)$, $V^\mu_\nu(x)$, $V(x)$ and $\tilde{V}^\mu_\nu(x)$ as well.

Eq. (3-2) includes various cases. For example, $U^\mu(x)$ can be independent of $\overline{\psi}(x)\Gamma_1\psi(x), \overline{\psi}(x)\Gamma_2\psi(x),\cdots$ and $A^\nu_{\perp G}(x)$, or can be only dependent of $\overline{\psi}(x)\Gamma_1\psi(x)$, $\overline{\psi}(x)\Gamma_2\psi(x),\cdots$ or $A^\nu_{\perp G}(x)$, or dependent of $\overline{\psi}(x)\Gamma_1\psi(x), \overline{\psi}(x)\Gamma_2\psi(x),\cdots$ and $A^\nu_{\perp G}(x)$, $V^\mu(x)$ as well. Of course, if both $U^\mu(x)$ and $V^\mu(x)$ are independent of $\overline{\psi}(x)\Gamma_1\psi(x), \overline{\psi}(x)\Gamma_2\psi(x),\cdots$ and $A^\nu_{\perp G}(x)$, then we retune to the case discussed in Sect. 2.

If both the 4-vector functions $U^\mu(x)$ and $V^\mu(x)$ in (2-12), (2-13) and (2-16), (2-17) are functions of $\overline{\psi}(x)\Gamma_1\psi(x), \overline{\psi}(x)\Gamma_2\psi(x),\cdots$ and $A^\nu_{\perp G}(x)$, then we can prove that the equations of motion (2-12), (2-13) and (2-16), (2-17) of charged particle and electromagnetic field are still gauge invariant due to all $U^\mu(x)$, $U^\mu_\nu(x)$, $U(x)$, $\tilde{U}^\mu_\nu(x)$, $V^\mu(x)$, $V^\mu_\nu(x)$, $V(x)$ and $\tilde{V}^\mu_\nu(x)$ are invariant under the gauge transformation (2-5) ~ (2-7), and the current conservation equation (2-23) holds yet.

On the other hand, if both the 4-vector functions $U^\mu(x)$ and $V^\mu(x)$ in (2-1), (2-2) and (2-4) are functions of $\overline{\psi}(x)\Gamma_1\psi(x), \overline{\psi}(x)\Gamma_2\psi(x),\cdots$ and $A^\nu_{\perp G}(x)$, then we can prove that the action given by (2-1), (2-2) and (2-4) is still gauge invariant.

However, the variational equations $\dfrac{\delta S}{\delta \overline{\psi}(x)} = 0$ and $\dfrac{\delta S}{\delta A_\mu(y)} = 0$ corresponding to the action (2-1), (2-2) and (2-4) now not only cannot lead to the equations of motion (2-12), (2-13) and (2-16), (2-17), but also lead to unacceptable equations of motion of charged particle and electromagnetic field (We don't write out the concrete calculation process here). On the other hand, it seems as if it is quite difficult to find out an action that can lead to the equations of motion (2-12), (2-13) and (2-16), (2-17) when $U^\mu(x)$ and $V^\mu(x)$ are dependent of $\overline{\psi}(x)\Gamma_1\psi(x), \overline{\psi}(x)\Gamma_2\psi(x), \cdots$ and $A^\nu_{\perp G}(x)$.

What we can do when we face this situation? If we give up the action principle and assume that the equations of motion of charged particle and electromagnetic field are still given by (2-12), (2-13) and (2-16), (2-17), respectively, then the theory still keeps integrity, for example, we can establish the corresponding quantum theory yet (See the discussion in Sect. 5). Hence, maybe this approach is not unacceptable.

In Sect. 4 we shall investigate a special case of (3-2) that both $f(x)$ and $U^\mu(x)$ are independent of $\psi(x)$ and $A^\nu(x)$, but $V^\mu(x)$ is a simple function of $A^\mu_{\perp G}(x)$. For this case, if we still choose (2-1), (2-2) and (2-4) as the action of the system and obey the action principle, then maybe the established theory is acceptable. For this purpose, we first discuss the construction of the general transverse four-vector $A^\mu_{\perp G}(x)$.

**3.2 Transverse and longitudinal four-vectors**

**3.2.1 The definitions of transverse and longitudinal four-vectors and the expansion of $A^\mu(x)$ according to the polarization vectors**

By $A^\mu_{//}(x)$ called the longitudinal four-vector corresponding to $A^\mu(x)$ we denote the part of the form $\partial^\mu \Xi(x)$ in a four-vector $A^\mu(x)$, based on $A^\mu_{//}(x)$, we define the transverse four-vector corresponding to $A^\mu(x)$:

$$A^\mu_\perp(x) = A^\mu(x) - A^\mu_{//}(x) \tag{3-3}$$

in which there is not function of the form $\partial^\mu \varphi(x)$.

The form of gauge transformation (2-5) only leads the change of $A^\mu_{//}(x)$, namely, $A^\mu_\perp(x)$ is invariant under the transformation (2-5). Hence, if we can segregate the longitudinal part $A^\mu_{//}(x)$ from $A^\mu(x)$, then by (3-3) we can construct the invariant components $A^\mu_\perp(x)$ under the transformation (2-5) of $A^\mu(x)$.

Formally, taking advantage of the Fourier transform, $A^\mu_{//}(x)$ and $A^\mu(x)$ can be written to the forms

$$A^\mu_{//}(x) = \partial^\mu \Xi(x) = \frac{\partial}{\partial x_\mu} \int \frac{d^4 k}{(2\pi)^4} \hat{\Xi}(k) e^{-ik\cdot x} = -i \int \frac{d^4 k}{(2\pi)^4} k^\mu \hat{\Xi}(k) e^{-ik\cdot x}, \tag{3-4}$$

$$A^\mu(x) = \int \frac{d^4k}{(2\pi)^4} \hat{A}^\mu(k) e^{-ik\cdot x}, \quad \hat{A}^\mu(k) = \int d^4x A^\mu(x) e^{ik\cdot x}, \quad \hat{A}^{\mu*}(k) = \hat{A}^\mu(-k). \tag{3-5}$$

The last formula in (3-5) holds since $A^\mu(x)$ is a real function.

From (3-4) and (3-5) we see that the longitudinal part $A_{//}^\mu(x)$ of $A^\mu(x)$ is the component of $\hat{A}^\mu(k)$ along the $k^\mu$ axis, which can be formally written to the form $\frac{k^\mu}{k^2} k_\nu \hat{A}^\nu(k)$; for segregating such part, we take advantage of the four polarization vectors[1] $\varepsilon_{(\lambda)}^\mu(k)$ ($\lambda = 0, 1, 2, 3$), which satisfy

$$\varepsilon_{(0)}^\mu(k) = n^\mu, \; n_\mu n^\mu = 1, \; n^0 > 0; \; k_\mu \varepsilon_{(1)}^\mu(k) = k_\mu \varepsilon_{(2)}^\mu(k) = 0; \; \varepsilon_{(3)}^\mu(k) = \frac{k^\mu - n^\mu k\cdot n}{\sqrt{(k\cdot n)^2 - k^2}};$$

$$g_{\mu\nu} \varepsilon_{(\lambda)}^\mu(k) \varepsilon_{(\lambda')}^\nu(k) = g_{(\lambda)(\lambda')}, \; \sum_{\lambda=0}^{3} \sum_{\lambda'=0}^{3} g_{(\lambda)(\lambda')} \varepsilon_{(\lambda)}^\mu(k) \varepsilon_{(\lambda')}^\nu(k) = g^{\mu\nu}; \tag{3-6}$$

$\hat{A}^\mu(k)$ thus has the following expansion:

$$\hat{A}^\mu(k) = \sum_{\lambda=0}^{3} \varepsilon_{(\lambda)}^\mu(k) \hat{A}_{(\lambda)}(k), \quad \hat{A}_{(\lambda)}(k) = \sum_{\lambda'=0}^{3} g_{(\lambda)(\lambda')} \varepsilon_{\nu(\lambda')}(k) \hat{A}^\nu(k). \tag{3-7}$$

Notice $\varepsilon_{(3)}^\mu(k) = \frac{k^\mu}{\sqrt{(k\cdot n)^2 - k^2}} - \frac{k\cdot n}{\sqrt{(k\cdot n)^2 - k^2}} \varepsilon_{(0)}^\mu(k)$, we can introduce

$$A_{(0)}^\mu(x) = \int \frac{d^4k}{(2\pi)^4} \varepsilon_{(0)}^\mu(k) \hat{A}_{(\bar{0})}(k) e^{-ik\cdot x} = n^\mu \int \frac{d^4k}{(2\pi)^4} \hat{A}_{(\bar{0})}(k) e^{-ik\cdot x},$$

$$\hat{A}_{(\bar{0})}(k) = \hat{A}_{(0)}(k) - \frac{k\cdot n}{\sqrt{(k\cdot n)^2 - k^2}} \hat{A}_{(3)}(k), \tag{3-8}$$

$$A_{(\bar{\lambda})}^\mu(x) = \int \frac{d^4k}{(2\pi)^4} \varepsilon_{(\lambda)}^\mu(k) \hat{A}_{(\lambda)}(k) e^{-ik\cdot x} \quad (\lambda = 1, 2), \tag{3-9}$$

$$A_{(3)}^\mu(x) = \int \frac{d^4k}{(2\pi)^4} \frac{k^\mu}{\sqrt{(k\cdot n)^2 - k^2}} \hat{A}_{(3)}(k) e^{-ik\cdot x} = \frac{\partial}{\partial x_\mu} \int \frac{d^4k}{(2\pi)^4} \frac{i}{\sqrt{(k\cdot n)^2 - k^2}} \hat{A}_{(3)}(k) e^{-ik\cdot x}; \tag{3-10}$$

From (3-8) ~ (3-11) we have

$$A^\mu(x) = A_{(0)}^\mu(x) + \sum_{\lambda=1}^{2} A_{(\bar{\lambda})}^\mu(x) + A_{(3)}^\mu(x). \tag{3-11}$$

**3.2.2 The characteristics of $A_{(\bar{\lambda})}^\mu(x)$ ($\lambda = 1, 2$) and $A_{(3)}^\mu(x)$**

Using the characteristics $k_\mu \varepsilon_{(\lambda)}^\mu(k) = 0$ ($\lambda = 1, 2$) given in (3-6), from (3-9) we see that the component along the $k^\mu$ axis of the Fourier transform $\varepsilon_{(\lambda)}^\mu(k)\hat{A}_{(\lambda)}(k)$ of $A_{(\bar{\lambda})}^\mu(x)$ $(\lambda = 1, 2)$ is

$$\frac{k^\mu}{k^2}k_\nu \varepsilon_{(\lambda)}^\nu(k)\hat{A}_{(\lambda)}(k) = 0 \quad (\lambda = 1, 2), \tag{3-12}$$

in other words, both $\varepsilon_{(\lambda)}^\mu(k)\hat{A}_{(\lambda)}(k)$ $(\lambda = 1, 2)$ are perpendicular to the $k^\mu$ axis, hence, both $A_{(\bar{\lambda})}^\mu(x)$ $(\lambda = 1, 2)$ are transverse four-vectors.

From (3-10) we see that the Fourier transform $\dfrac{k^\mu}{\sqrt{(k\cdot n)^2 - k^2}}\hat{A}_{(3)}(k)$ of $A_{(\bar{3})}^\mu(x)$ is fully parallel to the $k^\mu$ axis, this can be verified formally as follows:

$$\frac{k^\mu}{k^2}k_\nu \frac{k^\nu}{\sqrt{(k\cdot n)^2 - k^2}}\hat{A}_{(3)}(k) = \frac{k^\mu}{k^2}\frac{k^2}{\sqrt{(k\cdot n)^2 - k^2}}\hat{A}_{(3)}(k) = \frac{k^\mu}{\sqrt{(k\cdot n)^2 - k^2}}\hat{A}_{(3)}(k), \tag{3-13}$$

hence, $A_{(\bar{3})}^\mu(x)$ is longitudinal four-vector.

### 3.2.3 The longitudinal part $A_{(0)//}^\mu(x)$ of $A_{(0)}^\mu(x)$

Formally, the longitudinal part $A_{(0)//}^\mu(x)$ of $A_{(0)}^\mu(x)$ can be written to the form

$$\begin{aligned}A_{(0)//}^\mu(x) &= \int \frac{d^4k}{(2\pi)^4}\frac{k^\mu}{k^2}k_\nu \varepsilon_{(0)}^\nu(k)\hat{A}_{(\bar{0})}(k)\,e^{-ik\cdot x} = \int d^4y \int \frac{d^4k}{(2\pi)^4}\frac{k^\mu}{k^2}k_\nu A_{(0)}^\nu(y)\,e^{-ik\cdot(x-y)} \\ &= \frac{\partial^2}{\partial x_\mu \partial x^\nu}\int d^4y\, D(x-y)A_{(0)}^\nu(y),\end{aligned} \tag{3-14}$$

$$D(x) = \int \frac{d^4k}{(2\pi)^4}\frac{-1}{k^2}e^{-ik\cdot x} = \int \frac{d^3k}{(2\pi)^3}e^{i\mathbf{k}\cdot\mathbf{x}}\int \frac{dk_0}{2\pi}\frac{-1}{k_0^2 - \mathbf{k}\cdot\mathbf{k}}e^{-ik_0 x^0}; \tag{3-15}$$

However, (3-14) and (3-15) cause two questions: ① As well known, the function $D(x)$ varies with the integral paths of the variable $k_0$ bypassing the two points $\pm|\mathbf{k}|$; ② If $A_{(0)}^\mu(x)$ satisfies $\Box A_{(0)}^\mu(x) = 0$, where $\Box$ means the d'Alembertian operator $\partial_\lambda \partial^\lambda = g^{\mu\nu}\dfrac{\partial^2}{\partial x^\mu \partial x^\nu}$, then there are singularities in the integral $\int d^4y\, D(x-y)A_{(0)}^\nu(y)$.

For dealing with these two questions, according to the classification of $k^2 \neq 0$ and $k^2 = 0$, $A_{(0)}^\mu(x)$ is firstly divided into two parts:

$$A^\mu_{(\bar 0)}(x) = \int \frac{\mathrm{d}^4 k}{(2\pi)^4} \varepsilon^\mu_{(\bar 0)}(k) \hat{A}_{(\bar 0)}(k) \mathrm{e}^{-\mathrm{i}k\cdot x} = \sum_{k^0} \frac{\Delta k^0}{2\pi} \int \frac{\mathrm{d}^3 k}{(2\pi)^3} n^\mu \hat{A}_{(\bar 0)}(k) \mathrm{e}^{-\mathrm{i}k\cdot x}$$

$$= \sum_{\substack{k'^0 \\ k'^0 \neq \pm |k'|}} \frac{\Delta k^0}{2\pi} \int \frac{\mathrm{d}^3 k}{(2\pi)^3} n^\mu \hat{A}_{(\bar 0)}(k) \mathrm{e}^{-\mathrm{i}k\cdot x} \tag{3-16}$$

$$+ \frac{\Delta k^0}{2\pi} \int \frac{\mathrm{d}^3 k}{(2\pi)^3} n^\mu \hat{A}_{(\bar 0)}(k) \mathrm{e}^{-\mathrm{i}k\cdot x}\bigg|_{k^0 = |k|} + \frac{\Delta k^0}{2\pi} \int \frac{\mathrm{d}^3 k}{(2\pi)^3} n^\mu \hat{A}_{(\bar 0)}(k) \mathrm{e}^{-\mathrm{i}k\cdot x}\bigg|_{k^0 = -|k|}$$

$$= \overline{A}^\mu_{(\bar 0)}(x) + \widetilde{A}^\mu_{(\bar 0)}(x),$$

$$\overline{A}^\mu_{(\bar 0)}(x) = \sum_{\substack{k'^0 \\ k'^0 \neq \pm |k'|}} \frac{\Delta k^0}{2\pi} \int \frac{\mathrm{d}^3 k}{(2\pi)^3} n^\mu \hat{A}_{(\bar 0)}(k) \mathrm{e}^{-\mathrm{i}k\cdot x}, \tag{3-17}$$

$$\widetilde{A}^\mu_{(\bar 0)}(x) = \frac{\Delta k^0}{2\pi} \int \frac{\mathrm{d}^3 k}{(2\pi)^3} n^\mu \hat{A}_{(\bar 0)}(k) \mathrm{e}^{-\mathrm{i}k\cdot x}\bigg|_{k^0 = |k|} + \frac{\Delta k^0}{2\pi} \int \frac{\mathrm{d}^3 k}{(2\pi)^3} n^\mu \hat{A}_{(\bar 0)}(k) \mathrm{e}^{-\mathrm{i}k\cdot x}\bigg|_{k^0 = -|k|}$$

$$= \frac{\Delta k^0}{2\pi} \int \frac{\mathrm{d}^3 k}{(2\pi)^3} A^\mu_{(\bar 0)}(y) \mathrm{e}^{-\mathrm{i}k\cdot(x-y)}\bigg|_{k^0 = |k|} \mathrm{d}^4 y + \frac{\Delta k^0}{2\pi} \int \frac{\mathrm{d}^3 k}{(2\pi)^3} A^\mu_{(\bar 0)}(y) \mathrm{e}^{-\mathrm{i}k\cdot(x-y)}\bigg|_{k^0 = -|k|} \mathrm{d}^4 y \tag{3-18}$$

$$= \frac{\Delta k^0}{2\pi} \int \frac{\mathrm{d}^3 k}{(2\pi)^3} \left( \mathrm{e}^{-\mathrm{i}|k|(x^0 - y^0) + \mathrm{i}\boldsymbol{k}\cdot(\boldsymbol{x}-\boldsymbol{y})} + \mathrm{e}^{\mathrm{i}|k|(x^0 - y^0) + \mathrm{i}\boldsymbol{k}\cdot(\boldsymbol{x}-\boldsymbol{y})} \right) A^\mu_{(\bar 0)}(y) \mathrm{d}^4 y.$$

For $\overline{A}^\mu_{(\bar 0)}(x)$ we have $\Box \overline{A}^\mu_{(\bar 0)}(x) \neq 0$ unless $\hat{A}_{(\bar 0)}(k)\big|_{k'^0 \neq \pm |k'|} = 0$, $\widetilde{A}^\mu_{(\bar 0)}(x)$ satisfies

$$\Box \widetilde{A}^\mu_{(\bar 0)}(x) = 0. \tag{3-19}$$

① The longitudinal part $\overline{A}^\mu_{(\bar 0)//}(x)$ of $\overline{A}^\mu_{(\bar 0)}(x)$

As well-known, for the function $D(x)$ defined by (3-16), according to the different integral paths of the variable $k_0$ bypassing the two points $\pm|k|$, we can obtain four functions

$$D_{C_i}(x) = \int_{C_i} \frac{\mathrm{d}^4 k}{(2\pi)^4} \frac{-1}{k^2} \mathrm{e}^{-\mathrm{i}k\cdot x} = \int \frac{\mathrm{d}^3 k}{(2\pi)^3} \mathrm{e}^{\mathrm{i}\boldsymbol{k}\cdot\boldsymbol{x}} \int_{C_i} \frac{\mathrm{d}k_0}{2\pi} \frac{-1}{k_0^2 - \boldsymbol{k}\cdot\boldsymbol{k}} \mathrm{e}^{-\mathrm{i}k_0 x^0} \quad (i=1,2,3,4) \tag{3-20}$$

corresponding to the four paths $C_i$ ($i=1,2,3,4$) in Fig.1. Other forms of $D_{C_i}(x)$ ($i=1,2,3,4$) can be found in the Appendix A of this paper.

As well known, all $D_{C_i}(x)$ ($i=1,2,3,4$) satisfy

$$\Box D_{C_i}(x) = \delta^4(x) \quad (i=1,2,3,4); \tag{3-21}$$

$D_{C_1}(x) \equiv D_{\mathrm{ret}}(x)$, $D_{C_2}(x) \equiv D_{\mathrm{adv}}(x)$ and $D_{C_3}(x) \equiv D_{\mathrm{F}}(x)$ are so called the retarded, advanced Green functions and the Feynman propagator for the paths $C_1$, $C_2$, $C_3$ in Fig.1 (a), (b), (c), respectively[1, 4].

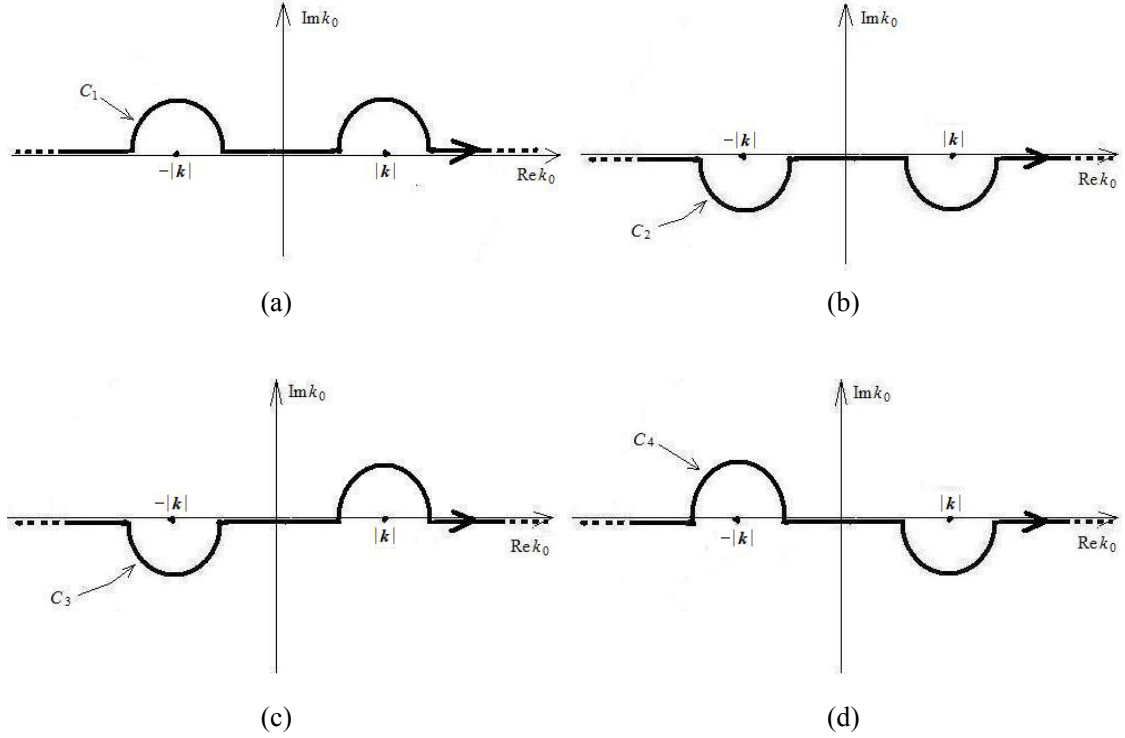

(a)    (b)    (c)    (d)

Fig.1   The path for the integral of the variable $k_0$

Hence, in spite of the physical meaning of the four functions $D_{C_i}(x)$ ($i=1,2,3,4$) and only from the point of view of pure mathematics, according to (3-13), the longitudinal part $\overline{A}^{\mu}_{(0)//}(x)$ of $\overline{A}^{\mu}_{(0)}(x)$ can be written to the form

$$\overline{A}^{\mu}_{(0)//}(x) = \frac{\partial^2}{\partial x_\mu \partial x^\nu}\int d^4 y D_{\text{Total}}(x-y)\overline{A}^{\nu}_{(0)}(y). \qquad (3\text{-}22)$$

The general form of the function $D_{\text{Total}}(x)$ is $D_{\text{Total}}(x) = \overline{C}_1 D_{C_1}(x) + \overline{C}_2 D_{C_2}(x) + \overline{C}_3 D_{C_3}(x) + \overline{C}_4 D_{C_4}(x)$, where all $\overline{C}_i$ ($i=1,2,3,4$) are constants. However, according to (A-7) we see that the four functions $D_{C_i}(x)$ ($i=1,2,3,4$) are not dependent, $D_{\text{Total}}(x)$ thus can be made up of arbitrary three of the four functions; here we choose $D_{C_1}(x)$, $D_{C_2}(x)$ and $D_{C_3}(x)$ as three dependent functions and, thus, $D_{\text{Total}}(x)$ becomes $D_{\text{Total}}(x) = \overline{C}_1 D_{\text{ret}}(x) + \overline{C}_2 D_{\text{adv}}(x) + \overline{C}_3 D_{\text{F}}(x)$. On the other hand, $\overline{A}^{\mu}_{(0)//}(x)$ obtained by (3-22) should be real, since $\overline{A}^{\mu}_{(0)}(x)$ is real, but the Feynman propagator $D_{\text{F}}(x)$ is a complex function, hence, we have to weed out $D_{\text{F}}(x)$ from the expression of $D_{\text{Total}}(x)$ and, thus, $D_{\text{Total}}(x) = \overline{C}_1 D_{\text{ret}}(x) + \overline{C}_2 D_{\text{adv}}(x)$. For ensuring

$$\Box D_{\text{Total}}(x) = \delta^4(x), \tag{3-23}$$

the two constants $\overline{C}_1$ and $\overline{C}_2$ must satisfy $\overline{C}_1 + \overline{C}_2 = 1$. Finally we have

$$D_{\text{Total}}(x) = \frac{1+\overline{C}}{2} D_{\text{ret}}(x) + \frac{1-\overline{C}}{2} D_{\text{adv}}(x), \tag{3-24}$$

where $\overline{C}$ is an arbitrary constant.

It seems as if $\overline{A}^{\mu}_{(0)//}(x)$ given by (3-22) is uncertain, since the constant $\overline{C}$ in (3-24) is uncertain. However, we can prove that $\overline{A}^{\mu}_{(0)//}(x)$ determined by (3-22) is independent of the choice of the constant $\overline{C}$.

In fact, for a different choice of the constant $\overline{C}'$, we have:

$$\overline{A}'^{\mu}_{(0)//}(x) = \frac{\partial^2}{\partial x_\mu \partial x^\nu} \int D'_{\text{Total}}(x-y) \overline{A}^{\nu}_{(0)}(y) \mathrm{d}^4 y, \quad D'_{\text{Total}}(x) = \frac{1+\overline{C}'}{2} D_{\text{ret}}(x) + \frac{1-\overline{C}'}{2} D_{\text{adv}}(x),$$

using (A-8) we have

$$D'_{\text{Total}}(x-y) - D_{\text{Total}}(x-y) = \frac{\overline{C}' - \overline{C}}{2} \int \frac{\mathrm{d}^3 k}{(2\pi)^3} \frac{\mathrm{i}}{2|\boldsymbol{k}|} \left( \mathrm{e}^{-\mathrm{i}k\cdot(x-y)} - \mathrm{e}^{\mathrm{i}k\cdot(x-y)} \right) \Big|_{k^0 = |\boldsymbol{k}|},$$

taking into account (3-18), the remainder of $\overline{A}'^{\mu}_{(0)//}(x)$ and $\overline{A}^{\mu}_{(0)//}(x)$ becomes

$$\overline{A}'^{\mu}_{(0)//}(x) - \overline{A}^{\mu}_{(0)//}(x) = \frac{\partial^2}{\partial x_\mu \partial x^\nu} \int \left( D'_{\text{Total}}(x-y) - D_{\text{Total}}(x-y) \right) \overline{A}^{\nu}_{(0)}(y) \mathrm{d}^4 y$$

$$= \frac{\overline{C}' - \overline{C}}{2} \frac{\partial^2}{\partial x_\mu \partial x^\nu} \int \frac{\mathrm{d}^3 k}{(2\pi)^3} \frac{\mathrm{i}}{2|\boldsymbol{k}|} \left( \mathrm{e}^{-\mathrm{i}k\cdot(x-y)} - \mathrm{e}^{\mathrm{i}k\cdot(x-y)} \right) \Big|_{k^0 = |\boldsymbol{k}|} \sum_{k'^0 \neq \pm |\boldsymbol{k}'|} \frac{\Delta k'^0}{2\pi} \int \frac{\mathrm{d}^3 k'}{(2\pi)^3} \hat{A}^{\nu}_{(0)}(k') \mathrm{e}^{-\mathrm{i}k'\cdot y} \mathrm{d}^4 y$$

$$= \frac{\overline{C}' - \overline{C}}{2} \frac{\partial^2}{\partial x_\mu \partial x^\nu} \int \frac{\mathrm{d}^3 k}{(2\pi)^3} \frac{\mathrm{i}}{2|\boldsymbol{k}|} \mathrm{e}^{-\mathrm{i}k\cdot x} \Big|_{k^0 = |\boldsymbol{k}|} \sum_{k'^0 \neq \pm |\boldsymbol{k}'|} \Delta k'^0 \int \mathrm{d}^3 k' \hat{A}^{\nu}_{(0)}(k') \left( \int \frac{\mathrm{d}^4 y}{(2\pi)^4} \mathrm{e}^{\mathrm{i}(k-k')\cdot y} \Big|_{k^0 = |\boldsymbol{k}|, k'^0 \neq \pm |\boldsymbol{k}'|} \right)$$

$$- \frac{\overline{C}' - \overline{C}}{2} \frac{\partial^2}{\partial x_\mu \partial x^\nu} \int \frac{\mathrm{d}^3 k}{(2\pi)^3} \frac{\mathrm{i}}{2|\boldsymbol{k}|} \mathrm{e}^{\mathrm{i}k\cdot x} \Big|_{k^0 = |\boldsymbol{k}|} \sum_{k'^0 \neq \pm |\boldsymbol{k}'|} \Delta k'^0 \int \mathrm{d}^3 k' \hat{A}^{\nu}_{(0)}(k') \left( \int \frac{\mathrm{d}^4 y}{(2\pi)^4} \mathrm{e}^{-\mathrm{i}(k+k')\cdot y} \Big|_{k^0 = |\boldsymbol{k}|, k'^0 \neq \pm |\boldsymbol{k}'|} \right)$$

$$= \frac{\overline{C}' - \overline{C}}{2} \frac{\partial^2}{\partial x_\mu \partial x^\nu} \int \frac{\mathrm{d}^3 k}{(2\pi)^3} \frac{\mathrm{i}}{2|\boldsymbol{k}|} \sum_{k'^0 \neq \pm |\boldsymbol{k}'|} \Delta k'^0 \int \mathrm{d}^3 k' \hat{A}^{\nu}_{(0)}(k')$$

$$\times \left( \mathrm{e}^{-\mathrm{i}k\cdot x} \delta(k^0 - k'^0) \delta^3(\boldsymbol{k} - \boldsymbol{k}') \Big|_{k^0 = |\boldsymbol{k}|, k'^0 \neq \pm |\boldsymbol{k}'|} - \mathrm{e}^{\mathrm{i}k\cdot x} \delta(k^0 + k'^0) \delta^3(\boldsymbol{k} + \boldsymbol{k}') \Big|_{k^0 = |\boldsymbol{k}|, k'^0 \neq \pm |\boldsymbol{k}'|} \right)$$

$$= 0,$$

since $\delta(k^0 \pm k'^0) \delta^3(\boldsymbol{k} \pm \boldsymbol{k}') \Big|_{k^0 = |\boldsymbol{k}|, k'^0 \neq \pm |\boldsymbol{k}'|} = 0$. Hence, what $\overline{A}^{\mu}_{(0)//}(x)$ we obtain by different choice of the constant $\overline{C}$ are the same.

② The longitudinal part $\tilde{A}^{\mu}_{(0)//}(x)$ of $\tilde{A}^{\mu}_{(0)}(x)$

We have several approaches to deal with the singularities in the integral

$\int d^4 y D(x-y)\widetilde{A}^\nu_{(0)}(y)$ arisen from $\widetilde{A}^\mu_{(0)}(x)$ satisfies (3-20), this question is essentially is only that of how to *define* this singular integral. Here we present a simple method.

According to (3-14) and (3-22), we first write the longitudinal part $\widetilde{A}^\mu_{(0)//}(x)$ of $\widetilde{A}^\mu_{(0)}(x)$ to the form $\widetilde{A}^\mu_{(0)//}(x) = \frac{\partial}{\partial x_\mu} \int d^4 y D_{\text{Total}}(x-y) \frac{\partial \widetilde{A}^\nu_{(0)}(y)}{\partial y^\nu}$, and then, using the formula

$$\Box\left(\frac{1}{2}x_\lambda K^\lambda(x)\right) = K^\lambda{}_{,\lambda}(x) + \frac{1}{2}x_\lambda \Box K^\lambda(x),$$

it is easy to prove that this formula holds for an arbitrary vector function $K^\mu(x)$, and taking into account (3-19) and (3-23), through integration by parts we have

$$\widetilde{A}^\mu_{(0)//}(x) = \frac{\partial}{\partial x_\mu} \int D_{\text{Total}}(x-y)\left(\frac{\partial^2}{\partial y_\tau \partial y^\tau}\left(\frac{1}{2}y_\lambda \widetilde{A}^\lambda_{(0)}(y)\right) - \frac{1}{2}y_\lambda \frac{\partial^2 \widetilde{A}^\lambda_{(0)}(y)}{\partial y_\tau \partial y^\tau}\right)d^4 y$$

$$= \frac{\partial}{\partial x_\mu} \int D_{\text{Total}}(x-y)\frac{\partial^2}{\partial y_\tau \partial y^\tau}\left(\frac{1}{2}y_\lambda \widetilde{A}^\lambda_{(0)}(y)\right)d^4 y = \frac{\partial}{\partial x_\mu} \int \frac{\partial^2 D_{\text{Total}}(x-y)}{\partial y_\tau \partial y^\tau}\frac{1}{2}y_\lambda \widetilde{A}^\lambda_{(0)}(y)d^4 y \quad (3\text{-}25)$$

$$= \frac{\partial}{\partial x_\mu} \int \delta^4(x-y)\frac{1}{2}y_\lambda \widetilde{A}^\lambda_{(0)}(y)d^4 y = \frac{\partial}{\partial x_\mu}\left(\frac{1}{2}x_\lambda \widetilde{A}^\lambda_{(0)}(x)\right).$$

We see that $\widetilde{A}^\mu_{(0)//}(x)$ determined by (3-25) is independent of the choice of the constants $\overline{C}$ in the function $D_{\text{Total}}(x)$ given by (3-24).

### 3.2.4 The formal formulas of longitudinal and transverse four-vectors

Summarizing (3-12), (3-13), (3-22) and (3-25), formally, the longitudinal four-vector $A^\mu_{//}(x)$ of $A^\mu(x)$ can be written to the following unified form:

$$A^\mu_{//}(x) = \frac{\partial^2}{\partial x_\mu \partial x^\nu} \int d^4 y D_{\text{Total}}(x-y) A^\nu(y), \quad (3\text{-}26)$$

of course, we have to follow the calculation method given by (3-12), (3-13), (3-22) and (3-25) when we calculate concretely the longitudinal four-vector $A^\mu_{//}(x)$ of $A^\mu(x)$.

According to (3-3), the transverse four-vector $A^\mu_\perp(x)$ corresponding to $A^\mu(x)$ is

$$A^\mu_\perp(x) = A^\mu(x) - \frac{\partial^2}{\partial x_\mu \partial x^\nu} \int d^4 y D_{\text{Total}}(x-y) A^\nu(y). \quad (3\text{-}27)$$

It has been proved that $A^\mu_{//}(x)$ is independent of the choice of the constant $\overline{C}$ in the function $D_{\text{Total}}(x)$ given by (3-24), $A^\mu_\perp(x)$ as well.

It is obvious that both $A^\mu_{//}(x)$ and $A^\mu_\perp(x)$ given by (3-26) and (3-27) respectively are Lorentz covariant, and, $A^\mu_\perp(x)$ given by (3-27) is invariant under the transformation (2-5). Especially, even if the function $\theta(x)$ in (2-5) satisfies $\Box \theta(x) = 0$, which is so called gauge transformation of the second kind, $A_\perp^\mu(x)$ given by (3-27) is still invariant.

### 3.2.5 The determination of transverse four-vector $A_\perp^\mu(x)$ for a special case

It is obvious that $A_\perp^\mu(x)$ and $A_\parallel^\mu(x)$ given by (3-27) and (3-26) satisfy

$$A_{\perp,\mu}^\mu(x) = 0, \quad A_\parallel^{\mu,\nu}(x) - A_\parallel^{\nu,\mu}(x) = 0, \tag{3-28}$$

respectively. However, even if a four-vector $A^\mu(x)$ satisfies

$$A^\mu{}_{,\mu}(x) = 0, \tag{3-29}$$

we cannot conclude that $A^\mu(x)$ is a transverse four-vector immediately, since maybe in $A^\mu(x)$ there is still a longitudinal four-vector $\partial^\mu \varphi(x)$, where $\varphi(x)$ satisfies $\Box \varphi(x) = 0$.

For this special case, when we want to determine such longitudinal term $\partial^\mu \varphi(x)$ in $A^\mu(x)$, similar to (3-16) ~ (3-18), $A^\mu(x)$ expressed by (3-5) is firstly divided into two parts:

$$A^\mu(x) = \int \frac{d^4k}{(2\pi)^4} \hat{A}^\mu(k) e^{-ik\cdot x} = \sum_{k^0} \frac{\Delta k^0}{2\pi} \int \frac{d^3k}{(2\pi)^3} \hat{A}^\mu(k) e^{-ik\cdot x}$$

$$= \sum_{\substack{k^0 \\ k^0 \neq \pm|\mathbf{k}|}} \frac{\Delta k^0}{2\pi} \int \frac{d^3k}{(2\pi)^3} \hat{A}^\mu(k) e^{-ik\cdot x} + \frac{\Delta k^0}{2\pi} \int \frac{d^3k}{(2\pi)^3} \hat{A}^\mu(k) e^{-ik\cdot x} \bigg|_{k^0=|\mathbf{k}|} + \frac{\Delta k^0}{2\pi} \int \frac{d^3k}{(2\pi)^3} \hat{A}^\mu(k) e^{-ik\cdot x} \bigg|_{k^0=-|\mathbf{k}|}$$

$$= \overline{A}^\mu(x) + \widetilde{A}^\mu(x), \tag{3-30}$$

$$\overline{A}^\mu(x) \equiv \sum_{\substack{k^0 \\ k^0 \neq \pm|\mathbf{k}|}} \frac{\Delta k^0}{2\pi} \int \frac{d^3k}{(2\pi)^3} \hat{A}^\mu(k) e^{-ik\cdot x}, \tag{3-31}$$

$$\begin{aligned}
\widetilde{A}^\mu(x) &\equiv \frac{\Delta k^0}{2\pi} \int \frac{d^3k}{(2\pi)^3} \hat{A}^\mu(k) e^{-ik\cdot x} \bigg|_{k^0=|\mathbf{k}|} + \frac{\Delta k^0}{2\pi} \int \frac{d^3k}{(2\pi)^3} \hat{A}^\mu(k) e^{-ik\cdot x} \bigg|_{k^0=-|\mathbf{k}|} \\
&= \frac{\Delta k^0}{2\pi} \int \frac{d^3k}{(2\pi)^3} \left( \hat{A}^\mu(k) e^{-ik\cdot x} + \hat{A}^{\mu*}(k) e^{ik\cdot x} \right) \bigg|_{k^0=|\mathbf{k}|} \\
&= \frac{\Delta k^0}{2\pi} \int \frac{d^3k}{(2\pi)^3} A^\mu(y) e^{-ik\cdot(x-y)} \bigg|_{k^0=|\mathbf{k}|} d^4y + \frac{\Delta k^0}{2\pi} \int \frac{d^3k}{(2\pi)^3} A^\mu(y) e^{-ik\cdot(x-y)} \bigg|_{k^0=-|\mathbf{k}|} d^4y \\
&= \frac{\Delta k^0}{2\pi} \int \frac{d^3k}{(2\pi)^3} \left( e^{-i|\mathbf{k}|(x^0-y^0)+i\mathbf{k}\cdot(\mathbf{x}-\mathbf{y})} + e^{i|\mathbf{k}|(x^0-y^0)+i\mathbf{k}\cdot(\mathbf{x}-\mathbf{y})} \right) A^\mu(y) d^4y,
\end{aligned} \tag{3-32}$$

where $\widetilde{A}^\mu(x)$ satisfies $\Box \widetilde{A}^\mu(x) = 0$.

And, further, for, according to the four polarization vectors $\varepsilon_{(\lambda)}^\mu(\mathbf{k})$ ($\lambda = 0, 1, 2, 3$) that satisfy (3-6) but in which $k^2 = 0$, $\widetilde{A}^\mu(x)$ has the expansion:

$$\widetilde{A}^{\mu}(x) = \frac{\Delta k^0}{2\pi} \int \frac{d^3 k}{(2\pi)^3} \left( e^{-i|k|(x^0 - y^0) + i\mathbf{k}\cdot(\mathbf{x}-\mathbf{y})} + e^{i|k|(x^0 - y^0) + i\mathbf{k}\cdot(\mathbf{x}-\mathbf{y})} \right) g^{\mu\nu} A_{\nu}(y) \, d^4 y$$

$$= \frac{\Delta k^0}{2\pi} \int \frac{d^3 k}{(2\pi)^3} \left( e^{-i|k|(x^0 - y^0) + i\mathbf{k}\cdot(\mathbf{x}-\mathbf{y})} + e^{i|k|(x^0 - y^0) + i\mathbf{k}\cdot(\mathbf{x}-\mathbf{y})} \right) \sum_{\lambda=0}^{3} \sum_{\lambda'=0}^{3} g_{(\lambda)(\lambda')} \varepsilon_{(\lambda)}^{\mu}(k) \varepsilon_{(\lambda')}^{\nu}(k) A_{\nu}(y) \, d^4 y \quad (3\text{-}33)$$

$$= \widetilde{A}_{(0)}^{\mu}(x) + \sum_{\lambda=1}^{2} \widetilde{A}_{(\lambda)}^{\mu}(x) + \widetilde{A}_{(3)}^{\mu}(x),$$

$$\widetilde{A}_{(0)}^{\mu}(x) = \frac{\Delta k^0}{2\pi} \int \frac{d^3 k}{(2\pi)^3} \left( e^{-i|k|(x^0 - y^0) + i\mathbf{k}\cdot(\mathbf{x}-\mathbf{y})} + e^{i|k|(x^0 - y^0) + i\mathbf{k}\cdot(\mathbf{x}-\mathbf{y})} \right) \varepsilon_{(0)}^{\mu}(k) \left( \varepsilon_{(0)}^{\nu}(k) + \varepsilon_{(3)}^{\nu}(k) \right) A_{\nu}(y) \, d^4 y$$

$$= \frac{\Delta k^0}{2\pi} \int \frac{d^3 k}{(2\pi)^3} \left( e^{-i|k|x^0 + i\mathbf{k}\cdot\mathbf{x}} + e^{i|k|x^0 + i\mathbf{k}\cdot\mathbf{x}} \right) n^{\mu} \frac{k^{\nu}}{k \cdot n} \int d^4 y \left( e^{i|k|y^0 - i\mathbf{k}\cdot\mathbf{y}} + e^{-i|k|y^0 - i\mathbf{k}\cdot\mathbf{y}} \right) A_{\nu}(y), \quad (3\text{-}34)$$

$$\widetilde{A}_{(\lambda)}^{\mu}(x) = -\frac{\Delta k^0}{2\pi} \int \frac{d^3 k}{(2\pi)^3} \left( e^{-i|k|(x^0 - y^0) + i\mathbf{k}\cdot(\mathbf{x}-\mathbf{y})} + e^{i|k|(x^0 - y^0) + i\mathbf{k}\cdot(\mathbf{x}-\mathbf{y})} \right)$$
$$\times \varepsilon_{(\lambda)}^{\mu}(k) \varepsilon_{(\lambda)}^{\nu}(k) A_{\nu}(y) \, d^4 y \qquad (\lambda = 1, 2), \quad (3\text{-}35)$$

$$\widetilde{A}_{(3)}^{\mu}(x) = -\frac{\Delta k^0}{2\pi} \int \frac{d^3 k}{(2\pi)^3} \left( e^{-i|k|(x^0 - y^0) + i\mathbf{k}\cdot(\mathbf{x}-\mathbf{y})} + e^{i|k|(x^0 - y^0) + i\mathbf{k}\cdot(\mathbf{x}-\mathbf{y})} \right) \frac{k^{\mu}}{k \cdot n} \varepsilon_{(3)}^{\nu}(k) A_{\nu}(y) \, d^4 y$$

$$= -\frac{\Delta k^0}{2\pi} \int \frac{d^3 k}{(2\pi)^3} \left( e^{-i|k|x^0 + i\mathbf{k}\cdot\mathbf{x}} + e^{i|k|x^0 + i\mathbf{k}\cdot\mathbf{x}} \right) \frac{k^{\mu}}{k \cdot n} \varepsilon_{(3)}^{\nu}(k) \int d^4 y \left( e^{i|k|y^0 - i\mathbf{k}\cdot\mathbf{y}} + e^{-i|k|y^0 - i\mathbf{k}\cdot\mathbf{y}} \right) A_{\nu}(y). \quad (3\text{-}36)$$

Notice $\dfrac{\partial \widetilde{A}_{(\lambda)}^{\mu}(x)}{\partial x^{\mu}} = 0$ $(\lambda = 1, 2)$, $\dfrac{\partial \widetilde{A}_{(3)}^{\mu}(x)}{\partial x^{\mu}} = 0$ (since $k^2 = 0$), the condition (3-29) leads to

$$\frac{\partial \widetilde{A}^{\mu}(x)}{\partial x^{\mu}} = \frac{\partial \widetilde{A}_{(0)}^{\mu}(x)}{\partial x^{\mu}} + \sum_{\lambda=1}^{2} \frac{\partial \widetilde{A}_{(\lambda)}^{\mu}(x)}{\partial x^{\mu}} + \frac{\partial \widetilde{A}_{(3)}^{\mu}(x)}{\partial x^{\mu}} = \frac{\partial \widetilde{A}_{(0)}^{\mu}(x)}{\partial x^{\mu}}$$

$$= \frac{\Delta k^0}{2\pi} \int \frac{d^3 k}{(2\pi)^3} \left( e^{-i|k|x^0 + i\mathbf{k}\cdot\mathbf{x}} + e^{i|k|x^0 + i\mathbf{k}\cdot\mathbf{x}} \right) k^{\nu} \int d^4 y \left( e^{i|k|y^0 - i\mathbf{k}\cdot\mathbf{y}} + e^{-i|k|y^0 - i\mathbf{k}\cdot\mathbf{y}} \right) A_{\nu}(y)$$

$$= 0,$$

this means

$$k^{\nu} \int d^4 y \left( e^{i|k|y^0 - i\mathbf{k}\cdot\mathbf{y}} + e^{-i|k|y^0 - i\mathbf{k}\cdot\mathbf{y}} \right) A_{\nu}(y) = 0; \quad (3\text{-}37)$$

(In fact, using (3-29) we can prove (3-37) directly.) And, further, we have $\widetilde{A}_{(0)}^{\mu}(x) = 0$.

Substituting $\varepsilon_{(3)}^{\mu}(k) = \dfrac{k^{\mu}}{k \cdot n} - \varepsilon_{(0)}^{\mu}(k)$ to (3-36) and using (3-37) we obtain

$$\widetilde{A}_{(3)}^{\mu}(x) = -\frac{\Delta k^0}{2\pi} \int \frac{d^3 k}{(2\pi)^3} \left( e^{-i|k|x^0 + i\mathbf{k}\cdot\mathbf{x}} + e^{i|k|x^0 + i\mathbf{k}\cdot\mathbf{x}} \right) \frac{k^{\mu}}{k \cdot n}$$
$$\times \left( \frac{k^{\nu}}{k \cdot n} - \varepsilon_{(0)}^{\nu}(k) \right) \int d^4 y \left( e^{i|k|y^0 - i\mathbf{k}\cdot\mathbf{y}} + e^{-i|k|y^0 - i\mathbf{k}\cdot\mathbf{y}} \right) A_{\nu}(y) \quad (3\text{-}38)$$

$$= \frac{\Delta k^0}{2\pi} \int \frac{d^3 k}{(2\pi)^3} \left( e^{-i|k|x^0 + i\mathbf{k}\cdot\mathbf{x}} + e^{i|k|x^0 + i\mathbf{k}\cdot\mathbf{x}} \right) \frac{k^{\mu}}{k \cdot n} \varepsilon_{(0)}^{\nu}(k) \int d^4 y \left( e^{i|k|y^0 - i\mathbf{k}\cdot\mathbf{y}} + e^{-i|k|y^0 - i\mathbf{k}\cdot\mathbf{y}} \right) A_{\nu}(y).$$

From the above discussion we see that even if $A^\mu(x)$ satisfies (3-29), maybe in which there is still a longitudinal four-vector $\widetilde{A}^\mu_{(3)}(x)$ expressed by (3-38); and only after removing this term can we obtain the corresponding transverse four-vector $A^\mu_\perp(x) = A^\mu(x) - \widetilde{A}^\mu_{(3)}(x)$.

We discuss two examples. One is

$$A^\mu(x) = \left(A^0(\boldsymbol{x}), \frac{1}{2}\boldsymbol{B}\times\boldsymbol{x}\right) = \left(A^0(\boldsymbol{x}), \frac{1}{2}(B^2 x^3 - B^3 x^2), \frac{1}{2}(B^3 x^1 - B^1 x^3), \frac{1}{2}(B^1 x^2 - B^2 x^1)\right), \quad (3\text{-}39)$$

where $A^0(\boldsymbol{x})$ is independent of time $x^0$ and $\boldsymbol{B}$ is a constant 3-vector, we have

$$A^\mu{}_{,\mu}(x) = \frac{\partial A^0(\boldsymbol{x})}{\partial t} + \nabla\cdot\left(\frac{1}{2}\boldsymbol{B}\times\boldsymbol{x}\right) = 0.$$

$A^\mu(x)$ given by (3-33) is independent of time $x^0$, (3-32) thus becomes

$$\widetilde{A}^\mu(x) = \frac{\Delta k^0}{2\pi}\int d^3y\, A^\mu(\boldsymbol{y})\int dy^0 \int \frac{d^3 k}{(2\pi)^3}\left(e^{-i|\boldsymbol{k}|(x^0-y^0)+i\boldsymbol{k}\cdot(\boldsymbol{x}-\boldsymbol{y})} + e^{i|\boldsymbol{k}|(x^0-y^0)+i\boldsymbol{k}\cdot(\boldsymbol{x}-\boldsymbol{y})}\right)$$

$$= \frac{\Delta k^0}{2\pi}\int d^3y\, A^\mu(\boldsymbol{y})\int dy^0 \int_0^\infty \frac{k^2 dk}{(2\pi)^3}\left(e^{-ik(x^0-y^0)} + e^{ik(x^0-y^0)}\right)\int_0^\pi \sin\theta\, d\theta\, e^{ik|\boldsymbol{x}-\boldsymbol{y}|\cos\theta}\int_0^{2\pi} d\varphi$$

$$= \frac{\Delta k^0}{2\pi}\int d^3y\, A^\mu(\boldsymbol{y})\int_0^\infty \frac{k^2 dk}{(2\pi)^2}\left(e^{-ikx^0}\int dy^0 e^{iky^0} + e^{ikx^0}\int dy^0 e^{-iky^0}\right)\frac{-1}{ik|\boldsymbol{x}-\boldsymbol{y}|}\left(e^{-ik|\boldsymbol{x}-\boldsymbol{y}|} - e^{ik|\boldsymbol{x}-\boldsymbol{y}|}\right)$$

$$= \frac{\Delta k^0}{2\pi}i\int d^3y\, \frac{A^\mu(\boldsymbol{y})}{|\boldsymbol{x}-\boldsymbol{y}|}\int_0^\infty \frac{dk}{2\pi} k\delta(k)\left(e^{-ikx^0} + e^{ikx^0}\right)\left(e^{-ik|\boldsymbol{x}-\boldsymbol{y}|} - e^{ik|\boldsymbol{x}-\boldsymbol{y}|}\right) = 0,$$

since $k\delta(k) = 0$. $A^\mu(x)$ given by (3-33) is thus a transverse four-vector.

Another example is

$$A^\mu(x) = \int D_{\text{ret}}(x-y)J^\mu(y)d^4 y, \quad \text{where } J^\mu(x) \text{ satisfies } \frac{\partial J^\mu(x)}{\partial x^\mu} = 0. \quad (3\text{-}40)$$

For this case, we have

$$\frac{\partial A^\mu(x)}{\partial x^\mu} = \int \frac{\partial D_{\text{ret}}(x-y)}{\partial x^\mu}J^\mu(y)d^4 y = -\int \frac{\partial D_{\text{ret}}(x-y)}{\partial y^\mu}J^\mu(y)d^4 y = \int D_{\text{ret}}(x-y)\frac{\partial J^\mu(y)}{\partial y^\mu}d^4 y = 0.$$

(3-38) becomes

$$\widetilde{A}^{\mu}_{(3)}(x) = \int \frac{d^3k}{(2\pi)^3} \left( e^{-i|\mathbf{k}|(x^0-y^0)+i\mathbf{k}\cdot(\mathbf{x}-\mathbf{y})} + e^{i|\mathbf{k}|(x^0-y^0)+i\mathbf{k}\cdot(\mathbf{x}-\mathbf{y})} \right) \frac{k^{\mu}}{k\cdot n} \varepsilon^{\nu}_{(0)}(\mathbf{k}) \int D_{\text{ret}}(y-z) J_{\nu}(z) d^4z \, d^4y$$

$$= \frac{\Delta k^0}{2\pi} \int \frac{d^3k}{(2\pi)^3} \left( e^{-i|\mathbf{k}|(x^0-y^0)+i\mathbf{k}\cdot(\mathbf{x}-\mathbf{y})} + e^{i|\mathbf{k}|(x^0-y^0)+i\mathbf{k}\cdot(\mathbf{x}-\mathbf{y})} \right) \frac{k^{\mu}}{k\cdot n} \varepsilon^{\nu}_{(0)}(\mathbf{k})$$
$$\times \theta(y^0-z^0) \int \frac{d^3k'}{(2\pi)^3} \frac{i}{2|\mathbf{k}'|} \left( e^{-i|\mathbf{k}'|(y^0-z^0)+i\mathbf{k}'\cdot(\mathbf{y}-\mathbf{z})} - e^{i|\mathbf{k}'|(y^0-z^0)+i\mathbf{k}'\cdot(\mathbf{y}-\mathbf{z})} \right) J_{\nu}(z) d^4z d^4y$$

$$= \frac{\Delta k^0}{2\pi} \int \frac{d^3k}{(2\pi)^3} \left( e^{-i|\mathbf{k}|(x^0-y^0)+i\mathbf{k}\cdot\mathbf{x}} + e^{i|\mathbf{k}|(x^0-y^0)+i\mathbf{k}\cdot\mathbf{x}} \right) \frac{k^{\mu}}{k\cdot n} \varepsilon^{\nu}_{(0)}(\mathbf{k})$$
$$\times \theta(y^0-z^0) \int d^3k' \frac{i}{2|\mathbf{k}'|} \left( e^{-i|\mathbf{k}'|(y^0-z^0)-i\mathbf{k}'\cdot\mathbf{z}} - e^{i|\mathbf{k}'|(y^0-z^0)-i\mathbf{k}'\cdot\mathbf{z}} \right) J_{\nu}(z) d^4z dy^0 \int \frac{d^3y}{(2\pi)^3} e^{i(\mathbf{k}'-\mathbf{k})\cdot\mathbf{y}}$$

$$= \frac{\Delta k^0}{2\pi} \int \frac{d^3k}{(2\pi)^3} \left( e^{-i|\mathbf{k}|(x^0-y^0)+i\mathbf{k}\cdot\mathbf{x}} + e^{i|\mathbf{k}|(x^0-y^0)+i\mathbf{k}\cdot\mathbf{x}} \right) \frac{k^{\mu}}{k\cdot n} \varepsilon^{\nu}_{(0)}(\mathbf{k})$$
$$\times \theta(y^0-z^0) \int d^3k' \delta^3(\mathbf{k}'-\mathbf{k}) \frac{i}{2|\mathbf{k}'|} \left( e^{-i|\mathbf{k}'|(y^0-z^0)-i\mathbf{k}'\cdot\mathbf{z}} - e^{i|\mathbf{k}'|(y^0-z^0)-i\mathbf{k}'\cdot\mathbf{z}} \right) J_{\nu}(z) d^4z dy^0$$

$$= \frac{\Delta k^0}{2\pi} \int \frac{d^3k}{(2\pi)^3} \left( e^{-i|\mathbf{k}|(x^0-y^0)+i\mathbf{k}\cdot\mathbf{x}} + e^{i|\mathbf{k}|(x^0-y^0)+i\mathbf{k}\cdot\mathbf{x}} \right) \frac{k^{\mu}}{k\cdot n} \varepsilon^{\nu}_{(0)}(\mathbf{k})$$
$$\times \theta(y^0-z^0) \frac{i}{2|\mathbf{k}|} \left( e^{-i|\mathbf{k}|(y^0-z^0)-i\mathbf{k}\cdot\mathbf{z}} - e^{i|\mathbf{k}|(y^0-z^0)-i\mathbf{k}\cdot\mathbf{z}} \right) J_{\nu}(z) d^4z dy^0$$

$$= \frac{\Delta k^0}{2\pi} \int \frac{d^3k}{(2\pi)^3} \frac{i}{2|\mathbf{k}|} \left( e^{-i|\mathbf{k}|(x^0-z^0)+i\mathbf{k}\cdot(\mathbf{x}-\mathbf{z})} - e^{i|\mathbf{k}|(x^0-z^0)+i\mathbf{k}\cdot(\mathbf{x}-\mathbf{z})} \right) \frac{k^{\mu}}{k\cdot n} \varepsilon^{\nu}_{(0)}(\mathbf{k}) J_{\nu}(z) d^4z \int dy^0 \theta(y^0-z^0)$$
$$+ \frac{\Delta k^0}{2\pi} \int \frac{d^3k}{(2\pi)^3} \frac{i}{2|\mathbf{k}|} \left( e^{i|\mathbf{k}|(x^0+z^0)+i\mathbf{k}\cdot(\mathbf{x}-\mathbf{z})} \int dy^0 e^{-2i|\mathbf{k}|y^0} \theta(y^0-z^0) \right.$$
$$\left. - e^{-i|\mathbf{k}|(x^0+z^0)+i\mathbf{k}\cdot(\mathbf{x}-\mathbf{z})} \int dy^0 e^{2i|\mathbf{k}|y^0} \theta(y^0-z^0) \right) \frac{k^{\mu}}{k\cdot n} \varepsilon^{\nu}_{(0)}(\mathbf{k}) J_{\nu}(z) d^4z$$

$$= \lim_{T\to\infty} \left[ \frac{\Delta k^0}{2\pi} \int \frac{d^3k}{(2\pi)^3} \frac{i}{2|\mathbf{k}|} \left( e^{-i|\mathbf{k}|(x^0-z^0)+i\mathbf{k}\cdot(\mathbf{x}-\mathbf{z})} - e^{i|\mathbf{k}|(x^0-z^0)+i\mathbf{k}\cdot(\mathbf{x}-\mathbf{z})} \right) \frac{k^{\mu}}{k\cdot n} \varepsilon^{\nu}_{(0)}(\mathbf{k}) J_{\nu}(z) d^4z \int_{z^0}^{T} dy^0 \right.$$
$$+ \frac{\Delta k^0}{2\pi} \int \frac{d^3k}{(2\pi)^3} \frac{i}{2|\mathbf{k}|} \left( e^{i|\mathbf{k}|(x^0+z^0)+i\mathbf{k}\cdot(\mathbf{x}-\mathbf{z})} \int_{z^0}^{T} dy^0 e^{-2i|\mathbf{k}|y^0} \right.$$
$$\left. \left. - e^{-i|\mathbf{k}|(x^0+z^0)+i\mathbf{k}\cdot(\mathbf{x}-\mathbf{z})} \int_{z^0}^{T} dy^0 e^{2i|\mathbf{k}|y^0} \right) \frac{k^{\mu}}{k\cdot n} \varepsilon^{\nu}_{(0)}(\mathbf{k}) J_{\nu}(z) d^4z \right]$$

$$= \lim_{T\to\infty} \left[ \frac{T \Delta k^0}{2\pi} \int \frac{d^3k}{(2\pi)^3} \frac{i}{2|\mathbf{k}|} \left( e^{-i|\mathbf{k}|(x^0-z^0)+i\mathbf{k}\cdot(\mathbf{x}-\mathbf{z})} - e^{i|\mathbf{k}|(x^0-z^0)+i\mathbf{k}\cdot(\mathbf{x}-\mathbf{z})} \right) \frac{k^{\mu}}{k\cdot n} \varepsilon^{\nu}_{(0)}(\mathbf{k}) J_{\nu}(z) d^4z \right.$$
$$+ z^0 \frac{\Delta k^0}{2\pi} \int \frac{d^3k}{(2\pi)^3} \frac{i}{2|\mathbf{k}|} \left( e^{-i|\mathbf{k}|(x^0-z^0)+i\mathbf{k}\cdot(\mathbf{x}-\mathbf{z})} - e^{i|\mathbf{k}|(x^0-z^0)+i\mathbf{k}\cdot(\mathbf{x}-\mathbf{z})} \right) \frac{k^{\mu}}{k\cdot n} \varepsilon^{\nu}_{(0)}(\mathbf{k}) J_{\nu}(z) d^4z$$
$$+ \frac{\Delta k^0}{2\pi} \int \frac{d^3k}{(2\pi)^3} \frac{i}{2|\mathbf{k}|} \left( e^{i|\mathbf{k}|(x^0+z^0)+i\mathbf{k}\cdot(\mathbf{x}-\mathbf{z})} \frac{1}{-2i|\mathbf{k}|} \left( e^{-2i|\mathbf{k}|T} - e^{-2i|\mathbf{k}|z^0} \right) \right.$$
$$\left. \left. - e^{-i|\mathbf{k}|(x^0+z^0)+i\mathbf{k}\cdot(\mathbf{x}-\mathbf{z})} \frac{1}{2i|\mathbf{k}|} \left( e^{2i|\mathbf{k}|T} - e^{2i|\mathbf{k}|z^0} \right) \right) \frac{k^{\mu}}{k\cdot n} \varepsilon^{\nu}_{(0)}(\mathbf{k}) J_{\nu}(z) d^4z \right],$$

notice $\Delta k^0 = \dfrac{2\pi}{T}$, we obtain

$$\widetilde{A}^{\mu}_{(3)}(x) = \int \frac{d^3k}{(2\pi)^3} \frac{i}{2|\boldsymbol{k}|} \left( e^{-i|\boldsymbol{k}|(x^0-z^0)+i\boldsymbol{k}\cdot(\boldsymbol{x}-\boldsymbol{z})} - e^{i|\boldsymbol{k}|(x^0-z^0)+i\boldsymbol{k}\cdot(\boldsymbol{x}-\boldsymbol{z})} \right) \frac{k^\mu}{k\cdot n} \varepsilon^\nu_{(0)}(\boldsymbol{k}) J_\nu(z) d^4z$$

$$+ \lim_{T\to\infty} \frac{1}{T} \left[ -z^0 \int \frac{d^3k}{(2\pi)^3} \frac{i}{2|\boldsymbol{k}|} \left( e^{-i|\boldsymbol{k}|(x^0-z^0)+i\boldsymbol{k}\cdot(\boldsymbol{x}-\boldsymbol{z})} - e^{i|\boldsymbol{k}|(x^0-z^0)+i\boldsymbol{k}\cdot(\boldsymbol{x}-\boldsymbol{z})} \right) \frac{k^\mu}{k\cdot n} \varepsilon^\nu_{(0)}(\boldsymbol{k}) J_\nu(z) d^4z \right.$$

$$+ \int \frac{d^3k}{(2\pi)^3} \frac{1}{(2|\boldsymbol{k}|)^2} \left( e^{i|\boldsymbol{k}|(x^0-z^0)+i\boldsymbol{k}\cdot(\boldsymbol{x}-\boldsymbol{z})} + e^{-i|\boldsymbol{k}|(x^0-z^0)+i\boldsymbol{k}\cdot(\boldsymbol{x}-\boldsymbol{z})} \right) \frac{k^\mu}{k\cdot n} \varepsilon^\nu_{(0)}(\boldsymbol{k}) J_\nu(z) d^4z$$

$$+ \int \frac{d^3k}{(2\pi)^3} \frac{1}{(2|\boldsymbol{k}|)^2} \left( -e^{i|\boldsymbol{k}|(x^0+z^0)+i\boldsymbol{k}\cdot(\boldsymbol{x}-\boldsymbol{z})} e^{-2i|\boldsymbol{k}|T} - e^{-i|\boldsymbol{k}|(x^0+z^0)+i\boldsymbol{k}\cdot(\boldsymbol{x}-\boldsymbol{z})} e^{2i|\boldsymbol{k}|T} \right)$$

$$\left. \times \frac{k^\mu}{k\cdot n} \varepsilon^\nu_{(0)}(\boldsymbol{k}) J_\nu(z) d^4z \right]$$

$$= \int \frac{d^3k}{(2\pi)^3} \frac{i}{2|\boldsymbol{k}|} \left( e^{-i|\boldsymbol{k}|(x^0-z^0)+i\boldsymbol{k}\cdot(\boldsymbol{x}-\boldsymbol{z})} - e^{i|\boldsymbol{k}|(x^0-z^0)+i\boldsymbol{k}\cdot(\boldsymbol{x}-\boldsymbol{z})} \right) \frac{k^\mu}{k\cdot n} \varepsilon^\nu_{(0)}(\boldsymbol{k}) J_\nu(z) d^4z . \qquad (3\text{-}41)$$

And, further, for $A^\mu(x)$ given by (3-40), we obtain the corresponding transverse four-vector

$$A^\mu_\perp(x) = A^\mu(x) - \widetilde{A}^\mu_{(3)}(x)$$
$$= \int D_{\text{ret}}(x-y) J^\mu(y) d^4y \qquad (3\text{-}42)$$
$$- \int \frac{d^3k}{(2\pi)^3} \frac{i}{2|\boldsymbol{k}|} \left( e^{-i|\boldsymbol{k}|(x^0-z^0)+i\boldsymbol{k}\cdot(\boldsymbol{x}-\boldsymbol{z})} - e^{i|\boldsymbol{k}|(x^0-z^0)+i\boldsymbol{k}\cdot(\boldsymbol{x}-\boldsymbol{z})} \right) \frac{k^\mu}{k\cdot n} \varepsilon^\nu_{(0)}(\boldsymbol{k}) J_\nu(z) d^4z .$$

### 3.2.6 A general transverse four-vector $A^\mu_{\perp\text{G}}(x)$ corresponding to $A^\mu(x)$

Based on $A^\mu_\perp(x)$ defined by (3-27), we can construct a more general transverse four-vector $A^\mu_{\perp\text{G}}(x)$ for $A^\mu(x)$ by the form

$$A^\mu_{\perp\text{G}}(x) = \int d^4y \int \frac{d^4k}{(2\pi)^4} \xi(k) e^{-ik\cdot(x-y)} A^\mu_\perp(y) , \qquad (3\text{-}43)$$

where $\xi(k)$ is a scalar function. For example, we can take $\xi(k) = c_1 \exp\!\left(-c_2(k^2)^2\right) + \dfrac{c_3}{1+c_4(k^2)^2}$, where all $c_i$ ($i=1,2,3,4$) are constants, etc; Of course, if we take $\xi(k)=1$, then $A^\mu_{\perp\text{G}}(x) = A^\mu_\perp(x)$. It is obvious that $A^\mu_{\perp\text{G}}(x)$ defined by (3-43) is invariant under the transformation (2-5).

Although $A^\mu_\perp(x)$ given by (3-27) is independent of the choice of the constant $\overline{C}$ in the function $D_{\text{Total}}(x)$ defined by (3-24), for the sake of clarity, we indicate $D_{\text{Total}}(x) = D_{\text{ret}}(x)$, the formal formulas (3-26) and (3-27) of $A^\mu_{/\!/}(x)$ and $A^\mu_\perp(x)$ can now be written to the forms

$$A^\mu_{/\!/}(x) = \frac{\partial^2}{\partial x_\mu \partial x^\nu} \int D_{\text{ret}}(x-y) A^\nu(y) d^4y ; \qquad (3\text{-}44)$$

$$A_\perp^\mu(x) = A^\mu(x) - A_\parallel^\mu(x) = A^\mu(x) - \frac{\partial^2}{\partial x_\mu \partial x^\nu} \int D_{\text{ret}}(x-y) A^\nu(y) \mathrm{d}^4 y, \tag{3-45}$$

respectively. Of course, $A_\perp^\mu(x)$ in the general transverse four-vector $A_{\perp G}^\mu(x)$ expressed by (3-43) is now given by (3-45).

## 4 The case that both the functions $f(x)$ and $U^\mu(x)$ are dependent of $\psi(x)$ and $A^\mu(x)$ but $V^\mu(x) = x^\mu + aA_{\perp G}^\mu(x)$

### 4.1 General case

We consider a function

$$V^\mu(x) = x^\mu + aA_{\perp G}^\mu(x), \tag{4-1}$$

where $a$ is a constant, the dimension of $aA_{\perp G}^\mu(x)$ is length, e.g., meter in the SI units; $A_{\perp G}^\mu(x)$ is defined by (3-43). The corresponding expression of $V_\nu^\mu(x)$ is

$$V_\nu^\mu(x) = \frac{\partial V^\mu(x)}{\partial x^\nu} = \delta_\nu^\mu + aA_{\perp G,\nu}^\mu(x); \tag{4-2}$$

And, further, the expressions of $V(x)$ and $\tilde{V}_\nu^\mu(x)$ are (See the discussion in the Appendix A.5 of this paper)

$$V(x) = 1 - \frac{a^2}{2} A_{\perp G,\nu}^\mu(x) A_{\perp G,\mu}^\nu(x) + \frac{a^3}{3} A_{\perp G,\beta}^\alpha(x) A_{\perp G,\gamma}^\beta(x) A_{\perp G,\alpha}^\gamma(x)$$
$$+ \frac{a^4}{4} \left( \frac{1}{2} A_{\perp G,\nu}^\mu(x) A_{\perp G,\mu}^\nu(x) A_{\perp G,\sigma}^\rho(x) A_{\perp G,\rho}^\sigma(x) - A_{\perp G,\nu}^\mu(x) A_{\perp G,\rho}^\nu(x) A_{\perp G,\sigma}^\rho(x) A_{\perp G,\mu}^\sigma(x) \right), \tag{4-3}$$

$$\tilde{V}_\nu^\mu(x) = \frac{1}{V(x)} \Bigg[ \delta_\nu^\mu - aA_{\perp G,\nu}^\mu(x) + a^2 \left( -\frac{1}{2} \delta_\nu^\mu A_{\perp G,\beta}^\alpha(x) A_{\perp G,\alpha}^\beta(x) + A_{\perp G,\lambda}^\mu(x) A_{\perp G,\nu}^\lambda(x) \right)$$
$$+ a^3 \bigg( \frac{1}{3} \delta_\nu^\mu A_{\perp G,\beta}^\alpha(x) A_{\perp G,\gamma}^\beta(x) A_{\perp G,\alpha}^\gamma(x) + \frac{1}{2} A_{\perp G,\nu}^\mu(x) A_{\perp G,\beta}^\alpha(x) A_{\perp G,\alpha}^\beta(x) \tag{4-4}$$
$$- A_{\perp G,\alpha}^\mu(x) A_{\perp G,\beta}^\alpha(x) A_{\perp G,\nu}^\beta(x) \bigg) \Bigg].$$

$V^\mu(x)$ given by (4-1) thus satisfies (1-6) and (1-8) since $V(x) \neq 0$ and $V^\mu(x) \to x^\mu$ when $a \to 0$, respectively.

For the case that both the functions $f(x)$ and $U^\mu(x)$ are dependent of $\psi(x)$ and $A^\mu(x)$ but $V^\mu(x)$ is given by (4-1), if we still choose (2-1), (2-2) and (2-4) as the action of the system, then this action is gauge invariant, since all $V^\mu(x)$, $V_\nu^\mu(x)$, $V(x)$ and $\tilde{V}_\nu^\mu(x)$ given by (4-1), (4-2), (4-3) and (4-4) respectively are invariant under the gauge transformation (2-5).

Obeying the action principle, what obtained the equation of motion of charged particle by the variational equation $\dfrac{\delta S}{\delta \psi(x)} = 0$ is

$$\left(i\gamma^\mu \frac{\partial}{\partial x^\mu} - m\right)\psi(x) = e\gamma^\mu U_\mu^\alpha(x)\int d^4 y f\left(U^\lambda(x) - \left(y^\lambda + aA_{\perp G}^\lambda(y)\right)\right) V(y)\widetilde{V}_\alpha^\beta(y)A_\beta(y)\psi(x); \tag{4-5}$$

The equation of motion of electromagnetic field obtained by the variational equation $\frac{\delta S}{\delta A_\mu(y)} = 0$

is (The concrete derivation is in the Appendix B of this paper)

$$\frac{\partial}{\partial y^\nu}\left(A^{\mu,\nu}(y) - A^{\nu,\mu}(y)\right) = eW_{(1)}^\mu(y) + aeW_{(2)\perp G}^\mu(y), \tag{4-6}$$

where

$$W_{(1)}^\mu(y) = V(y)\widetilde{V}_\nu^\mu(y)\int d^4 x j^\alpha(x) U_\alpha^\nu(x) f\left(U^\lambda(x) - \left(y^\lambda + aA_{\perp G}^\lambda(y)\right)\right), \tag{4-7}$$

$V(x)$ and $\widetilde{V}_\nu^\mu(x)$ are given by (4-3) and (4-4), respectively;

$$W_{(2)\perp G}^\mu(y) = \int d^4 z \int \frac{d^4 k}{(2\pi)^4} \xi(k) e^{-ik\cdot(y-z)} W_{(2)\perp}^\mu(z), \tag{4-8}$$

$$W_{(2)\perp}^\mu(z) = W_{(2)}^\mu(z) - \frac{\partial^2}{\partial z_\mu \partial z^\nu}\int D_{\text{ret}}(z-z') W_{(2)}^\nu(z') d^4 z', \tag{4-9}$$

$$W_{(2)\nu}(z) = V(z)\widetilde{V}_\nu^\rho(z)\widetilde{V}_\alpha^\sigma(z)\left(\frac{\partial A_\rho(z)}{\partial z^\sigma} - \frac{\partial A_\sigma(z)}{\partial z^\rho}\right)\int d^4 x j^\beta(x) U_\beta^\alpha(x) f\left(U^\lambda(x) - \left(z^\lambda + aA_{\perp G}^\lambda(z)\right)\right). \tag{4-10}$$

Both the obtained equations of motion of charged particle and electromagnetic field (4-5) and (4-6) ~ (4-10) respectively are still gauge invariant under the gauge transformation (2-5) ~ (2-7), and lead to that the current conservation equation (2-23) holds. Hence, the established theory for this special case is perhaps acceptable.

If we take $\xi(k) = 1$ in (3-43), namely, $A_{\perp G}^\mu(x) = A_\perp^\mu(x)$, where $A_\perp^\mu(x)$ is given by (3-43), then (4-1) and (4-2) become

$$V^\mu(x) = x^\mu + aA_\perp^\mu(x), \tag{4-11}$$

$$V_\nu^\mu(x) = \frac{\partial V^\mu(x)}{\partial x^\nu} = \delta_\nu^\mu + aA_{\perp,\nu}^\mu(x). \tag{4-12}$$

The expressions of $V(x)$ and $\widetilde{V}_\nu^\mu(x)$ are (4-3) and (4-4) but in which $A_{\perp G}^\mu(x)$ is replaced with $A_\perp^\mu(x)$.

For this special case, the equation (4-5) becomes

$$\left(i\gamma^\mu \frac{\partial}{\partial x^\mu} - m\right)\psi(x) = e\gamma^\mu U_\mu^\alpha(x)\int d^4 y f\left(U^\lambda(x) - \left(y^\lambda + aA_\perp^\lambda(y)\right)\right) V(y)\widetilde{V}_\alpha^\beta(y)A_\beta(y)\psi(x); \tag{4-13}$$

On the other hand, we have

$$\frac{\partial A_\rho(z)}{\partial z^\sigma} - \frac{\partial A_\sigma(z)}{\partial z^\rho} = A_{\perp\rho,\sigma}(z) - A_{\perp\sigma,\rho}(z) = \frac{1}{a}\left[g_{\tau\rho}\left(\delta_\sigma^\tau + aA_{\perp,\sigma}^\tau(z)\right) - g_{\tau\sigma}\left(\delta_\rho^\tau + aA_{\perp,\rho}^\tau(z)\right)\right]$$
$$= \frac{1}{a}\left(g_{\tau\rho}V_\sigma^\tau(z) - g_{\tau\sigma}V_\rho^\tau(z)\right), \tag{4-14}$$

the equations (4-6) ~ (4-10) of motion of electromagnetic field can be simplified to the form (The concrete derivation is in the Appendix B of this paper):

$$\frac{\partial}{\partial y^\nu}\left(A^{\mu,\nu}(y) - A^{\nu,\mu}(y)\right) = eW^\mu_{(3)}(y) - aeW^\mu_{//(4)}(y), \qquad (4\text{-}15)$$

where

$$W^\mu_{(3)}(y) = g^{\mu\nu} g_{\alpha\gamma} V(y) \widetilde{V}^\gamma_\nu(y) \int \mathrm{d}^4 x j^\beta(x) U^\alpha_\beta(x) f\left(U^\lambda(x) - \left(y^\lambda + aA^\lambda_\perp(y)\right)\right), \qquad (4\text{-}16)$$

$$W^\mu_{//(4)}(y) = \frac{\partial^2}{\partial y_\mu \partial y^\nu} \int \mathrm{d}^4 z D_{\text{ret}}(y-z) W^\nu_{(4)}(z), \qquad (4\text{-}17)$$

$$W^\mu_{(4)}(z) = \frac{1}{a}\left(W^\mu_{(3)}(z) - W^\mu_{(1)}(z)\right) = W^\mu_{\alpha(5)}(z)\int \mathrm{d}^4 x j^\beta(x) U^\alpha_\beta(x) f\left(U^\lambda(x) - \left(z^\lambda + aA^\lambda_\perp(z)\right)\right), \qquad (4\text{-}18)$$

$$\begin{aligned}W^\mu_{\nu(5)}(z) &= \frac{1}{a} g^{\mu\sigma}\left(g_{\nu\rho} V(z)\widetilde{V}^\rho_\sigma(z) - g_{\sigma\rho} V(z)\widetilde{V}^\rho_\nu(z)\right) \\ &= g^{\mu\sigma}\left(A_{\perp\sigma,\nu}(z) - A_{\perp\nu,\sigma}(z)\right) + ag^{\mu\sigma}\left(A_{\perp\nu,\lambda}(z)A^\lambda_{\perp,\sigma}(z) - A_{\perp\sigma,\lambda}(z)A^\lambda_{\perp,\nu}(z)\right) \\ &\quad + a^2 g^{\mu\sigma}\left[\frac{1}{2}\left(A_{\perp\nu,\sigma}(z) - A_{\perp\sigma,\nu}(z)\right)A^\alpha_{\perp,\beta}(z)A^\beta_{\perp,\alpha}(z)\right. \\ &\quad\left. + A_{\perp\sigma,\alpha}(z)A^\alpha_{\perp,\beta}(z)A^\beta_{\perp,\nu}(z) - A_{\perp\nu,\alpha}(z)A^\alpha_{\perp,\beta}(z)A^\beta_{\perp,\sigma}(z)\right].\end{aligned} \qquad (4\text{-}19)$$

In (4-17), we have taken into account the expression (3-42) of longitudinal four-vector; In (4-19), we have used the formula (4-4) but in which $A^\mu_{\perp G}(x)$ is replaced with $A^\mu_\perp(x)$.

## 4.2 A fully determined example

In the theory whose action is given by (2-1), (2-2) and (2-4) in which the function $V^\mu(x)$ is given by (4-11) and the corresponding equations of motion of charged particle and electromagnetic field are given by (4-13), (4-15) ~ (4-19), the two functions $f(x)$ and $U^\mu(x)$ have not been determined. For the sake of simplicity, we choose

$$f(x) = \delta^4(x), \ U^\mu(x) = x^\mu, \qquad (4\text{-}20)$$

we therefore obtain a fully determined example.

According to (4-20), $U^\mu_\nu(x) = \delta^\mu_\nu$, $\widetilde{U}^\mu_\nu(x) = \delta^\mu_\nu$; the equation of motion of charged particle (4-13) becomes

$$\left(i\gamma^\mu \frac{\partial}{\partial x^\mu} - m\right)\psi(x) = e\gamma^\mu \int \mathrm{d}^4 y \delta^4\left(x^\lambda - \left(y^\lambda + aA^\lambda_\perp(y)\right)\right) V(y)\widetilde{V}^\nu_\mu(y) A_\nu(y)\psi(x); \qquad (4\text{-}21)$$

the equation of motion of electromagnetic field is still of the form (4-15), but in which

$$W^\mu_{(3)}(y) = g^{\mu\nu} V(y)\widetilde{V}^\gamma_\nu(y) \int \mathrm{d}^4 x j_\gamma(x)\delta^4\left(x^\lambda - \left(y^\lambda + aA^\lambda_\perp(y)\right)\right), \qquad (4\text{-}22)$$

$$W^\mu_{(4)}(z) = W^\mu_{\nu(5)}(z) \int \mathrm{d}^4 x j^\nu(x)\delta^4\left(x^\lambda - \left(z^\lambda + aA^\lambda_\perp(z)\right)\right). \qquad (4\text{-}23)$$

Of course, $W^\mu_{//(4)}(y)$ and $W^\mu_{\nu(5)}(z)$ are still given by (4-17) and (4-19), respectively.

For the function $\delta^4\left(x^\lambda - \left(y^\lambda + aA^\lambda_\perp(y)\right)\right)$ we have

$$\delta^4\!\left(x^\lambda - \left(y^\lambda + aA_\perp^\lambda(y)\right)\right) = \int\frac{\mathrm{d}^4k}{(2\pi)^4}\mathrm{e}^{\mathrm{i}k_\lambda\left(x^\lambda - \left(y^\lambda + aA_\perp^\lambda(y)\right)\right)} = \int\frac{\mathrm{d}^4k}{(2\pi)^4}\mathrm{e}^{\mathrm{i}k_\lambda\left(x^\lambda - y^\lambda\right)}\mathrm{e}^{-\mathrm{i}ak_\lambda A_\perp^\lambda(y)}$$
$$= \int\frac{\mathrm{d}^4k}{(2\pi)^4}\mathrm{e}^{\mathrm{i}k_\lambda\left(x^\lambda - y^\lambda\right)}\left(1 + \sum_{n=1}^{\infty}\frac{(-\mathrm{i}a)^n}{n!}\left(k_\lambda A_\perp^\lambda(y)\right)^n\right).$$
(4-24)

Taking advantage of the expression (4-24), we can deal with $\delta^4\!\left(x^\lambda - \left(y^\lambda + aA_\perp^\lambda(y)\right)\right)$ when $A_\perp^\lambda(y)$ is an operator in quantum theory.

Of course, we can choose different expressions of $\delta(x)$, for example, $\delta(x) = \lim_{\varepsilon\to 0}\frac{1}{\pi}\frac{\varepsilon}{\varepsilon^2 + x^2}$, $\delta(x) = \lim_{\varepsilon\to 0}\frac{\sin(x/\varepsilon)}{\pi x}$, $\delta(x) = \lim_{\varepsilon\to 0}\frac{1}{\sqrt{2\pi}\varepsilon}\mathrm{e}^{-\frac{x^2}{2\varepsilon^2}}$, etc. A common characteristic of those expressions is that if they are written as power series expansions, then what we obtain are the form of infinite series but not polynomials in which there are only finite terms; this characteristic causes the tremendous complexities of calculation in quantum theory.

**4.3 The equation of motion of charged particle in a special external (classical) electromagnetic field**

We now investigate the form of the equation of motion (4-21) of charged particle in a special external (classical) electromagnetic field $A^\mu(x)$ described by (3-37), in Sect. 3.2.5 we have prove that $A^\mu(x)$ given by (3-37) is a transverse four-vector, thus, for this case $V^\mu(x)$ in (4-21) becomes $V^\mu(x) = x^\mu + aA^\mu(x)$, where $A^\mu(x)$ is given by (3-37). And, further, we can calculate the corresponding $V_\nu^\mu(x)$, $V(x)$ and $\widetilde{V}_\nu^\mu(x)$ according to the formulas (4-12), (4-3) and (4-4). On the other hand, since the form of $A^\mu(x)$ described by (3-37) is very simple, we can employ directly the definitions of $V_\nu^\mu(x)$, $V(x)$ and $\widetilde{V}_\nu^\mu(x)$ and obtain

$$V_0^\lambda = \delta_0^\lambda,\ V_i^0(y) = a\frac{\partial A^0(y)}{\partial y^i},\ V_j^i = \begin{bmatrix} 1 & -\frac{a}{2}B^3 & \frac{a}{2}B^2 \\ \frac{a}{2}B^3 & 1 & -\frac{a}{2}B^1 \\ -\frac{a}{2}B^2 & \frac{a}{2}B^1 & 1 \end{bmatrix}_{ij};$$
(4-25)
$$V = \left\|V_\beta^\alpha\right\| = \left\|V_j^i\right\| = 1 + \frac{a^2}{4}\boldsymbol{B}\cdot\boldsymbol{B}.$$

$$\widetilde{V}_0^\lambda = \delta_0^\lambda, \quad \widetilde{V}_i^0(y) = -a\widetilde{V}_i^j \frac{\partial A^0(y)}{\partial y^j},$$

$$\widetilde{V}_j^i = \frac{1}{1+\frac{a^2}{4}\mathbf{B}\cdot\mathbf{B}} \begin{bmatrix} 1+\frac{a^2}{4}(B^1)^2 & \frac{a}{2}B^3 + \frac{a^2}{4}B^2B^1 & -\frac{a}{2}B^2 + \frac{a^2}{4}B^3B^1 \\ -\frac{a}{2}B^3 + \frac{a^2}{4}B^1B^2 & 1+\frac{a^2}{4}(B^2)^2 & \frac{a}{2}B^1 + \frac{a^2}{4}B^3B^2 \\ \frac{a}{2}B^2 + \frac{a^2}{4}B^1B^3 & -\frac{a}{2}B^1 + \frac{a^2}{4}B^2B^3 & 1+\frac{a^2}{4}(B^3)^2 \end{bmatrix}_{ij}. \tag{4-26}$$

From (4-25) and (4-26) we see that both $V_j^i$ and $\widetilde{V}_j^i$ are constant matrixes.

According to (4-25) we have

$$V^i(y) = y^i + aA^i(y) = V_j^i y^j. \tag{4-27}$$

Hence,

$$\delta^4\left(x^\lambda - \left(y^\lambda + aA_\perp^\lambda(y)\right)\right) = \delta\left(x^0 - \left(y^0 + aA^0(y)\right)\right)\delta^3\left(x^i - V_j^i y^j\right); \tag{4-28}$$

and, further, for $\delta^3\left(x^i - V_j^i y^j\right)$ we have

$$\delta^3\left(x^i - V_j^i y^j\right) = \int \frac{d^3k}{(2\pi)^3} e^{ik_i\left(x^i - V_j^i y^j\right)} = \int \frac{d^3k}{(2\pi)^3} e^{ik_i V_j^i\left(\widetilde{V}_k^j x^k - y^j\right)}$$

$$\overset{k_i V_j^i = k_j'}{=\!=\!=\!=\!=\!=\!=\!=} \int \frac{d^3k'}{(2\pi)^3} \frac{1}{\|V_j^i\|} e^{ik_j'\left(\widetilde{V}_k^j x^k - y^j\right)} = \frac{1}{V}\delta^3\left(y^i - \widetilde{V}_j^i x^j\right). \tag{4-29}$$

In the above calculation, we have used $d^3k' = \|V_j^i\| d^3k = V d^3k$ and taken account of the fact that $V$ is a constant.

Using the above formulas (4-25) ~ (4-29) and notice that all $V(y)$, $\widetilde{V}_\beta^\alpha(y)$ and $A^\mu(x)$ given by (4-25) are independent of time $y^0$, we therefore have

$$\int d^4y \delta^4\left(x^\lambda - \left(y^\lambda + aA_\perp^\lambda(y)\right)\right) V(y)\widetilde{V}_\mu^\nu(y) A_\nu(y)$$

$$= \int dy^0 \delta\left(x^0 - \left(y^0 + aA^0(y)\right)\right) \int d^3y \frac{1}{V}\delta^3\left(y^i - \widetilde{V}_j^i x^j\right) V(y)\widetilde{V}_\mu^\nu(y) A_\nu(y) \tag{4-30}$$

$$= \widetilde{V}_\mu^\nu(y) A_\nu(y)\Big|_{\substack{y^0=x^0-aA^0(y) \\ y^i=\widetilde{V}_j^i x^j}} = \begin{cases} A_0(y)\big|_{y^l=\widetilde{V}_m^l x^m}, & \mu=0; \\ \left(-a\widetilde{V}_i^j \frac{\partial A^0(y)}{\partial y^j} A_0(y) + \widetilde{V}_i^j A_j(y)\right)\Big|_{y^l=\widetilde{V}_m^l x^m}, & \mu=i. \end{cases}$$

If $y^i = \widetilde{V}_j^i x^j$ and $\widetilde{V}_j^i$ is a constant matrix, then $\frac{\partial y^i}{\partial x^j} = \widetilde{V}_j^i$ and, thus,

$$\widetilde{V}_i^j \frac{\partial A^0(y)}{\partial y^j} A_0(y)\Big|_{y^i=\widetilde{V}_j^i x^j} = \frac{1}{2}\frac{\partial y^j}{\partial x^i}\frac{\partial\{[A^0(y)]^2\}}{\partial y^j}\Big|_{y^l=\widetilde{V}_m^l x^m} = \frac{1}{2}\frac{\partial\{[A^0(\widetilde{V}_m^l x^m)]^2\}}{\partial x^i}; \tag{4-31}$$

According to (4-26) and $A_i = g_{ij}A^j = -A^i$ we calculate $\widetilde{V}_i^{\ j} A_j(y)\big|_{y^l = \widetilde{V}_m^l x^m}$ and obtain

$$\widetilde{V}_i^{\ j} A_j(y)\Big|_{y^l = \widetilde{V}_m^l x^m} = \widetilde{V}_i^{\ j}\left(-\frac{1}{2}\varepsilon_{jkl}B^k y^l\right)\bigg|_{y^l = \widetilde{V}_m^l x^m} = -\frac{1}{2}\widetilde{V}_i^{\ j}\varepsilon_{jkl}B^k \widetilde{V}_m^l x^m$$

$$= \left(-\frac{\frac{1}{2}\boldsymbol{B}\times\boldsymbol{x}}{1+\frac{a^2}{4}\boldsymbol{B}\cdot\boldsymbol{B}}\right)_i = \left(-\frac{\boldsymbol{A}(\boldsymbol{x})}{1+\frac{a^2}{4}\boldsymbol{B}\cdot\boldsymbol{B}}\right)_i = \frac{A_i(\boldsymbol{x})}{1+\frac{a^2}{4}\boldsymbol{B}\cdot\boldsymbol{B}}.$$ (4-32)

Substituting (4-31) and (4-32) to (4-30), we obtain

$$\int d^4 y\, \delta^4\!\left(x^\lambda - \left(y^\lambda + aA_\perp^\lambda(y)\right)\right) V(y)\widetilde{V}_\mu^{\ \nu}(y) A_\nu(y) = \begin{cases} A^0\!\left(\widetilde{V}_m^l x^m\right), & \mu = 0; \\[6pt] -\dfrac{a}{2}\dfrac{\partial\!\left\{\left[A^0\!\left(\widetilde{V}_m^l x^m\right)\right]^2\right\}}{\partial x^i} + \dfrac{A_i(\boldsymbol{x})}{1+\dfrac{a^2}{4}\boldsymbol{B}\cdot\boldsymbol{B}}, & \mu = i. \end{cases}$$ (4-33)

Substituting (4-33) to (4-21), the equation of motion (4-21) of charged particle thus becomes

$$\left(i\gamma^\mu \frac{\partial}{\partial x^\mu} - m\right)\psi(x) = e\left(\gamma^0 A^0\!\left(\widetilde{V}_m^l x^m\right) + \gamma^i\left(-\frac{a}{2}\frac{\partial\!\left\{\left[A^0\!\left(\widetilde{V}_m^l x^m\right)\right]^2\right\}}{\partial x^i} + \frac{A_i(\boldsymbol{x})}{1+\dfrac{a^2}{4}\boldsymbol{B}\cdot\boldsymbol{B}}\right)\right)\psi(x).$$ (4-34)

Setting

$$\psi(x) = e^{i\frac{ae}{2}\left[A^0\left(\widetilde{V}_m^l x^m\right)\right]^2} \widetilde{\psi}(x),$$ (4-35)

for which we have

$$\frac{\partial}{\partial t}\psi(x) = e^{i\frac{ae}{2}\left[A^0\left(\widetilde{V}_m^l x^m\right)\right]^2}\frac{\partial}{\partial t}\widetilde{\psi}(x),$$ (4-36)

$$\frac{\partial}{\partial x^i}\psi(x) = i\frac{ae}{2}\frac{\partial\!\left\{\left[A^0\!\left(\widetilde{V}_m^l x^m\right)\right]^2\right\}}{\partial x^i} e^{i\frac{ae}{2}\left[A^0\left(\widetilde{V}_m^l x^m\right)\right]^2}\widetilde{\psi}(x) + e^{i\frac{ae}{2}\left[A^0\left(\widetilde{V}_m^l x^m\right)\right]^2}\frac{\partial}{\partial x^i}\widetilde{\psi}(x).$$ (4-37)

Substituting (4-35) ~ (4-37) to (4-34), we obtain the equation of motion of $\widetilde{\psi}(x)$:

$$\left(i\gamma^\mu \frac{\partial}{\partial x^\mu} - m\right)\widetilde{\psi}(x) = e\left(\gamma^0 A^0\!\left(\widetilde{V}_m^l x^m\right) + \frac{\gamma^i A_i(\boldsymbol{x})}{1+\dfrac{a^2}{4}\boldsymbol{B}\cdot\boldsymbol{B}}\right)\widetilde{\psi}(x).$$ (4-38)

If $\boldsymbol{B} = 0$, then $\widetilde{V}_m^l = \delta_m^l$ and $A^i = 0$, (4-38) thus becomes

$$\left(i\gamma^\mu \frac{\partial}{\partial x^\mu} - m\right)\widetilde{\psi}(x) = e\gamma^0 A^0(x)\widetilde{\psi}(x).$$ (4-39)

Eq. (4-39) is just the same as the Dirac equation with potential in conventional relativistic quantum mechanics, and from (4-35) we see that the difference between $\psi(x)$ and $\widetilde{\psi}(x)$ is only a phase factor, hence, for this case we obtain the same conclusions with conventional relativistic quantum mechanics. For example, we obtain the same energy levels of hydrogen atom without the Lamb shift when $A^0(x) = -\dfrac{e}{|x|}$.

If $\boldsymbol{B} \neq 0$ in (4-38), then we take

$$\gamma^0 = \beta = \begin{pmatrix} I & 0 \\ 0 & -I \end{pmatrix}, \quad \gamma^i = \beta \alpha^i, \quad \alpha^i = \begin{pmatrix} 0 & \sigma^i \\ \sigma^i & 0 \end{pmatrix}; \quad \widetilde{\psi}(x) = e^{-imt} \begin{pmatrix} \widetilde{\psi}_{\text{big}}(x) \\ \widetilde{\psi}_{\text{small}}(x) \end{pmatrix}. \tag{4-40}$$

Substituting (4-40) to (4-38), we have

$$\left( i \frac{\partial}{\partial t} - eA^0 \left( \widetilde{V}_j^i x^j \right) \right) \widetilde{\psi}_{\text{big}}(x) = \boldsymbol{\sigma} \cdot \left( \boldsymbol{p} - e \frac{\boldsymbol{A}(x)}{1 + \frac{a^2}{4} \boldsymbol{B} \cdot \boldsymbol{B}} \right) \widetilde{\psi}_{\text{small}}(x), \tag{4-41}$$

$$\left( i \frac{\partial}{\partial t} - eA^0 \left( \widetilde{V}_j^i x^j \right) + 2m \right) \widetilde{\psi}_{\text{small}}(x) = \boldsymbol{\sigma} \cdot \left( \boldsymbol{p} - e \frac{\boldsymbol{A}(x)}{1 + \frac{a^2}{4} \boldsymbol{B} \cdot \boldsymbol{B}} \right) \widetilde{\psi}_{\text{big}}(x). \tag{4-42}$$

Substituting $\widetilde{\psi}_{\text{small}}(x) \approx \frac{1}{2m} \boldsymbol{\sigma} \cdot \left( \boldsymbol{p} - e \frac{\boldsymbol{A}(x)}{1 + \frac{a^2}{4} \boldsymbol{B} \cdot \boldsymbol{B}} \right) \widetilde{\psi}_{\text{big}}(x)$ obtained by (4-42) to (4-41), we obtain

$$\left( i \frac{\partial}{\partial t} - eA^0 \left( \widetilde{V}_j^i x^j \right) \right) \widetilde{\psi}_{\text{big}}(x) = \frac{1}{2m} \left( \boldsymbol{\sigma} \cdot \left( \boldsymbol{p} - e \frac{\boldsymbol{A}(x)}{1 + \frac{a^2}{4} \boldsymbol{B} \cdot \boldsymbol{B}} \right) \right)^2 \widetilde{\psi}_{\text{big}}(x)$$

$$= \left( \frac{1}{2m} \left( \boldsymbol{p} - e \frac{\frac{1}{2} \boldsymbol{B} \times \boldsymbol{x}}{1 + \frac{a^2}{4} \boldsymbol{B} \cdot \boldsymbol{B}} \right)^2 - \frac{e}{2m} \frac{1}{1 + \frac{a^2}{4} \boldsymbol{B} \cdot \boldsymbol{B}} \boldsymbol{\sigma} \cdot \boldsymbol{B} \right) \widetilde{\psi}_{\text{big}}(x). \tag{4-43}$$

In (4-43), the difference between the factor $\dfrac{e}{2m} \dfrac{1}{1 + \dfrac{a^2}{4} \boldsymbol{B} \cdot \boldsymbol{B}}$ and the Bohr magneton $\dfrac{e}{2m}$ is

$\dfrac{e}{2m} \left( \dfrac{1}{1 + \dfrac{a^2}{4} \boldsymbol{B} \cdot \boldsymbol{B}} - 1 \right)$. However, this difference cannot be regarded as the anomalous magnetic moment of the electron observed by experiment due to the two reasons: ① $\dfrac{e}{2m} \dfrac{1}{1 + \dfrac{a^2}{4} \boldsymbol{B} \cdot \boldsymbol{B}} < \dfrac{e}{2m}$ but, as well-known, the sum of the Bohr magneton and the anomalous magnetic moment of electron is $\dfrac{e}{2m} \left( 1 + \dfrac{\alpha}{2\pi} + \cdots \right) > \dfrac{e}{2m}$; ② The term $\dfrac{e}{2m} \left( \dfrac{1}{1 + \dfrac{a^2}{4} \boldsymbol{B} \cdot \boldsymbol{B}} - 1 \right)$ is dependent of $\boldsymbol{B} \cdot \boldsymbol{B}$.

Hence, we cannot obtain the Lamb shift and the anomalous magnetic moment of the electron from the equation (4-34), the reason is that (4-34) is an equation of motion of charged particle in a given external (classic) electromagnetic field. If we want to obtain the Lamb shift and the anomalous magnetic moment of the electron, then we have to investigate the action of quantized radiation field. This is in accord with the point of view of conventional local QED that the Lamb shift and the anomalous magnetic moment of the electron are caused by quantized radiation field.

## 5  Quantization of the theory

**5.1 Some difficulties of canonical quantization of nonlocal field theory**

We have investigated some cases of quantum electrodynamics with nonlocal interaction, which are summarized in Tab.1.

**Tab.1  The investigated cases**

| | |
|---|---|
| Case1 (discussed in §2) | All $f(x)$, $U^\mu(x)$ and $V^\mu(x)$ are independent of $\psi(x)$ and $A^\mu(x)$ |
| Case2.1 (discussed in §3.1) | Both $f(x)$ and $U^\mu(x)$ are independent of $\psi(x)$ and $A^\mu(x)$, $V^\mu(x) = x^\mu + aA_{\perp G}^\mu(x)$ |
| Case2.2 (discussed in §3.1) | A special form of Case2.1: $A_{\perp G}^\mu(x) = A_\perp^\mu(x)$ |
| Case2.3 (discussed in §3.2) | A fully determined example of Case2.2: $f(x) = \delta^4(x)$, $U^\mu(x) = x^\mu$ |

If we employ the standard procedure of canonical quantization based on the Lagrangian-Hamiltonian approach and variation principle to establish the corresponding quantum theory for the cases listed in Tab. 1, then we shall meet some difficulties, since the Lagrangian-Hamiltonian approach and variation principle of nonlocal field theory are censured to be ambiguous[5]. In fact, even if for the Case1 in Tab. 1 in which there is not any time derivative term of $\psi(x)$ and $A^\mu(x)$ in the interaction term $S_\mathrm{I}$ given by (2-4), canonical quantization method still shows some uncertainty, say nothing of the cases2.1 ~ 2.3 in Tab. 1 in which there are time derivative terms of $A^\mu(x)$ in $S_\mathrm{I}$.

It seems as if some other methods of quantization, e.g., the Dirac-Bargmann method for a strange Lagrangian system, and that of path integral and of quantization for theories including higher-order derivative of field variables, are still very difficult for the cases listed in Tab. 1.

Here we employ the method based on the Yang-Feldman equations and the Lehmann-Symanzik-Zimmermann formalism to realize quantization of the cases listed in Tab. 1. This method skirts the Lagrangian-Hamiltonian approach and can also be used to those theories for which only the equations of motion are known but the corresponding action is unknown.

**5.2 Quantization of the theory under the Lorenz gauge condition**

We shall deal with the cases listed in Tab. 1 uniformly under the Lorenz gauge condition and, without more explanation, present the main contexts here. (Some discussions for the

Yang-Feldman equations and the Lehmann-Symanzik-Zimmermann formalism can be found in [1, 4, 6, 7].)

① The equations of motion of the field operators

We expect that, just as in the conventional local QED, the Lorenz gauge does not generate new term, e.g., the Faddeev-Popov ghost, in the quantum theory. The equations of motion of the field operators $\psi(x)$ and $A^\mu(y)$ thus are

$$\left(i\gamma^\mu \frac{\partial}{\partial x^\mu} - m\right)\psi(x) = e\gamma^\mu \int d^4y\, \Theta^\nu_\mu(x,y) A_\nu(y)\psi(x) + e\gamma^\mu \int d^4y\, \left(\Theta^\nu_\mu(x,y)\right)_{\text{ext}} A_{\nu\,\text{ext}}(y)\psi(x), \quad (5\text{-}1)$$

$$\Theta^\nu_\mu(x,y) = \begin{cases} U^\gamma_\mu(x) f\!\left(U^\lambda(x) - V^\lambda(y)\right) V(y)\widetilde{V}^\nu_\gamma(y), & \text{for the Case 1 in Tab.1;} \\ U^\gamma_\mu(x) f\!\left(U^\lambda(x) - \left(y^\lambda + aA^\lambda_{\perp\text{G}}(y)\right)\right) V(y)\widetilde{V}^\nu_\gamma(y), & \text{for the Case 2.1 in Tab.1;} \\ U^\gamma_\mu(x) f\!\left(U^\lambda(x) - \left(y^\lambda + aA^\lambda_\perp(y)\right)\right) V(y)\widetilde{V}^\nu_\gamma(y), & \text{for the Case 2.2 in Tab.1;} \\ \delta^4\!\left(x^\lambda - \left(y^\lambda + aA^\lambda_\perp(y)\right)\right) V(y)\widetilde{V}^\nu_\mu(y), & \text{for the Case 2.3 in Tab.1.} \end{cases} \quad (5\text{-}2)$$

In (5-1), $\left(\Theta^\nu_\mu(x,y)\right)_{\text{ext}}$ is still expressed by (5-2) but in which $A^\mu(x)$ is replaced with a given external field $A^\mu_{\text{ext}}(x)$.

$$\Box A^\mu(y) = e\int d^4x\, \Omega^\mu_\nu(y,x) j^\nu(x), \quad (5\text{-}3)$$

$$\Omega^\mu_\nu(y,x) = \begin{cases}
U^\gamma_\nu(x) f\!\left(U^\lambda(x) - V^\lambda(y)\right) V(y)\widetilde{V}^\mu_\gamma(y), & \text{for the Case 1 in Tab.1;} \\
V(y)\widetilde{V}^\mu_\gamma(y) U^\gamma_\nu(x) f\!\left(U^\lambda(x) - \left(y^\lambda + aA^\lambda_{\perp\text{G}}(y)\right)\right) \\
\quad - aU^\alpha_\nu(x)\int d^4z \int \dfrac{d^4k}{(2\pi)^4} \xi(k)\,e^{-ik\cdot(y-z)} f\!\left(U^\lambda(x) - \left(z^\lambda + aA^\lambda_{\perp\text{G}}(z)\right)\right) \\
\qquad \times g^{\mu\beta} V(z)\widetilde{V}^\rho_\alpha(z)\widetilde{V}^\sigma_\beta(z)\left(\dfrac{\partial A_\rho(z)}{\partial z^\sigma} - \dfrac{\partial A_\sigma(z)}{\partial z^\rho}\right) \\
\quad + aU^\alpha_\nu(x)\int d^4z \int \dfrac{d^4k}{(2\pi)^4} \xi(k)\,e^{-ik\cdot(y-z)} \dfrac{\partial^2}{\partial z_\mu \partial z_\beta} \int d^4z'\, D_{\text{ret}}(z-z') \\
\qquad \times f\!\left(U^\lambda(x) - \left(z'^\lambda + aA^\lambda_{\perp\text{G}}(z')\right)\right) V(z')\widetilde{V}^\rho_\alpha(z')\widetilde{V}^\sigma_\beta(z')\left(\dfrac{\partial A_\rho(z')}{\partial z'^\sigma} - \dfrac{\partial A_\sigma(z')}{\partial z'^\rho}\right), \\
\hfill \text{for the Case 2.1 in Tab.1;} \\
g^{\mu\alpha} g_{\beta\gamma} U^\gamma_\nu(x) f\!\left(U^\lambda(x) - \left(y^\lambda + aA^\lambda_\perp(y)\right)\right) V(y)\widetilde{V}^\beta_\alpha(y) \\
\quad - aU^\alpha_\nu(x)\dfrac{\partial^2}{\partial y_\mu \partial y^\beta} \int d^4z\, D_{\text{ret}}(y-z) W^\beta_{\alpha(5)}(z) f\!\left(U^\lambda(x) - \left(z^\lambda + aA^\lambda_\perp(z)\right)\right), \\
\hfill \text{for the Case 2.2 in Tab.1;} \\
g^{\mu\alpha} g_{\beta\nu} \delta^4\!\left(x^\lambda - \left(y^\lambda + aA^\lambda_\perp(y)\right)\right) V(y)\widetilde{V}^\beta_\alpha(y) \\
\quad - a\dfrac{\partial^2}{\partial y_\mu \partial y^\gamma} \int d^4z\, D_{\text{ret}}(y-z) W^\gamma_{\nu(5)}(z) \delta^4\!\left(x^\lambda - \left(z^\lambda + aA^\lambda_\perp(z)\right)\right), \\
\hfill \text{for the Case 2.3 in Tab.1.}
\end{cases}$$

$$(5\text{-}4)$$

For the Case1, both $\Theta^\nu_\mu(x,y)$ and $\Omega^\mu_\nu(y,x)$ are independent of $\psi(x)$ and $A^\mu(x)$; for the Cases2.1 ~ 2.3, both $\Theta^\nu_\mu(x, y)$ and $\Omega^\mu_\nu(y, x)$ are only dependent the corresponding transverse four-vector of $A^\mu(x)$ but independent of $\psi(x)$.

Replacing $S_{EM}$ in (2-2) with $S'_{EM} = -\frac{1}{2}\int d^4 x A_{\mu,\nu}(x) A^{\mu,\nu}(x)$, we can obtain the equations of motion (5-1) ~ (5-4) of charged particle and electromagnetic field by the variational equations $\frac{\delta S}{\delta \overline{\psi}(x)} = 0$ and $\frac{\delta S}{\delta A_\mu(y)} = 0$, respectively.

② "In" fields and the Fock space

The field operators $\psi_{in}(x)$, $\overline{\psi}_{in}(x)$ and $A^\mu_{in}(y)$ of "In" fields satisfy the corresponding equations $\left(i\gamma^\mu \frac{\partial}{\partial x^\mu} - m\right)\psi_{in}(x) = 0$ and $\Box A^\mu_{in}(y) = 0$ of free fields, respectively; And which can be written to the forms:

$$\psi_{in}(x) = \int d^3 p \sum_{\pm r} \left(\psi^{(-)}_r(x, p) b_{rin}(p) + \psi^{(+)}_r(x, p) d^+_{rin}(p)\right), \tag{5-5}$$

$$\overline{\psi}_{in}(x) = \int d^3 p \sum_{\pm r} \left(\overline{\psi}^{(+)}_r(x, p) b^+_{rin}(p) + \overline{\psi}^{(-)}_r(x, p) d_{rin}(p)\right); \tag{5-6}$$

$$A^\mu_{in}(y) = \int \frac{d^3 k}{(2\pi)^{3/2}\sqrt{2|k|}} \left(e^{-ik\cdot y}\sum_{\lambda=0}^{3}\varepsilon^\mu_{(\lambda)}(k) a_{(\lambda)in}(k) + e^{ik\cdot y}\sum_{\lambda=0}^{3}\varepsilon^\mu_{(\lambda)}(k) a^+_{(\lambda)in}(k)\right). \tag{5-7}$$

In (5-7), the four polarization vectors $\varepsilon^\mu_{(\lambda)}(k)$ ($\lambda = 0, 1, 2, 3$) satisfy (3-6) but in which $k^2 = 0$; other quantities appearing in (5-5) ~ (5-7) can be found in the Appendix C of this paper.

We can generate all states of the Fock space by acting with the creation operators $b^+_{rin}(p)$, $d^+_{rin}(p)$ and $a^+_{in(\lambda)}(k)$ on the vacuum state $|0\rangle$.

Since $A^\mu_{in}(y)$ given by (5-7) satisfies $\Box A^\mu_{in}(y) = 0$, similar to (3-33) ~ (3-36), $A^\mu_{in}(y)$ has the expansion:

$$A^\mu_{in}(y) = A^\mu_{in(\overline{0})}(y) + \sum_{\lambda=1}^{2} A^\mu_{in(\overline{\lambda})}(y) + A^\mu_{in(\overline{3})}(y);$$

$$A^\mu_{in(\overline{0})}(y) = \int \frac{d^3 k}{(2\pi)^{3/2}\sqrt{2|k|}} \varepsilon^\mu_{(0)}(k)\left(e^{-ik\cdot y}\left(a_{(0)in}(k) - a_{(3)in}(k)\right) + e^{ik\cdot y}\left(a^+_{(0)in}(k) - a^+_{(3)in}(k)\right)\right),$$

$$A^\mu_{in(\overline{\lambda})}(y) = \int \frac{d^3 k}{(2\pi)^{3/2}\sqrt{2|k|}} \left(e^{-ik\cdot y}\varepsilon^\mu_{(\lambda)}(k) a_{(\lambda)in}(k) + e^{ik\cdot y}\varepsilon^\mu_{(\lambda)}(k) a^+_{(\lambda)in}(k)\right) \quad (\lambda = 1, 2),$$

$$A^\mu_{in(\overline{3})}(y) = \int \frac{d^3 k}{(2\pi)^{3/2}\sqrt{2|k|}} \frac{k^\mu}{k\cdot n}\left(e^{-ik\cdot y} a_{(3)in}(k) + e^{ik\cdot y} a^+_{(3)in}(k)\right).$$

(5-8)

And, further, according to (3-13), (3-25) and (3-3) we obtain the longitudinal and transverse parts $A^\mu_{in//}(y)$ and $A^\mu_{in\perp}(y)$ of $A^\mu_{in}(y)$:

$$A^{\mu}_{\text{in}//}(y) = A^{\mu}_{\text{in}(\bar{0})//}(y) + A^{\mu}_{\text{in}(\bar{3})}(y) = \frac{\partial}{\partial y_{\mu}}\left(\frac{1}{2}y_{\lambda}A^{\lambda}_{\text{in}(\bar{0})}(y)\right) + A^{\mu}_{\text{in}(\bar{3})}(y)$$

$$= \frac{1}{2}A^{\mu}_{\text{in}(\bar{0})}(y) + \frac{1}{2}y_{\lambda}A^{\lambda,\mu}_{\text{in}(\bar{0})}(y) + A^{\mu}_{\text{in}(\bar{3})}(y), \qquad (5\text{-}9)$$

$$A^{\mu}_{\text{in}\perp}(y) = A^{\mu}_{\text{in}}(y) - A^{\mu}_{\text{in}//}(y) = \frac{1}{2}A^{\mu}_{\text{in}(\bar{0})}(y) - \frac{1}{2}y_{\lambda}A^{\lambda,\mu}_{\text{in}(\bar{0})}(y) + \sum_{\lambda=1}^{2}A^{\mu}_{\text{in}(\bar{\lambda})}(y),$$

respectively.

③ The Lorenz gauge condition

The Lorenz gauge condition holds in the mean

$$\left\langle \left| \frac{\partial A^{\mu}(y)}{\partial y^{\mu}} \right| \right\rangle = 0. \qquad (5\text{-}10)$$

From (5-1) ~ (5-4) and the current conservation equation (2-23) we can prove $\Box \frac{\partial A^{\mu}(y)}{\partial y^{\mu}} = 0$, hence, (5-10) holds at any moment only if it holds at initial time.

For ensuring the state vector describing a given physical situation has positive square norm, indefinite metric and the Gupta-Bleuler formalism are needed. For example,

$$\left(a_{(0)\text{in}}(\mathbf{k}) - a_{(3)\text{in}}(\mathbf{k})\right)\big|\ \big\rangle = 0. \qquad (5\text{-}11)$$

④ The asymptotic conditions and the Yang-Feldman equations

Using the idea of "weak convergence", we have the asymptotic conditions for the field operators:

$$\psi(x) \to \sqrt{Z_2}\psi_{\text{in}}(x),\ \overline{\psi}(x) \to \sqrt{Z_2}\overline{\psi}_{\text{in}}(x),\ A^{\mu}(x) \to \sqrt{Z_3}A^{\mu}_{\text{in}}(x),\ \text{when}\ x^0 \to -\infty. \qquad (5\text{-}12)$$

And, further, considering the asymptotic conditions (5-12), the equations of motion (5-1) and (5-3) can be written to the following integral forms:

$$\psi(x) = \sqrt{Z_2}\psi_{\text{in}}(x) + e\int d^4x' S_{\text{ret}}(x-x')\gamma^{\mu}\int d^4y\, \Theta^{\nu}_{\mu}(x',y)A_{\nu}(y)\psi(x')$$
$$+ e\int d^4x' S_{\text{ret}}(x-x')\gamma^{\mu}\int d^4y\left(\Theta^{\nu}_{\mu}(x',y)\right)_{\text{ext}} A_{\nu\,\text{ext}}(y)\psi(x')\ , \qquad (5\text{-}13)$$

$$\overline{\psi}(x) = \sqrt{Z_2}\overline{\psi}_{\text{in}}(x) + e\int d^4x'\int d^4y\, \Theta^{\nu}_{\mu}(x',y)A_{\nu}(y)\psi(x')\gamma^{\mu}S_{\text{adv}}(x'-x)$$
$$+ e\int d^4x'\int d^4y\left(\Theta^{\nu}_{\mu}(x',y)\right)_{\text{ext}} A_{\nu\,\text{ext}}(y)\overline{\psi}(x')\,\gamma^{\mu}S_{\text{adv}}(x'-x), \qquad (5\text{-}14)$$

$$A^{\mu}(y) = \sqrt{Z_3}A^{\mu}_{\text{in}}(y) + e\int d^4y' D_{\text{ret}}(y-y')\int d^4x\, \Omega^{\mu}_{\nu}(y',x)\frac{\overline{\psi}(x)\gamma^{\nu}\psi(x) - \psi^{\text{T}}(x)\left(\gamma^{\nu}\right)^{\text{T}}\overline{\psi}^{\text{T}}(x)}{2}. \qquad (5\text{-}15)$$

In (5-15), we have used $\dfrac{\overline{\psi}(x)\gamma^{\nu}\psi(x) - \psi^{\text{T}}(x)\left(\gamma^{\nu}\right)^{\text{T}}\overline{\psi}^{\text{T}}(x)}{2}$ to replace $\overline{\psi}(x)\gamma^{\nu}\psi(x)$.

For the Case1 in Tab.1, both $\Theta^{\nu}_{\mu}(x,y)$ and $\Omega^{\mu}_{\nu}(y,x)$ are independent of $\psi(x)$ and $A^{\mu}(x)$; however, for the Cases2.1 ~ 2.3 in Tab.1, there are some uncertainty arisen from operator ordering, since both $\Theta^{\nu}_{\mu}(x,y)$ and $\Omega^{\mu}_{\nu}(y,x)$ are functions of the operator $A^{\mu}(x)$. We can add some stipulations to avoid such uncertainty; when we do this, what principles we follow are: ① The theories must recur the conventional local QED under the limit case indicated by (1-3) and (1-8); ② On the premise of ensuring the principle ①, we choose simplest stipulations.

Following the above two principles, we substitute (5-15) to (5-13) and (5-14), and put the normal product $\mathbf{N}(\cdots)$ [1] in place; then replacing (5-13) ~ (5-15), the integral forms of the equations of motion (5-1) and (5-3) are now written to

$$\psi(x) = \sqrt{Z_2}\psi_{\text{in}}(x) + e\sqrt{Z_3}\int d^4x' \int d^4y \, S_{\text{ret}}(x-x')\gamma^\mu \mathbf{N}\big(\Theta^\nu_\mu(x',y)A_{\nu\text{in}}(y)\big)\psi(x')$$

$$+ e^2 \int d^4x' \int d^4y \int d^4y' \int d^4x'' S_{\text{ret}}(x-x')D_{\text{ret}}(y-y')g_{\nu\rho}\gamma^\mu \mathbf{N}\big(\Theta^\nu_\mu(x',y)\Omega^\rho_\sigma(y',x'')\big)$$

$$\times \frac{\overline{\psi}(x'')\gamma^\sigma\psi(x'') - \psi^{\text{T}}(x'')\big(\gamma^\sigma\big)^{\text{T}}\overline{\psi}^{\text{T}}(x'')}{2}\psi(x')$$

$$+ e\int d^4x' S_{\text{ret}}(x-x')\gamma^\mu \int d^4y \big(\Theta^\nu_\mu(x',y)\big)_{\text{ext}} A_{\nu\,\text{ext}}(y)\psi(x') \,,$$

(5-16)

$$\overline{\psi}(x) = \sqrt{Z_2}\overline{\psi}_{\text{in}}(x) + e\sqrt{Z_3}\int d^4x' \int d^4y \, \overline{\psi}(x')\mathbf{N}\big(A_{\nu\text{in}}(y)\Theta^\nu_\mu(x',y)\big)\gamma^\mu S_{\text{adv}}(x'-x)$$

$$+ e^2 \int d^4x' \int d^4y \int d^4y' \int d^4x'' \overline{\psi}(x')\frac{\overline{\psi}(x'')\gamma^\sigma\psi(x'') - \psi^{\text{T}}(x'')\big(\gamma^\sigma\big)^{\text{T}}\overline{\psi}^{\text{T}}(x'')}{2}$$

$$\times g_{\nu\rho}\mathbf{N}\big(\Omega^\rho_\sigma(y',x'')\Theta^\nu_\mu(x',y)\big)\gamma^\mu D_{\text{ret}}(y-y')S_{\text{adv}}(x'-x)$$

$$+ e\int d^4x' \int d^4y \big(\Theta^\nu_\mu(x',y)\big)_{\text{ext}} A_{\nu\,\text{ext}}(y)\overline{\psi}(x')\gamma^\mu S_{\text{adv}}(x'-x);$$

(5-17)

$$A^\mu(y) = \sqrt{Z_3}A^\mu_{\text{in}}(y) + e\int d^4y' D_{\text{ret}}(y-y')J^\mu(y') \,,$$ (5-18)

$$J^\mu(y') = \int d^4x \mathbf{N}\big(\Omega^\mu_\nu(y',x)\big)\frac{\overline{\psi}(x)\gamma^\nu\psi(x) - \psi^{\text{T}}(x)\big(\gamma^\nu\big)^{\text{T}}\overline{\psi}^{\text{T}}(x)}{2} \,.$$ (5-19)

We now can substitute so called the Yang-Feldman ansatz

$$\psi(x) = \sum_{n=0}^{\infty} e^n \psi_{(n)}(x), \quad A^\mu(y) = \sum_{n=0}^{\infty} e^n A^\mu_{(n)}(y)$$ (5-20)

to the operator equations (5-16) ~ (5-19) and obtain recursively the field operators $\psi(x)$ and $A^\mu(y)$.

Notice that the normal product $\mathbf{N}(\cdots)$ is only defined on the operators $b_{r\text{in}}(\mathbf{p})$, $d_{r\text{in}}(\mathbf{p})$, $a_{\text{in}(\lambda)}(\mathbf{k})$, $b^+_{r\text{in}}(\mathbf{p})$, $d^+_{r\text{in}}(\mathbf{p})$ and $a^+_{\text{in}(\lambda)}(\mathbf{k})$, hence, for obtaining $\psi_{(n)}(x)$ and $A^\mu_{(n)}(y)$, we use the normal product after all $\psi_{(i)}(x), A^\mu_{(i)}(y)$ $(i = 0,1,2,\cdots,n-1)$ are expressed by $b_{r\text{in}}(\mathbf{p})$, $d_{r\text{in}}(\mathbf{p})$, $a_{\text{in}(\lambda)}(\mathbf{k})$, $b^+_{r\text{in}}(\mathbf{p})$, $d^+_{r\text{in}}(\mathbf{p})$ and $a^+_{\text{in}(\lambda)}(\mathbf{k})$ in (5-16) ~ (5-19).

The iterative process of calculations in solving the operator equations (5-16) ~ (5-19) is thus determined fully. For example, we obtain

$$\psi_{(0)}(x) = \sqrt{Z_2}\psi_{\text{in}}(x), \quad \overline{\psi}_{(0)}(x) = \sqrt{Z_2}\overline{\psi}_{\text{in}}(x), \quad A^\mu_{(0)}(y) = \sqrt{Z_3}A^\mu_{\text{in}}(y) \,;$$ (5-21)

$$\psi_{(1)}(x) = \int d^4x' S_{\text{ret}}(x-x')\gamma^\mu \int d^4y \, \mathbf{N}\big(\Theta^\nu_{\mu(0)}(x',y)A_{\nu(0)}(y)\big)\psi_{(0)}(x')$$

$$+ \int d^4x' S_{\text{ret}}(x-x')\gamma^\mu \int d^4y \big(\Theta^\nu_\mu(x',y)\big)_{\text{ext}} A_{\nu\,\text{ext}}(y)\psi_{(0)}(x'),$$

(5-22)

$$A^\mu_{(1)}(y) = \int d^4y' D_{\text{ret}}(y-y') \int d^4x \, \mathbf{N}\big(\Omega^\mu_{\nu(0)}(y',x)\big)\mathbf{N}\big(\overline{\psi}_{(0)}(x)\gamma^\nu\psi_{(0)}(x)\big) \,,$$ (5-23)

where $\Theta^\nu_{\mu(0)}(x',y)$ and $\Omega^\mu_{\nu(0)}(y',x)$ are expressed by (5-2) and (5-4), respectively, but in which $A^\mu(x)$ is replaced with $A^\mu_{(0)}(x)$.

⑤ The determination of the transverse four-vector $A^\mu_\perp(x)$ of $A^\mu(y)$

According to the current conservation equation (2-23), we can prove that $J^\mu(y')$ given by (5-19) satisfies $\dfrac{\partial J^\mu(y)}{\partial y^\mu} = 0$, hence, by comparing (5-18) and (3-40) and using (5-21) and (3-42), we can determine the transverse four-vector $A^\mu_\perp(x)$ corresponding to $A^\mu(y)$ given by (5-18) as follows.

$$A^\mu_\perp(x) = \sqrt{Z_3}\, A^\mu_{\text{in}\perp}(y) + e\int D_{\text{ret}}(x-y) J^\mu(y)\mathrm{d}^4 y$$
$$- e\int \frac{\mathrm{d}^3 k}{(2\pi)^3} \frac{\mathrm{i}}{2|\boldsymbol{k}|}\left(\mathrm{e}^{-\mathrm{i}|\boldsymbol{k}|(x^0-z^0)+\mathrm{i}\boldsymbol{k}\cdot(\boldsymbol{x}-\boldsymbol{z})} - \mathrm{e}^{\mathrm{i}|\boldsymbol{k}|(x^0-z^0)+\mathrm{i}\boldsymbol{k}\cdot(\boldsymbol{x}-\boldsymbol{z})}\right)\frac{k^\mu}{k\cdot n}\varepsilon^\nu_{(0)}(\boldsymbol{k})\, J_\nu(z)\mathrm{d}^4 z, \tag{5-24}$$

where $A^\mu_{\text{in}\perp}(y)$ and $J^\mu(y)$ are given by (5-9) and (5-19), respectively.

Of course, for concrete calculation, according to (5-20), we have to calculate $A^\mu_{(n)\perp}(y)$ $(n=1,2,3,\cdots)$ in steps.

⑥ "Out" fields and the Lehmann-Symanzik-Zimmermann reduction formulas

We introduces the field operators $\psi_{\text{out}}(x)$, $\overline{\psi}_{\text{out}}(x)$ and $A^\mu_{\text{out}}(y)$ of "Out" fields, which satisfy the corresponding equations $\left(\mathrm{i}\gamma^\mu\dfrac{\partial}{\partial x^\mu} - m\right)\psi_{\text{out}}(x) = 0$ and $\Box A^\mu_{\text{out}}(y) = 0$ of free fields, respectively; and write $\psi_{\text{out}}(x)$, $\overline{\psi}_{\text{out}}(x)$ and $A^\mu_{\text{out}}(y)$ to the forms:

$$\psi_{\text{out}}(x) = \int \mathrm{d}^3 p \sum_{\pm r}\left(\psi^{(-)}_r(x,\boldsymbol{p}) b_{r\text{out}}(\boldsymbol{p}) + \psi^{(+)}_r(x,\boldsymbol{p}) d^+_{r\text{out}}(\boldsymbol{p})\right), \tag{5-25}$$

$$\overline{\psi}_{\text{out}}(x) = \int \mathrm{d}^3 p \sum_{\pm r}\left(\overline{\psi}^{(+)}_r(x,\boldsymbol{p}) b^+_{r\text{out}}(\boldsymbol{p}) + \overline{\psi}^{(-)}_r(x,\boldsymbol{p}) d_{r\text{out}}(\boldsymbol{p})\right); \tag{5-26}$$

$$A^\mu_{\text{out}}(y) = \int \mathrm{d}^3 k \sum_{\lambda=0}^{3}\left(A^{\mu(-)}_{(\lambda)}(y,\boldsymbol{k}) a_{(\lambda)\text{out}}(\boldsymbol{k}) + A^{\mu(+)}_{(\lambda)}(y,\boldsymbol{k}) a^+_{(\lambda)\text{out}}(\boldsymbol{k})\right). \tag{5-27}$$

Using the idea of "weak convergence", we have the asymptotic conditions for the field operators:

$$\psi(x) \to \sqrt{Z_2}\,\psi_{\text{out}}(x),\ \overline{\psi}(x) \to \sqrt{Z_2}\,\overline{\psi}_{\text{out}}(x),\ A^\mu(x) \to \sqrt{Z_3}\,A^\mu_{\text{out}}(x),\ \text{when}\ x^0 \to +\infty. \tag{5-28}$$

And, further, corresponding to (5-16) ~ (5-19), we have the concrete forms:

$$\psi(x) = \sqrt{Z_2}\psi_{\text{out}}(x) + e\sqrt{Z_3}\int d^4x' \int d^4y\, S_{\text{adv}}(x-x')\gamma^\mu \mathbf{N}\!\left(\Theta_\mu^\nu(x',y)A_{\nu\text{out}}(y)\right)\psi(x')$$

$$+ e^2 \int d^4x' \int d^4y \int d^4y' \int d^4x'' S_{\text{adv}}(x-x')D_{\text{adv}}(y-y')g_{\nu\rho}\gamma^\mu \mathbf{N}\!\left(\Theta_\mu^\nu(x',y)\Omega_\sigma^\rho(y',x'')\right)$$

$$\times \frac{\overline{\psi}(x'')\gamma^\sigma \psi(x'') - \psi^{\mathrm{T}}(x'')\left(\gamma^\sigma\right)^{\mathrm{T}}\overline{\psi}^{\mathrm{T}}(x'')}{2}\psi(x') \quad (5\text{-}29)$$

$$+ e\int d^4x' S_{\text{adv}}(x-x')\gamma^\mu \int d^4y\left(\Theta_\mu^\nu(x',y)\right)_{\text{ext}} A_{\nu\,\text{ext}}(y)\psi(x')\ ,$$

$$\overline{\psi}(x) = \sqrt{Z_2}\,\overline{\psi}_{\text{out}}(x) + e\sqrt{Z_3}\int d^4x' \int d^4y\, \overline{\psi}(x')\mathbf{N}\!\left(A_{\nu\text{out}}(y)\Theta_\mu^\nu(x',y)\right)\gamma^\mu S_{\text{ret}}(x'-x)$$

$$+ e^2 \int d^4x' \int d^4y \int d^4y' \int d^4x'' \overline{\psi}(x')\frac{\overline{\psi}(x'')\gamma^\sigma\psi(x'') - \psi^{\mathrm{T}}(x'')\left(\gamma^\sigma\right)^{\mathrm{T}}\overline{\psi}^{\mathrm{T}}(x'')}{2} \quad (5\text{-}30)$$

$$\times g_{\nu\rho}\mathbf{N}\!\left(\Omega_\sigma^\rho(y',x'')\Theta_\mu^\nu(x',y)\right)\gamma^\mu D_{\text{adv}}(y-y')S_{\text{ret}}(x'-x)$$

$$+ e\int d^4x' \int d^4y\left(\Theta_\mu^\nu(x',y)\right)_{\text{ext}} A_{\nu\,\text{ext}}(y)\overline{\psi}(x')\gamma^\mu S_{\text{ret}}(x'-x);$$

$$A^\mu(y) = \sqrt{Z_3}\,A_{\text{out}}^\mu(y) + e\int d^4y' D_{\text{adv}}(y-y')J^\mu(y'). \quad (5\text{-}31)$$

In (5-31), $J^\mu(y')$ is given by (5-19).

Since all $\psi_{\text{in}}(x)$, $\overline{\psi}_{\text{in}}(x)$, $A_{\text{in}}^\mu(y)$, $\psi_{\text{out}}(x)$, $\overline{\psi}_{\text{out}}(x)$ and $A_{\text{out}}^\mu(y)$ satisfy the corresponding equations of free fields, we can prove that the Lehmann-Symanzik-Zimmermann reduction formulas hold yet for the theory given in this paper and we can, thus, take advantage of such reduction formulas to calculate transition amplitude conveniently.

For example, for the case those initial and final states are an electron, the corresponding transition amplitude is

$$\langle (\boldsymbol{p}',r')_{\text{out}} | (\boldsymbol{p},r)_{\text{in}} \rangle = \langle 0 | b_{r'\text{out}}(\boldsymbol{p}')b_{r\text{in}}^+(\boldsymbol{p}) | 0 \rangle$$

$$= \delta_{r'r}\delta^3(\boldsymbol{p}'-\boldsymbol{p}) - \int d^4z_1 \int d^4z_2\, \overline{\psi}_{r'}^{(+)}(z_1,\boldsymbol{p}')\left(i\gamma^\mu \overrightarrow{\frac{\partial}{\partial z_1^\mu}} - m\right) \quad (5\text{-}32)$$

$$\times \langle 0_{\text{out}} | T\!\left(\frac{\psi(z_1)\overline{\psi}(z_2)}{Z_2}\right) | 0_{\text{in}} \rangle \left(-i\gamma^\nu \overleftarrow{\frac{\partial}{\partial z_2^\nu}} - m\right)\psi_r^{(-)}(z_2,\boldsymbol{p}).$$

Substituting the field operator $\psi(x)$ obtained from (5-16) ~ (5-19) by iterative method to (5-32), we can calculate the transition amplitude $\langle (\boldsymbol{p}',r')_{\text{out}} | (\boldsymbol{p},r)_{\text{in}} \rangle$.

We therefore obtain a complete quantum theory of the model of quantum electrodynamics with nonlocal interaction discussed in Sect. 2 and Sect. 4.

## 6  Summary

In this paper, we investigate some cases summarized in Tab.1 of a kind of model of quantum electrodynamics with nonlocal interaction, of which the main characteristics are as follows: The theories obey the action principle; free charged particle and free electromagnetic field obey the Dirac equation and the Maxwell equation of free fields, respectively; for the case with interaction, both the equations of motion of charged particle and electromagnetic field lead to the normal current conservation (2-23) naturally; all the action and the equations of motion are invariant under the gauge transformation (2-5) ~ (2-7). Besides these conclusions, the Lorentz invariance of the theories is obvious, and the theories return to the conventional local QED under the limit case indicated by (1-3) and (1-8). Hence, all the Cases in Tab.1 satisfy all what requirements a generalized theory must satisfy.

For steering clear of some difficulties and reprehensions of the Lagrangian-Hamiltonian approach and the action principle of nonlocal theory, we take advantage of the Yang-Feldman equations and the Lehmann-Symanzik-Zimmermann formalism to establish the corresponding quantum theory of all the theories in Tab.1.

We therefore have presented intact theoretical frameworks of classical and quantum cases for all theories in Tab.1 of a kind of model of quantum electrodynamics with nonlocal interaction; from the theoretical frameworks we see that, in principle, we can calculate any question of classical and quantum electrodynamics by the theories in Tab.1.

Some other questions will be studied further. For example, we can write out the Yang-Feldman equations and the Lehmann-Symanzik-Zimmermann formalism under the temporal gauge $A^0(x) = 0$ for the theories in Tab.1, thus, can we prove the equivalence of the formalism under the temporal gauge and that under the Lorenz gauge? Etc.

On the other hand, how to establish systematic and convenient approach to calculate transition amplitudes (Can we obtain some rules calculating transition amplitudes similar to the Feynman diagram)? Especially, by choosing appropriate functions $f(x)$, $U^\mu(x)$ and $V^\mu(x)$, e.g., for the fully determined example discussed in Sect. 4.2, can we obtain a theory of quantum electrodynamics in which there is not any divergence?

# Appendix A

**A.1 The proof of (2-10)**

According to the characteristics of determinant we have

$$\delta V(y) = V(y)\widetilde{V}_\alpha^\beta(y)\delta V_\beta^\alpha(y), \tag{A-1}$$

from (1-6) we have $\left(\delta V_\lambda^\mu(y)\right)\widetilde{V}_\nu^\lambda(y) + V_\lambda^\mu(y)\delta\widetilde{V}_\nu^\lambda(y) = 0$, i.e.,

$$\delta \widetilde{V}^{\mu}_{\nu}(y) = -\widetilde{V}^{\mu}_{\alpha}(y)\widetilde{V}^{\beta}_{\nu}(y)\delta V^{\alpha}_{\beta}(y), \tag{A-2}$$

we therefore obtain

$$\frac{\partial V(y)}{\partial y^{\rho}} = V(y)\widetilde{V}^{\beta}_{\alpha}(y)\frac{\partial V^{\alpha}_{\beta}(y)}{\partial y^{\rho}}, \quad \frac{\partial \widetilde{V}^{\mu}_{\nu}(y)}{\partial y^{\rho}} = -\widetilde{V}^{\mu}_{\alpha}(y)\widetilde{V}^{\beta}_{\nu}(y)\frac{\partial V^{\alpha}_{\beta}(y)}{\partial y^{\rho}}; \tag{A-3}$$

And, further, we have

$$\frac{\partial \left(V(y)\widetilde{V}^{\rho}_{\sigma}(y)\right)}{\partial y^{\rho}} = \frac{\partial V(y)}{\partial y^{\rho}}\widetilde{V}^{\rho}_{\sigma}(y) + V(y)\frac{\partial \widetilde{V}^{\rho}_{\sigma}(y)}{\partial y^{\rho}}$$

$$= V(y)\widetilde{V}^{\beta}_{\alpha}(y)\frac{\partial V^{\alpha}_{\beta}(y)}{\partial y^{\rho}}\widetilde{V}^{\rho}_{\sigma}(y) - V(y)\widetilde{V}^{\rho}_{\alpha}(y)\widetilde{V}^{\beta}_{\sigma}(y)\frac{\partial V^{\alpha}_{\beta}(y)}{\partial y^{\rho}}$$

$$= V(y)\widetilde{V}^{\rho}_{\alpha}(y)\widetilde{V}^{\beta}_{\sigma}(y)\left(\frac{\partial V^{\alpha}_{\rho}(y)}{\partial y^{\beta}} - \frac{\partial V^{\alpha}_{\beta}(y)}{\partial y^{\rho}}\right),$$

according to (1-5) we obtain (2-10).

### A.2 The proof of (2-11)

We use two different methods to prove (2-11). The first is based on the Fourier transform of $f(x)$. From (1-2) we have

$$\frac{\partial f\left(U^{\lambda}(x) - V^{\lambda}(y)\right)}{\partial x^{\alpha}} = \frac{\partial}{\partial x^{\alpha}}\int \frac{d^4k}{(2\pi)^4}\hat{f}(k)e^{ik_{\lambda}\left(U^{\lambda}(x) - V^{\lambda}(y)\right)}$$

$$= \frac{\partial U^{\beta}(x)}{\partial x^{\alpha}}\int \frac{d^4k}{(2\pi)^4}ik_{\beta}\hat{f}(k)e^{ik_{\lambda}\left(U^{\lambda}(x) - V^{\lambda}(y)\right)}$$

$$= U^{\beta}_{\alpha}(x)\int \frac{d^4k}{(2\pi)^4}ik_{\beta}\hat{f}(k)e^{ik_{\lambda}\left(U^{\lambda}(x) - V^{\lambda}(y)\right)},$$

$$\frac{\partial f\left(U^{\lambda}(x) - V^{\lambda}(y)\right)}{\partial y^{\alpha}} = \frac{\partial}{\partial y^{\alpha}}\int \frac{d^4k}{(2\pi)^4}\hat{f}(k)e^{ik_{\lambda}\left(U^{\lambda}(x) - V^{\lambda}(y)\right)}$$

$$= -\frac{\partial V^{\beta}(y)}{\partial y^{\alpha}}\int \frac{d^4k}{(2\pi)^4}ik_{\beta}\hat{f}(k)e^{ik_{\lambda}\left(U^{\lambda}(x) - V^{\lambda}(y)\right)}$$

$$= -V^{\beta}_{\alpha}(y)\int \frac{d^4k}{(2\pi)^4}ik_{\beta}\hat{f}(k)e^{ik_{\lambda}\left(U^{\lambda}(x) - V^{\lambda}(y)\right)},$$

from the above two expressions we obtain

$$\widetilde{U}^{\nu}_{\mu}(x)\frac{\partial f\left(U^{\lambda}(x) - V^{\lambda}(y)\right)}{\partial x^{\nu}} = -\widetilde{V}^{\nu}_{\mu}(y)\frac{\partial f\left(U^{\lambda}(x) - V^{\lambda}(y)\right)}{\partial y^{\nu}}. \tag{A-4}$$

The second method is using function algorithm directly but needn't to depend on the Fourier transform of $f(x)$. According to function algorithm we have

$$\frac{\partial f\left(U^{\lambda}(x) - V^{\lambda}(y)\right)}{\partial x^{\nu}} = \frac{\partial U^{\sigma}(x)}{\partial x^{\nu}}\frac{\partial f\left(U^{\lambda}(x) - V^{\lambda}(y)\right)}{\partial U^{\sigma}(x)} = U^{\sigma}_{\nu}(x)\frac{\partial \left(U^{\mu}(x) - V^{\mu}(y)\right)}{\partial U^{\sigma}(x)}\frac{\partial f\left(U^{\lambda}(x) - V^{\lambda}(y)\right)}{\partial \left(U^{\mu}(x) - V^{\mu}(y)\right)}$$

$$= U^{\sigma}_{\nu}(x)\delta^{\mu}_{\sigma}\frac{\partial f\left(U^{\lambda}(x) - V^{\lambda}(y)\right)}{\partial \left(U^{\mu}(x) - V^{\mu}(y)\right)} = U^{\mu}_{\nu}(x)\frac{\partial f\left(U^{\lambda}(x) - V^{\lambda}(y)\right)}{\partial \left(U^{\mu}(x) - V^{\mu}(y)\right)},$$

$$\frac{\partial f(U^\lambda(x)-V^\lambda(y))}{\partial y^\nu} = \frac{\partial V^\sigma(y)}{\partial y^\nu}\frac{\partial f(U^\lambda(x)-V^\lambda(y))}{\partial V^\sigma(y)} = V^\sigma_\nu(y)\frac{\partial(U^\mu(x)-V^\mu(y))}{\partial V^\sigma(y)}\frac{\partial f(U^\lambda(x)-V^\lambda(y))}{\partial(U^\mu(x)-V^\mu(y))}$$
$$= V^\sigma_\nu(y)(-\delta^\mu_\sigma)\frac{\partial f(U^\lambda(x)-V^\lambda(y))}{\partial(U^\mu(x)-V^\mu(y))} = -V^\mu_\nu(y)\frac{\partial f(U^\lambda(x)-V^\lambda(y))}{\partial(U^\mu(x)-V^\mu(y))},$$

from the above two expressions we obtain

$$\frac{\partial f(U^\lambda(x)-V^\lambda(y))}{\partial(U^\mu(x)-V^\mu(y))} = \widetilde{U}^\nu_\mu(x)\frac{\partial f(U^\lambda(x)-V^\lambda(y))}{\partial x^\nu} = -\widetilde{V}^\nu_\mu(y)\frac{\partial f(U^\lambda(x)-V^\lambda(y))}{\partial y^\nu}; \quad \text{(A-5)}$$

This is just (A-4).

According to (A-4), $\dfrac{\partial f(U^\lambda(x)-V^\lambda(y))}{\partial y^\alpha}$ and $\dfrac{\partial f(U^\lambda(x)-V^\lambda(y))}{\partial x^\beta}$ can be transformed each other, for example, we can obtain (2-11) by multiplying (A-4) with $V^\mu_\tau(y)$.

### A.3 The proof of (2-19)

We first calculate $\int d^4 y \delta^4(K^\mu(x)-K^\mu(y))g(y)$, where $g(y)$ is an arbitrary (scalar, vector or tensor) function, $K^\mu(x)$ is an arbitrary vector function, for which we set $\xi^\mu = K^\mu(y)$ and denote $\xi^\mu_0 = K^\mu(x)$, the corresponding inverse function is $y^\mu = y^\mu(\xi)$ and we have $y^\mu(\xi_0) = x^\mu$. Introducing $K^\alpha_\beta(y) \equiv \dfrac{\partial K^\alpha(y)}{\partial y^\beta}$ and the corresponding determinant $K(y) \equiv \left\|K^\alpha_\beta(y)\right\|$, and taking account of $d^4\xi = K(y)d^4 y$, we obtain

$$\int d^4 y \delta^4(K^\mu(x)-K^\mu(y))g(y) = \int K(y)d^4 y \delta^4(K^\mu(x)-K^\mu(y))\frac{g(y)}{K(y)}$$
$$= \int d^4\xi \delta^4(\xi - \xi_0)\frac{g(y(\xi))}{K(y(\xi))} = \frac{g(y(\xi_0))}{K(y(\xi_0))} = \frac{g(x)}{K(x)}.$$
(A-6)

According to (1-1) and (A-6) we have

$$\int d^4 z U(x)V(z)\widetilde{f}(U^\rho(x)-V^\rho(z))f(U^\sigma(y)-V^\sigma(z)) = U(x)\int d^4 V \widetilde{f}(U^\rho(x)-V^\rho)f(U^\sigma(y)-V^\sigma)$$
$$= U(x)\delta^4(U^\lambda(x)-U^\lambda(y)) = \delta^4(x-y).$$

This is just (2-19).

### A.4 The four functions $D_{C_i}(x)$ $(i=1,2,3,4)$

Defining[1, 4]

$$D_{(\pm)}(x) = \int \frac{d^3 k}{(2\pi)^3}\frac{i}{2|\boldsymbol{k}|}e^{\mp i|\boldsymbol{k}|x^0 + i\boldsymbol{k}\cdot\boldsymbol{x}},$$

the four functions $D_{C_i}(x)$ $(i=1,2,3,4)$ can be written to the following forms

$$D_{C_1}(x) = D_{\text{ret}}(x) = \begin{cases} D_{(+)}(x) - D_{(-)}(x), & x^0 > 0; \\ 0, & x^0 < 0, \end{cases} \quad D_{C_2}(x) = D_{\text{adv}}(x) = \begin{cases} 0, & x^0 > 0; \\ -D_{(+)}(x) + D_{(-)}(x), & x^0 < 0, \end{cases}$$

$$D_{C_3}(x) = D_{\text{F}}(x) = \begin{cases} D_{(+)}(x), & x^0 > 0; \\ D_{(-)}(x), & x^0 < 0, \end{cases} \quad D_{C_4}(x) = \begin{cases} -D_{(-)}(x), & x^0 > 0; \\ -D_{(+)}(x), & x^0 < 0. \end{cases}$$

From the above expressions we can prove

$$D_{C_1}(x) + D_{C_2}(x) = D_{C_3}(x) + D_{C_4}(x). \tag{A-7}$$

and we have

$$D_{\text{ret}}(x) - D_{\text{adv}}(x) = D_{(+)}(x) - D_{(-)}(x)$$
$$= \int \frac{d^3k}{(2\pi)^3} \frac{i}{2|\bm{k}|} \left( e^{-i|\bm{k}|x^0 + i\bm{k}\cdot\bm{x}} - e^{i|\bm{k}|x^0 + i\bm{k}\cdot\bm{x}} \right) = \int \frac{d^3k}{(2\pi)^3} \frac{i}{2|\bm{k}|} \left( e^{-ik\cdot x} - e^{ik\cdot x} \right)\Big|_{k^0 = |\bm{k}|}, \tag{A-8}$$

### A.5 A vector function

We consider a form of

$$K^\mu(x) = x^\mu + a\Lambda^\mu(x),$$

where $a$ is a constant, $\Lambda^\mu(x)$ is an arbitrary vector function. We have

$$K^\alpha_\beta(x) = \frac{\partial(x^\alpha + a\Lambda^\alpha(x))}{\partial x^\beta} = \delta^\alpha_\beta + a\Lambda^\alpha{}_{,\beta}(x);$$

the determinant $K(x)$ of $K^\alpha_\beta(x)$ is

$$K(x) \equiv \left\| \delta^\alpha_\beta + a\frac{\partial \Lambda^\alpha(x)}{\partial x^\beta} \right\| = \begin{vmatrix} 1+a\Lambda^0{}_{,0}(x) & a\Lambda^0{}_{,1}(x) & a\Lambda^0{}_{,2}(x) & a\Lambda^0{}_{,3}(x) \\ a\Lambda^1{}_{,0}(x) & 1+a\Lambda^1{}_{,1}(x) & a\Lambda^1{}_{,2}(x) & a\Lambda^1{}_{,3}(x) \\ a\Lambda^2{}_{,0}(x) & a\Lambda^2{}_{,1}(x) & 1+a\Lambda^2{}_{,2}(x) & a\Lambda^2{}_{,3}(x) \\ a\Lambda^3{}_{,0}(x) & a\Lambda^3{}_{,1}(x) & a\Lambda^3{}_{,2}(x) & 1+a\Lambda^3{}_{,3}(x) \end{vmatrix}$$

$$\equiv 1 + \sum_{n=1}^{4} \frac{a^n}{n!} K_{(n)}(x);$$

$$K_{(1)}(x) = \Lambda^\lambda{}_{,\lambda}(x),$$

$$K_{(2)}(x) = \Lambda^\mu{}_{,\mu}(x)\Lambda^\nu{}_{,\nu}(x) - \Lambda^\mu{}_{,\nu}(x)\Lambda^\nu{}_{,\mu}(x),$$

$$K_{(3)}(x) = \Lambda^\alpha{}_{,\alpha}(x)\Lambda^\beta{}_{,\beta}(x)\Lambda^\gamma{}_{,\gamma}(x) + 2\Lambda^\alpha{}_{,\beta}(x)\Lambda^\beta{}_{,\gamma}(x)\Lambda^\gamma{}_{,\alpha}(x)$$
$$- 3\Lambda^\lambda{}_{,\lambda}(x)\Lambda^\mu{}_{,\nu}(x)\Lambda^\nu{}_{,\mu}(x),$$

$$K_{(4)}(x) = \Lambda^\mu{}_{,\mu}(x)\Lambda^\nu{}_{,\nu}(x)\Lambda^\rho{}_{,\rho}(x)\Lambda^\sigma{}_{,\sigma}(x) + 3\Lambda^\mu{}_{,\nu}(x)\Lambda^\nu{}_{,\mu}(x)\Lambda^\rho{}_{,\sigma}(x)\Lambda^\sigma{}_{,\rho}(x)$$
$$+ 8\Lambda^\lambda{}_{,\lambda}(x)\Lambda^\alpha{}_{,\beta}(x)\Lambda^\beta{}_{,\gamma}(x)\Lambda^\gamma{}_{,\alpha}(x) - 6\Lambda^\mu{}_{,\mu}(x)\Lambda^\nu{}_{,\nu}(x)\Lambda^\rho{}_{,\sigma}(x)\Lambda^\sigma{}_{,\rho}(x)$$
$$- 6\Lambda^\mu{}_{,\nu}(x)\Lambda^\nu{}_{,\rho}(x)\Lambda^\rho{}_{,\sigma}(x)\Lambda^\sigma{}_{,\mu}(x).$$

The corresponding inverse functions $\widetilde{K}^\mu_\nu(x) \equiv \dfrac{\partial x^\mu}{\partial K^\nu(x)}$ is

$$\widetilde{K}_\nu^\mu(x) = \frac{1}{K(x)}\left(\delta_\nu^\mu + \sum_{n=1}^{3}\frac{a^n}{n!}\widetilde{K}_{\nu\,(n)}^\mu(x)\right),$$

$$\widetilde{K}_{\nu\,(1)}^\mu(x) = \delta_\nu^\mu K_{(1)}(x) - \Lambda^\mu{}_{,\nu}(x),$$

$$\widetilde{K}_{\nu\,(2)}^\mu(x) = \delta_\nu^\mu K_{(2)}(x) - 2\Lambda^\mu{}_{,\nu}(x)K_{(1)}(x) + 2\Lambda^\mu{}_{,\lambda}(x)\Lambda^\lambda{}_{,\nu}(x),$$

$$\widetilde{K}_{\nu\,(3)}^\mu(x) = \delta_\nu^\mu K_{(3)}(x) - 3\Lambda^\mu{}_{,\nu}(x)K_{(2)}(x) + 6\Lambda^\mu{}_{,\lambda}(x)\Lambda^\lambda{}_{,\nu}(x)K_{(1)}(x)$$
$$- 6\Lambda^\mu{}_{,\alpha}(x)\Lambda^\alpha{}_{,\beta}(x)\Lambda^\beta{}_{,\nu}(x).$$

We can verify that the above functions $K_\beta^\alpha(x)$ and $\widetilde{K}_\beta^\alpha(x)$ satisfy (1-5), (A-3) and (2-10), and we can prove that $\widetilde{K}_{\nu\,(3)}^\mu(x)$ satisfies

$$\widetilde{K}_{\lambda\,(3)}^\mu(x)\Lambda^\lambda{}_{,\nu}(x) = \frac{1}{4}\delta_\nu^\mu K_{(4)}(x).$$

If we take $\Lambda^\mu(x) = A_{\perp G}^\mu(x)$ in the above formulas, where $A_{\perp G}^\mu(x)$ are given by (3-43), then according to $A_{\perp G,\lambda}^\lambda(x) = 0$, from the above formulas we obtain the expressions of $V(x)$ and $\widetilde{V}_\nu^\mu(x)$ given by (4-3) and (4-4).

## Appendix B

For the case that both $f(x)$ and $U^\mu(x)$ are independent of $\psi(x)$ and $A^\mu(x)$, $V^\mu(x)$, $V_\nu^\mu(x)$, $V(x)$ and $\widetilde{V}_\nu^\mu(x)$ are expressed by (4-1), (4-2), (4-3) and (4-4), respectively, since all the functions $f(x)$, $U^\mu(x)$ and $V^\mu(x)$ are independent of wave function $\psi(x)$ of charged particle, from the action (2-1), (2-2) and (2-4) and the variational equation $\frac{\delta S}{\delta\psi(x)} = 0$ we obtain immediately the equation of motion of charged particle (4-5).

We now derivate the equation of motion of electromagnetic field by the variational equations $\frac{\delta S}{\delta A_\mu(y)} = 0$. For the sake of simpleness, notice[1, 4] $D_{\mathrm{ret}}(x) = \int\frac{\mathrm{d}^4k}{(2\pi)^4}\frac{-1}{(k_0 + \mathrm{i}\varepsilon)^2 - \boldsymbol{k}\cdot\boldsymbol{k}}\mathrm{e}^{-\mathrm{i}k\cdot x}$, we write formally $A_\perp^\mu(x)$ and $A_{\perp G}^\mu(x)$ to the forms

$$A_\perp^\mu(x) = \int d^4y \int \frac{d^4k}{(2\pi)^4} \varepsilon_\perp^{\mu\nu}(k) e^{-ik\cdot(x-y)} A_\nu(y),$$

$$A_{\perp G}^\mu(x) = \int d^4y \int \frac{d^4k}{(2\pi)^4} \xi(k) \varepsilon_\perp^{\mu\nu}(k) e^{-ik\cdot(x-y)} A_\nu(y), \tag{B-1}$$

$$\varepsilon_\perp^{\mu\nu}(k) = \frac{k^2 g^{\mu\nu} - k^\mu k^\nu}{(k_0 + i\varepsilon)^2 - \bm{k}\cdot\bm{k}}.$$

Formally, from (4-1) we see that $V^\mu(x)$ is only a function of $A_{\perp G}^\mu(x)$ and independent of $A_{\perp G,\nu}^\mu(x)$, and from the forms of (4-3) and (4-4) we see that both $V(x)$ and $\widetilde{V}_\nu^\mu(x)$ are only functions of $A_{\perp G,\nu}^\mu(x)$ and independent of $A_{\perp G}^\mu(x)$, we therefore have

$$\begin{aligned}
\frac{\delta S_I}{\delta A_\mu(y)} &= -e\int d^4z_1 d^4z_2 j^\alpha(z_1) U_\alpha^\tau(z_1) f\big(U^\lambda(z_1) - V^\lambda(z_2)\big) V(z_2) \widetilde{V}_\tau^\beta(z_2) \frac{\delta A_\beta(z_2)}{\delta A_\mu(y)} \\
&\quad - e\int d^4z_1 d^4z_2 j^\alpha(z_1) U_\alpha^\tau(z_1) \frac{\delta\big[f\big(U^\lambda(z_1) - V^\lambda(z_2)\big) V(z_2) \widetilde{V}_\tau^\beta(z_2)\big]}{\delta A_\mu(y)} A_\beta(z_2) \\
&= -e\int d^4z_1 d^4z_2 j^\alpha(z_1) U_\alpha^\tau(z_1) f\big(U^\lambda(z_1) - V^\lambda(z_2)\big) V(z_2) \widetilde{V}_\tau^\beta(z_2) \delta_\beta^\mu \delta^4(z_2 - y) \\
&\quad - e\int d^4z_1 d^4z_2 j^\alpha(z_1) U_\alpha^\gamma(z_1) \frac{\partial f\big(U^\lambda(z_1) - V^\lambda(z_2)\big)}{\partial A_{\perp G}^\rho(z_2)} \frac{\delta A_{\perp G}^\rho(z_2)}{\delta A_\mu(y)} V(z_2) \widetilde{V}_\gamma^\beta(z_2) A_\beta(z_2) \\
&\quad - e\int d^4z_1 d^4z_2 j^\alpha(z_1) U_\alpha^\gamma(z_1) f\big(U^\lambda(z_1) - V^\lambda(z_2)\big) \frac{\partial\big(V(z_2)\widetilde{V}_\gamma^\beta(z_2)\big)}{\partial A_{\perp G,\sigma}^\rho(z_2)} \frac{\delta A_{\perp G,\sigma}^\rho(z_2)}{\delta A_\mu(y)} A_\beta(z_2) \\
&= -e\int d^4z_1 j^\alpha(z_1) U_\alpha^\tau(z_1) f\big(U^\lambda(z_1) - V^\lambda(y)\big) V(y) \widetilde{V}_\tau^\mu(y) \\
&\quad - e\int d^4z_1 d^4z_2 j^\alpha(z_1) U_\alpha^\gamma(z_1) \frac{\partial f\big(U^\lambda(z_1) - V^\lambda(z_2)\big)}{\partial A_{\perp G}^\rho(z_2)} V(z_2) \widetilde{V}_\gamma^\beta(z_2) A_\beta(z_2) \\
&\quad \times \int \frac{d^4k}{(2\pi)^4} \xi(k) \varepsilon_\perp^{\rho\omega}(k) e^{-ik\cdot(z_2-z)} \frac{\delta A_\omega(z)}{\delta A_\mu(y)} d^4z \\
&\quad - e\int d^4z_1 d^4z_2 j^\alpha(z_1) U_\alpha^\gamma(z_1) f\big(U^\lambda(z_1) - V^\lambda(z_2)\big) \frac{\partial\big(V(z_2)\widetilde{V}_\gamma^\beta(z_2)\big)}{\partial A_{\perp G,\sigma}^\rho(z_2)} A_\beta(z_2) \\
&\quad \times \frac{\partial}{\partial z_2^\sigma} \int \frac{d^4k}{(2\pi)^4} \xi(k) \varepsilon_\perp^{\rho\omega}(k) e^{-ik\cdot(z_2-z)} \frac{\delta A_\omega(z)}{\delta A_\mu(y)} d^4z \\
&= -e W_{(1)}^\mu(y) \\
&\quad - e\int d^4z_1 d^4z_2 j^\alpha(z_1) U_\alpha^\gamma(z_1) \frac{\partial f\big(U^\lambda(z_1) - V^\lambda(z_2)\big)}{\partial A_{\perp G}^\rho(z_2)} V(z_2) \widetilde{V}_\gamma^\beta(z_2) A_\beta(z_2) \\
&\quad \times \int \frac{d^4k}{(2\pi)^4} \xi(k) \varepsilon_\perp^{\rho\mu}(k) e^{-ik\cdot(z_2-y)} \\
&\quad - e\int d^4z_1 d^4z_2 j^\alpha(z_1) U_\alpha^\gamma(z_1) f\big(U^\lambda(z_1) - V^\lambda(z_2)\big) \frac{\partial\big(V(z_2)\widetilde{V}_\gamma^\beta(z_2)\big)}{\partial A_{\perp G,\sigma}^\rho(z_2)} A_\beta(z_2) \\
&\quad \times \frac{\partial}{\partial z_2^\sigma} \int \frac{d^4k}{(2\pi)^4} \xi(k) \varepsilon_\perp^{\rho\mu}(k) e^{-ik\cdot(z_2-y)}.
\end{aligned} \tag{B-2}$$

In (B-2), $W_{(1)}^\mu$ is expressed by (4-7). In the above derivation, we have used the expression (B-1)

of $A^\mu_{\perp G}(x)$.

Similar to the derivation of (A-5), we have

$$\frac{\partial f\left(U^\lambda(z_1) - V^\lambda(z_2)\right)}{\partial A^\rho_{\perp G}(z_2)} = \frac{\partial f\left(U^\lambda(z_1) - V^\lambda(z_2)\right)}{\partial \left(U^\lambda(z_1) - V^\lambda(z_2)\right)} \frac{\partial \left(U^\lambda(z_1) - V^\lambda(z_2)\right)}{\partial A^\rho_{\perp G}(z_2)}$$
$$= -\widetilde{V}^\nu_\mu(z_2) \frac{\partial f\left(U^\lambda(z_1) - V^\lambda(z_2)\right)}{\partial z_2^\nu} \left(-\frac{\partial V^\mu(z_2)}{\partial A^\rho_{\perp G}(z_2)}\right) = a \frac{\partial f\left(U^\lambda(z_1) - V^\lambda(z_2)\right)}{\partial z_2^\sigma} \widetilde{V}^\sigma_\rho(z_2); \quad \text{(B-3)}$$

from (A-1) and (A-2) and considering (4-1), we have

$$\frac{\partial V(z_2)}{\partial A^\rho_{\perp G,\sigma}(z_2)} = V(z_2)\widetilde{V}^\theta_\varphi(z_2)\frac{\partial V^\varphi_\theta(z_2)}{\partial A^\rho_{\perp G,\sigma}(z_2)} = V(z_2)\widetilde{V}^\theta_\varphi(z_2) a \delta^\varphi_\rho \delta^\sigma_\theta = aV(z_2)\widetilde{V}^\sigma_\rho(z_2),$$
$$\frac{\partial \widetilde{V}^\beta_\gamma(z_2)}{\partial A^\rho_{\perp G,\sigma}(z_2)} = -\widetilde{V}^\beta_\varphi(z_2)\widetilde{V}^\theta_\gamma(z_2)\frac{\partial V^\varphi_\theta(z_2)}{\partial A^\rho_{\perp G,\sigma}(z_2)} = -\widetilde{V}^\beta_\varphi(z_2)\widetilde{V}^\theta_\gamma(z_2) a \delta^\varphi_\rho \delta^\sigma_\theta = -a\widetilde{V}^\beta_\rho(z_2)\widetilde{V}^\sigma_\gamma(z_2); \quad \text{(B-4)}$$

we therefore obtain

$$\frac{\partial \left(V(z_2)\widetilde{V}^\beta_\gamma(z_2)\right)}{\partial A^\rho_{\perp G,\sigma}(z_2)} = \frac{\partial V(z_2)}{\partial A^\rho_{\perp G,\sigma}(z_2)}\widetilde{V}^\beta_\gamma(z_2) + V(z_2)\frac{\partial \widetilde{V}^\beta_\gamma(z_2)}{\partial A^\rho_{\perp G,\sigma}(z_2)}$$
$$= aV(z_2)\widetilde{V}^\sigma_\rho(z_2)\widetilde{V}^\beta_\gamma(z_2) - V(z_2)a\widetilde{V}^\beta_\rho(z_2)\widetilde{V}^\sigma_\gamma(z_2) \quad \text{(B-5)}$$
$$= aV(z_2)\left(\widetilde{V}^\sigma_\rho(z_2)\widetilde{V}^\beta_\gamma(z_2) - \widetilde{V}^\beta_\rho(z_2)\widetilde{V}^\sigma_\gamma(z_2)\right).$$

Substituting (B-3) and (B-5) to (B-2), and using integration by parts for the last term in (B-2), we have

$$\frac{\delta S_\mathrm{I}}{\delta A_\mu(y)} = -eW^\mu_{(1)}(y)$$

$$-ae\int d^4z_1 d^4z_2 j^\alpha(z_1) U^\gamma_\alpha(z_1) \frac{\partial f\big(U^\lambda(z_1)-V^\lambda(z_2)\big)}{\partial z_2^\sigma} \widetilde{V}^\sigma_\rho(z_2) V(z_2) \widetilde{V}^\beta_\gamma(z_2) A_\beta(z_2)$$

$$\times \int \frac{d^4k}{(2\pi)^4} \xi(k) \varepsilon^{\rho\mu}_\perp(k) e^{-ik\cdot(z_2-y)}$$

$$+ae\int d^4z_1 d^4z_2 j^\alpha(z_1) U^\gamma_\alpha(z_1) \frac{\partial}{\partial z_2^\sigma}\Big[ f\big(U^\lambda(z_1)-V^\lambda(z_2)\big) V(z_2)$$

$$\times \big(\widetilde{V}^\sigma_\rho(z_2)\widetilde{V}^\beta_\gamma(z_2) - \widetilde{V}^\beta_\rho(z_2)\widetilde{V}^\sigma_\gamma(z_2)\big) A_\beta(z_2)\Big] \int \frac{d^4k}{(2\pi)^4} \xi(k) \varepsilon^{\rho\mu}_\perp(k) e^{-ik\cdot(z_2-y)}$$

$$= -eW^\mu_{(1)}(y)$$

$$-ae\int d^4z_1 d^4z_2 j^\alpha(z_1) U^\gamma_\alpha(z_1) \frac{\partial f\big(U^\lambda(z_1)-V^\lambda(z_2)\big)}{\partial z_2^\sigma} \widetilde{V}^\sigma_\rho(z_2) V(z_2) \widetilde{V}^\beta_\gamma(z_2) A_\beta(z_2)$$

$$\times \int \frac{d^4k}{(2\pi)^4} \xi(k) \varepsilon^{\rho\mu}_\perp(k) e^{-ik\cdot(z_2-y)}$$

$$+ae\int d^4z_1 d^4z_2 j^\alpha(z_1) U^\gamma_\alpha(z_1) \frac{\partial f\big(U^\lambda(z_1)-V^\lambda(z_2)\big)}{\partial z_2^\sigma} V(z_2)\big(\widetilde{V}^\sigma_\rho(z_2)\widetilde{V}^\beta_\gamma(z_2) - \widetilde{V}^\beta_\rho(z_2)\widetilde{V}^\sigma_\gamma(z_2)\big) A_\beta(z_2)$$

$$\times \int \frac{d^4k}{(2\pi)^4} \xi(k) \varepsilon^{\rho\mu}_\perp(k) e^{-ik\cdot(z_2-y)}$$

$$+ae\int d^4z_1 d^4z_2 j^\alpha(z_1) U^\gamma_\alpha(z_1) f\big(U^\lambda(z_1)-V^\lambda(z_2)\big) \frac{\partial\big[V(z_2)\big(\widetilde{V}^\sigma_\rho(z_2)\widetilde{V}^\beta_\gamma(z_2) - \widetilde{V}^\beta_\rho(z_2)\widetilde{V}^\sigma_\gamma(z_2)\big) A_\beta(z_2)\big]}{\partial z_2^\sigma}$$

$$\times \int \frac{d^4k}{(2\pi)^4} \xi(k) \varepsilon^{\rho\mu}_\perp(k) e^{-ik\cdot(z_2-y)}$$

$$= -eW^\mu_{(1)}(y)$$

$$-ae\int d^4z_1 d^4z_2 j^\alpha(z_1) U^\gamma_\alpha(z_1) \frac{\partial f\big(U^\lambda(z_1)-V^\lambda(z_2)\big)}{\partial z_2^\sigma} V(z_2) \widetilde{V}^\beta_\rho(z_2) \widetilde{V}^\sigma_\gamma(z_2) A_\beta(z_2)$$

$$\times \int \frac{d^4k}{(2\pi)^4} \xi(k) \varepsilon^{\rho\mu}_\perp(k) e^{-ik\cdot(z_2-y)}$$

$$+ae\int d^4z_1 d^4z_2 j^\alpha(z_1) U^\gamma_\alpha(z_1) f\big(U^\lambda(z_1)-V^\lambda(z_2)\big)$$

$$\times \frac{\partial\big[V(z_2)\big(\widetilde{V}^\sigma_\rho(z_2)\widetilde{V}^\beta_\gamma(z_2) - \widetilde{V}^\beta_\rho(z_2)\widetilde{V}^\sigma_\gamma(z_2)\big) A_\beta(z_2)\big]}{\partial z_2^\sigma} \int \frac{d^4k}{(2\pi)^4} \xi(k) \varepsilon^{\rho\mu}_\perp(k) e^{-ik\cdot(z_2-y)}. \qquad (\text{B-6})$$

Using (2-10), (A-3) and (1-5) we have

$$\frac{\partial\left[V(z_2)\left(\widetilde{V}_\rho^\sigma(z_2)\widetilde{V}_\gamma^\beta(z_2)-\widetilde{V}_\rho^\beta(z_2)\widetilde{V}_\gamma^\sigma(z_2)\right)A_\beta(z_2)\right]}{\partial z_2^\sigma}$$

$$=V(z_2)\widetilde{V}_\rho^\sigma(z_2)\frac{\partial \widetilde{V}_\gamma^\beta(z_2)}{\partial z_2^\sigma}A_\beta(z_2)-V(z_2)\frac{\partial \widetilde{V}_\rho^\beta(z_2)}{\partial z_2^\sigma}\widetilde{V}_\gamma^\sigma(z_2)A_\beta(z_2)$$

$$+V(z_2)\left(\widetilde{V}_\rho^\sigma(z_2)\widetilde{V}_\gamma^\beta(z_2)-\widetilde{V}_\rho^\beta(z_2)\widetilde{V}_\gamma^\sigma(z_2)\right)A_{\beta,\sigma}(z_2)$$

$$=V(z_2)\widetilde{V}_\rho^\sigma(z_2)\left(-\widetilde{V}_\varphi^\beta(z_2)\widetilde{V}_\gamma^\theta(z_2)\frac{\partial V_\theta^\varphi(z_2)}{\partial z_2^\sigma}\right)A_\beta(z_2) \qquad\text{(B-7)}$$

$$-V(z_2)\left(-\widetilde{V}_\varphi^\beta(z_2)\widetilde{V}_\rho^\theta(z_2)\frac{\partial V_\theta^\varphi(z_2)}{\partial z_2^\sigma}\right)\widetilde{V}_\gamma^\sigma(z_2)A_\beta(z_2)$$

$$+V(z_2)\widetilde{V}_\rho^\sigma(z_2)\widetilde{V}_\gamma^\beta(z_2)\left(A_{\beta,\sigma}(z_2)-A_{\sigma,\beta}(z_2)\right)$$

$$=V(z_2)\widetilde{V}_\rho^\sigma(z_2)\widetilde{V}_\gamma^\beta(z_2)\left(A_{\beta,\sigma}(z_2)-A_{\sigma,\beta}(z_2)\right).$$

Substituting (B-7) to (B-6) and considering (2-11), (B-6) becomes

$$\frac{\delta S_\text{I}}{\delta A_\mu(y)}=-eW_{(1)}^\mu(y)$$

$$+ae\int d^4z_1 d^4z_2\, j^\alpha(z_1)U_\alpha^\gamma(z_1)\widetilde{U}_\varphi^\theta(x)\frac{\partial f\left(U^\lambda(z_1)-V^\lambda(z_2)\right)}{\partial z_1^\theta}V_\sigma^\varphi(y)V(z_2)\widetilde{V}_\rho^\beta(z_2)\widetilde{V}_\gamma^\sigma(z_2)A_\beta(z_2)$$

$$\times \int\frac{d^4k}{(2\pi)^4}\xi(k)\varepsilon_\perp^{\rho\mu}(k)e^{-ik\cdot(z_2-y)}$$

$$+ae\int d^4z_1 d^4z_2\, j^\alpha(z_1)U_\alpha^\gamma(z_1)f\left(U^\lambda(z_1)-V^\lambda(z_2)\right)V(z_2)\widetilde{V}_\rho^\sigma(z_2)\widetilde{V}_\gamma^\beta(z_2)\left(A_{\beta,\sigma}(z_2)-A_{\sigma,\beta}(z_2)\right)$$

$$\times \int\frac{d^4k}{(2\pi)^4}\xi(k)\varepsilon_\perp^{\rho\mu}(k)e^{-ik\cdot(z_2-y)}$$

$$=-eW_{(1)}^\mu(y)$$

$$+ae\int d^4z_1 d^4z_2\, j^\alpha(z_1)\frac{\partial f\left(U^\lambda(z_1)-V^\lambda(z_2)\right)}{\partial z_1^\alpha}V(z_2)\widetilde{V}_\rho^\beta(z_2)A_\beta(z_2)\int\frac{d^4k}{(2\pi)^4}\xi(k)\varepsilon_\perp^{\rho\mu}(k)e^{-ik\cdot(z_2-y)}$$

$$+ae\int d^4z_1 d^4z_2\, j^\alpha(z_1)U_\alpha^\gamma(z_1)f\left(U^\lambda(z_1)-V^\lambda(z_2)\right)V(z_2)\widetilde{V}_\rho^\sigma(z_2)\widetilde{V}_\gamma^\beta(z_2)\left(A_{\beta,\sigma}(z_2)-A_{\sigma,\beta}(z_2)\right)$$

$$\times \int\frac{d^4k}{(2\pi)^4}\xi(k)\varepsilon_\perp^{\rho\mu}(k)e^{-ik\cdot(z_2-y)}$$

$$=-eW_{(1)}^\mu(y)$$

$$-ae\int d^4z_1 d^4z_2 \frac{\partial j^\alpha(z_1)}{\partial z_1^\alpha}f\left(U^\lambda(z_1)-V^\lambda(z_2)\right)V(z_2)\widetilde{V}_\rho^\beta(z_2)A_\beta(z_2)\int\frac{d^4k}{(2\pi)^4}\xi(k)\varepsilon_\perp^{\rho\mu}(k)e^{-ik\cdot(z_2-y)}$$

$$=-eW_{(1)}^\mu(y)-aeW_{(2)\perp}^\mu(y) \qquad\text{(B-8)}$$

$$-ae\int d^4z_1 d^4z_2 \frac{\partial j^\alpha(z_1)}{\partial z_1^\alpha}f\left(U^\lambda(z_1)-V^\lambda(z_2)\right)V(z_2)\widetilde{V}_\rho^\beta(z_2)A_\beta(z_2)\int\frac{d^4k}{(2\pi)^4}\xi(k)\varepsilon_\perp^{\rho\mu}(k)e^{ik\cdot(z_2-y)},$$

where we have used the integration by parts for $z_1^\alpha$, and $W_{(2)\perp\text{G}}^\mu(y)$ is given by (4-8) ~ (4-10).

On the other hand, as well-known,

$$\frac{\delta S_\text{EM}}{\delta A_\mu(y)}=-\frac{1}{4}\frac{\delta}{\delta A_\mu(y)}\int d^4x\left(A_{\alpha,\beta}(x)-A_{\beta,\alpha}(x)\right)\left(A^{\alpha,\beta}(x)-A^{\beta,\alpha}(x)\right)=\frac{\partial\left(A^{\mu,\nu}(y)-A^{\nu,\mu}(y)\right)}{\partial y^\nu}. \qquad\text{(B-9)}$$

Hence, the variational equation $\frac{\delta S}{\delta A_\mu(y)} = \frac{\delta S_{EM}}{\delta A_\mu(y)} + \frac{\delta S_I}{\delta A_\mu(y)} = 0$ leads to the equation of motion of electromagnetic field:

$$\frac{\partial}{\partial y^\nu}\left(A^{\mu,\nu}(y) - A^{\nu,\mu}(y)\right) = eW_{(1)}^\mu(y) + aeW_{(2)\perp}^\mu(y)$$
$$+ ae\int d^4z_1 d^4z_2 \frac{\partial j^\alpha(z_1)}{\partial z_1^\alpha} f\left(U^\lambda(z_1) - V^\lambda(z_2)\right) V(z_2)\widetilde{V}_\rho^\beta(z_2) A_\beta(z_2) \int \frac{d^4k}{(2\pi)^4} \xi(k)\varepsilon_\perp^{\rho\mu}(k)e^{-ik\cdot(z_2-y)}. \quad (B\text{-}10)$$

Using the method shown in §2.3, it is easy to prove that (B-10) leads to the current conservation equation (2-23); and, further, substituting the current conservation equation $\frac{\partial j^\alpha(z_1)}{\partial z_1^\alpha} = 0$ to (B-10), we obtain (4-6).

If $\xi(k)=1$ in (3-43), namely, $A_{\perp G}^\mu(x) = A_\perp^\mu(x)$, where $A_\perp^\mu(x)$ is given by (3-43), then according to (4-14), $W_{(2)\nu}(z)$ given by (4-10) becomes

$$W_{(2)}^\mu(z) = g^{\mu\nu}W_{(2)\nu}(z)$$
$$= g^{\mu\nu}V(z)\widetilde{V}_\nu^\rho(z)\widetilde{V}_\alpha^\sigma(z)\frac{1}{a}\left(g_{\tau\rho}V_\sigma^\tau(z) - g_{\tau\sigma}V_\rho^\tau(z)\right)\int d^4x j^\beta(x)U_\beta^\alpha(x)f\left(U^\lambda(x) - V^\lambda(z)\right)$$
$$= \frac{1}{a}g^{\mu\nu}V(z)\left(g_{\alpha\rho}\widetilde{V}_\nu^\rho(z) - g_{\nu\sigma}\widetilde{V}_\alpha^\sigma(z)\right)\int d^4x j^\beta(x)U_\beta^\alpha(x)f\left(U^\lambda(x) - V^\lambda(z)\right)$$
$$= \frac{1}{a}\left(W_{(3)}^\mu(z) - W_{(1)}^\mu(z)\right) \equiv W_{(4)}^\mu(z),$$

where $W_{(3)}^\mu(z)$ and $W_{(4)}^\mu(z)$ are expressed by (4-16), (4-18) and (4-19), respectively; $W_{(1)}^\mu(z)$ is still expressed by (4-7) but in which $A_{\perp G}^\mu(x)$ is replaced with $A_\perp^\mu(x)$. And, further, since $\xi(k)=1$ in (4-8), it becomes

$$W_{(2)\perp G}^\mu(y) = W_{(2)\perp}^\mu(y) = W_{(2)}^\mu(y) - W_{(2)//}^\mu(y) = \frac{1}{a}\left(W_{(3)}^\mu(y) - W_{(1)}^\mu(y)\right) - W_{(4)//}^\mu(y). \quad (B\text{-}11)$$

Substituting (B-11) to (4-6), we obtain

$$\frac{\partial}{\partial y^\nu}\left(A^{\mu,\nu}(y) - A^{\nu,\mu}(y)\right) = eW_{(1)}^\mu(y) + ae\left(\frac{1}{a}\left(W_{(3)}^\mu(y) - W_{(1)}^\mu(y)\right) - W_{(4)//}^\mu(y)\right)$$
$$= eW_{(3)}^\mu(y) - aeW_{(4)//}^\mu(y).$$

This is just (4-15).

# Appendix C

All the quantities appearing in (5-5) ~ (5-7) can be found in standard textbooks; for the integrity of the paper, we list out these quantities and some characteristics here.

$$\psi_r^{(-)}(x,\boldsymbol{p}) = \frac{1}{(2\pi)^{3/2}}\sqrt{\frac{m}{E_{\boldsymbol{p}}}}\mathrm{e}^{-ip\cdot x}u_r(\boldsymbol{p})\,,\quad \psi_r^{(+)}(x,\boldsymbol{p}) = \frac{1}{(2\pi)^{3/2}}\sqrt{\frac{m}{E_{\boldsymbol{p}}}}\mathrm{e}^{ip\cdot x}v_r(\boldsymbol{p})\,,$$

$$\overline{\psi}_r^{(+)}(x,\boldsymbol{p}) = \frac{1}{(2\pi)^{3/2}}\sqrt{\frac{m}{E_{\boldsymbol{p}}}}\mathrm{e}^{ip\cdot x}\overline{u}_r(\boldsymbol{p})\,,\quad \overline{\psi}_r^{(-)}(x,\boldsymbol{p}) = \frac{1}{(2\pi)^{3/2}}\sqrt{\frac{m}{E_{\boldsymbol{p}}}}\mathrm{e}^{-ip\cdot x}\overline{v}_r(\boldsymbol{p})\,,$$

$$\overline{u}_r(\boldsymbol{p})u_r(\boldsymbol{p})=1\,,\quad \overline{v}_r(\boldsymbol{p})v_r(\boldsymbol{p})=1\,,\quad \sum_{r=1}^{2}u_r(\boldsymbol{p})\overline{u}_r(\boldsymbol{p})=\frac{\gamma_\lambda p^\lambda+m}{2m}\,,\quad \sum_{r=1}^{2}v_r(\boldsymbol{p})\overline{v}_r(\boldsymbol{p})=\frac{\gamma_\lambda p^\lambda-m}{2m}\,;$$

$$A_{(\lambda)}^{\mu(-)}(x,\boldsymbol{k}) = \frac{1}{(2\pi)^{3/2}\sqrt{2|\boldsymbol{k}|}}\mathrm{e}^{-ik\cdot x}\varepsilon_{(\lambda)}^{\mu}(\boldsymbol{k})\,,\quad A_{(\lambda)}^{\mu(+)}(x,\boldsymbol{k}) = \frac{1}{(2\pi)^{3/2}\sqrt{2|\boldsymbol{k}|}}\mathrm{e}^{ik\cdot x}\varepsilon_{(\lambda)}^{\mu*}(\boldsymbol{k})\,.$$

$$\{b_{r\,\mathrm{in}}(\boldsymbol{p}),b_{r'\,\mathrm{in}}^{+}(\boldsymbol{p}')\}=\delta_{rr'}\delta^3(\boldsymbol{p}-\boldsymbol{p}')\,,\quad \{d_{r\,\mathrm{in}}(\boldsymbol{p}),d_{r'\,\mathrm{in}}^{+}(\boldsymbol{p}')\}=\delta_{rr'}\delta^3(\boldsymbol{p}-\boldsymbol{p}')\,,$$

$$\left[a_{(\lambda)\mathrm{in}}(\boldsymbol{k}),a_{(\lambda')\mathrm{in}}^{+}(\boldsymbol{k}')\right]=-g_{(\lambda)(\lambda')}\delta^3(\boldsymbol{k}-\boldsymbol{k}')\,;\quad \text{others}=0.$$